\documentclass[11pt, reqno]{article}

\pdfoutput=1

\usepackage{graphicx}
\usepackage{jheppub}
\usepackage{amssymb}
\usepackage{amsmath}
\usepackage[usenames,dvipsnames]{xcolor}
\usepackage{epsfig}
\usepackage{dcolumn}
\usepackage{tikz}
\usetikzlibrary{shapes.geometric, arrows}
\usepackage{upgreek}
\usepackage{setspace}
\usepackage{enumitem}
\usepackage{array,multirow,bigdelim,arydshln}
\usepackage{appendix}
\usepackage{xparse}
\usepackage[utf8]{inputenc}
\usepackage{hyperref}
\hypersetup{
	colorlinks,
	urlcolor=Maroon,
	linkcolor=Maroon,
	citecolor=Maroon
}

\usepackage{amsthm}
\usepackage{amsmath}

\theoremstyle{definition}

\theoremstyle{remark}

\usepackage{mathtools}
\usepackage{scalerel}

\usepackage{float}
\restylefloat{table}

\NewDocumentCommand{\binomial}{omm}
{%
	\genfrac(){0pt}{}{#2}{#3}%
	\IfValueT{#1}{_{\!#1}}%
}
\NewDocumentCommand{\eulerian}{omm}
{%
	\genfrac<>{0pt}{}{#2}{#3}%
	\IfValueT{#1}{_{\!#1}}%
}

\usepackage{underscore}

\usepackage{latexsym}
\usepackage{tikz}

\usepackage[percent]{overpic}

\usepackage{multirow} 
\usepackage{slashed}

\usepackage[]{shuffle}

\def\bea#1\eea{\begin{eqnarray}#1\end{eqnarray}}
\def\be#1\ee{\begin{equation}#1\end{equation}}
\def\ba#1\ea{\begin{align}#1\end{align}}

\def \ln {\\\nonumber}
\def \nl {\nonumber\\}

\def\yz#1\yz {{\color{blue} [[YZ: #1]] }}

\def\yzg#1\yzg {{\color{gray} [[YZ: #1]] }}

\def\yzz#1\yzz {{\color{gray} [[To be verified: #1]] }}

\def\ne#1\ne {{\color{green} [[NE: #1]] }}

\usepackage{longtable}

\usepackage{cleveref}

\usepackage{diagbox}

\usepackage{blkarray}

\usepackage{amsmath}

	\def \be {\begin{equation}}
		\def \ee {\end{equation}}
	\def \bg {\begin{gather}}
		\def \eeg {\end{gather}}

	\def \be {\begin{equation}}
		\def \en {\end{equation}}
	\def \bes {\begin{eqnarray}}
		\def \ens {\end{eqnarray}}

	\def \e {\epsilon}
	
	\numberwithin{equation}{section}

\usepackage{orcidlink}

\title{On the New Factorizations of  Yang-Mills Amplitudes}

\author[]{Yong Zhang \!\orcidlink{0000-0002-3522-0885}}
\affiliation[a]{Perimeter Institute for Theoretical Physics, Waterloo, ON N2L 2Y5, Canada.}
\affiliation[b]{School of Physical Science and Technology, Ningbo University, Ningbo 315211, China.}
\emailAdd{yzhang@perimeterinstitute.ca}

\abstract{

In this work, we prove the new factorization pattern for tree-level Yang-Mills (YM) amplitudes proposed in the companion paper \cite{Zhang:2024iun}. This pattern reveals a decomposition of amplitudes into a sum of gluings of lower-point amplitudes under specific kinematic constraints, making the hidden zeros of YM amplitudes manifest. Utilizing the Cachazo-He-Yuan (CHY) formalism, we rigorously derive these factorizations by systematically analyzing the contributions of singular solutions to the scattering equations. Through the identification and application of key algebraic identities, we demonstrate how cancellations among terms uncover a recursive structure intricately tied to the hidden zeros. This work not only conclusively validates the proposed factorization but also provides new insights into the geometric and algebraic organization of YM amplitudes within the CHY framework.

}

\begin{document}
	\maketitle
	\addtocontents{toc}{\protect\setcounter{tocdepth}{1}}
		
\allowdisplaybreaks[1]
\section{Introduction}

Scattering amplitudes are central to quantum field theory, encoding the probabilities of particle interactions while reflecting key principles like locality, unitarity, and gauge invariance. Beyond their physical interpretations, amplitudes exhibit a rich mathematical structure, with factorization as a cornerstone property. In specific kinematic regimes, amplitudes decompose into products of lower-point amplitudes, corresponding to the exchange of on-shell particles. This behavior underpins many foundational tools in amplitude computations and provides deep insights into their analytic and recursive structures.

Factorization manifests in forms such as collinear limits, where two external particles become collinear, producing universal splitting functions critical to perturbative QCD and precision collider physics \cite{Altarelli:1977zs, Catani:1998nv}. These limits also appear in celestial amplitudes, revealing holographic features tied to conformal symmetry \cite{Strominger:2017zoo, Pasterski:2021raf}. Similarly, soft theorems describe amplitude behavior as external momenta vanish, connecting higher-point amplitudes to their lower-point counterparts through a summation over multiple two-particle factorization channels. This relationship highlights their nature as an extension of traditional factorization principles, offering insights into infrared divergences and asymptotic symmetries \cite{Weinberg:1965nx, Strominger:2014pwa}.  Together, these forms of factorization—collinear limits, soft theorems, and traditional residues—have significantly advanced the computation and interpretation of scattering amplitudes.

Recent developments have extended these classical ideas, uncovering novel splitting patterns that go beyond traditional factorization. The 2-split framework has emerged as a unifying perspective, demonstrating how amplitudes factorize into two currents under specific kinematic constraints \cite{Cao:2024gln, Cao:2024qpp}. Initially studied in scalar theories like the biadjoint \(\phi^3\) model,  the Nonlinear Sigma Model (NLSM), and a special Galileon (sGal) theory, the 2-split has been extended to gauge and gravity theories, including Yang-Mills (YM) and General Relativity (GR). It has also found applications in string theory, providing a bridge between field-theoretic amplitudes and their string-theoretic counterparts. This versatility underscores the 2-split framework's utility in exploring amplitude structures across diverse contexts.

Before the 2-split, two distinct types of three-part splittings were proposed, laying the foundation for modern splitting patterns. The first, introduced in \cite{Cachazo:2021wsz}, employs semi-locality to smoothly split amplitudes into three currents without computing residues, providing insights into the analytic structure of scalar amplitudes, including 
$\phi^3$
  and  $\phi^p$ theories \cite{GimenezUmbert:2024jjn,GimenezUmbert:2025ech} \footnote{See their generalization to generalized biadjoint amplitudes in \cite{GimenezUmbert:2025ech}.}. The second, presented in \cite{Arkani-Hamed:2023swr}, focuses on rectangular configurations of Mandelstam variables, yielding three-part decompositions near specific zeros. These splitting patterns have motivated further studies in both scalar and string theories, linking amplitude structures to moduli-space integrals and recursive properties \cite{Mafra:2011nw, Arkani-Hamed:2024fyd}.

The curve integral formalism provides another complementary framework, recasting amplitudes as integrals over moduli spaces of algebraic curves. Initially developed for scalar theories such as \(\text{Tr}(\phi^3)\) \cite{Arkani-Hamed:2023swr}, this approach extends to YM and GR amplitudes via the scaffolding procedure, systematically reducing integrals while preserving recursive structures \cite{Arkani-Hamed:2023jry}. This method connects amplitude geometry to analytic properties, offering deep insights into loop-level extensions and string-theoretic formulations \cite{
Arkani-Hamed:2023lbd,  Arkani-Hamed:2023mvg,Arkani-Hamed:2024nhp, Arkani-Hamed:2024yvu,  Arkani-Hamed:2024tzl}.

At the core of these developments lies the concept of hidden zeros—specific kinematic constraints where amplitudes vanish. Initially identified in scalar theories, hidden zeros provided the foundation for the 2-split and curve integral frameworks, linking kinematics to amplitude structures. Their extension to YM \cite{Arkani-Hamed:2023swr} and GR amplitudes  demonstrates their universality as fundamental features encoding algebraic and geometric principles. Recent studies, including bootstrap techniques \cite{Rodina:2024yfc} \footnote{See recent progress at loop level in \cite{Backus:2025hpn}.}, Feynman diagrammatic analyses  \cite{Zhou:2024ddy,Huang:2025blb}, and double copy relations \cite{Kawai:1985xq, Bern:2008qj,Bartsch:2024amu, Li:2024qfp}, have further enriched our understanding of hidden zeros as both local cancellations and global structural features of amplitudes. These findings have established hidden zeros as a fundamental element in unraveling the recursive and algebraic foundations of scattering amplitudes.

In this paper, we study a new type of factorization for YM amplitudes, originally proposed in the companion paper \cite{Zhang:2024iun}. This factorization arises under specific kinematic constraints, where certain two-point Mandelstam variables in a rectangular configuration are set to zero. Under these conditions, YM amplitudes decompose into sums of gluings of lower-point amplitudes, with contributions from gluon pairs tied to the vanishing Mandelstam variables. This structure extends the hidden zeros conjectured in \cite{Arkani-Hamed:2023swr}, embedding them into a broader framework of amplitude decomposition. Unlike traditional factorizations based on residues at poles, this approach uncovers deeper algebraic and geometric structures while making the hidden zeros manifest.

We rigorously establish the new factorization using the Cachazo-He-Yuan (CHY) formalism \cite{Cachazo:2013hca, Cachazo:2013iea}, which encodes amplitudes as integrals over moduli spaces of punctured Riemann spheres. By analyzing the singular solutions of the scattering equations, we validate the recursive structures tied to the hidden zeros and reinterpret these zeros as natural outcomes of the structure of the moduli-space integral. This perspective reveals a striking connection between singularities and symmetries within the CHY framework. The new factorization formula not only deepens our understanding of YM amplitudes but also bridges hidden zeros with recursive amplitude construction, providing a unified framework for amplitude decomposition under kinematic constraints.

This paper is structured as follows. In \cref{sec:sec2}, we review the novel factorization patterns for YM amplitudes proposed in the companion paper and introduce the relevant CHY formalism, laying the foundation for the current study. Section \ref{sec:sec3} explores the singular solutions of the scattering equations that play a pivotal role in deriving these factorization patterns. Section \ref{sec:sec4} focuses on the induced lower-point scattering equations, which are essential for understanding the recursive structure of the amplitudes under specific constraints. In Sections \ref{sec:sec5} through \ref{sec:sec7}, we analyze progressively complex cases of turning on one, two, or three entries in the kinematic configurations that are not aligned within the same row or column, demonstrating how the factorization patterns emerge in these settings. Section \ref{sec:sec8} generalizes the analysis to arbitrary configurations, extending the factorization results to a broader class of scenarios. Finally, \cref{sec:sec9} provides a discussion of the implications of our findings and potential directions for future research. Additional technical details, including coefficients in BCJ relations, the proof of key identities, and explicit examples, are provided in appendices \ref{BCJco}, \ref{appb22}, and \ref{appproof}.

\section{Review of the New Factorization and the CHY Formulas \label{sec:sec2}}

In this section, we review the new factorization patterns for YM amplitudes introduced in \cite{Zhang:2024iun} and the CHY formalism, which provides the analytical framework for these patterns.

\subsection{Review of the New Factorization}

We begin by considering a matrix of Mandelstam variables, \( s_{ab} := k_a \cdot k_b \), denoted as \( h_m \), for \( 1 \leq m \leq n-3 \). This matrix is defined as 
\begin{align}
    h_m:=\begin{Bmatrix} 
    \hspace{-0.0cm} s_{1\,m+2}\, , \,s_{1\,m+3}\, , \ldots , s_{1\,n-1}\,\, , \\
    \hspace{-0.0cm} s_{2\,m+2}\, , \,s_{2\,m+3}\, , \ldots ,  s_{2\,n-1}\,\, , \\
    \hspace{-0.5cm}
    \vdots \hspace{1.1cm} \vdots \hspace{1.7cm} \vdots\\
    \hspace{-0.3cm}\hspace{0.3cm}s_{m\,m+2} , s_{m\,m+3} , \ldots , s_{m\,n-1}  ~~
    \end{Bmatrix}\,.
    \label{eq:constrhm}
\end{align}

As proposed in \cite{Zhang:2024iun}, the \( n \)-point YM amplitudes factorize into sums of gluings of a three-point and an \((n-1)\)-point YM amplitude in the subspace where \( h_m = 0 \),
\ba
\label{totalcase}
&A({\mathbb I}_n)\! \xrightarrow{h_m=0}\!\! \!\!
\sum_{\substack{1\leq i \leq m <
m+2 \leq j \leq n-1}}
\!\!\!
F_{m,n}(i,j)\,,
\ea
where the gluon pair contribution \( F_{m,n}(i,j) \) is given by
\footnote{The relative sign between YM amplitudes for different multiplicities can be adjusted by the definition of the YM amplitudes. In this paper as well as the companion paper \cite{Zhang:2024iun}, we fix these relative signs by defining the YM amplitudes using the CHY integral \eqref{chyintegral} with the integrand specified in \eqref{YMintegranddef}, following the conventions established in \cite{Cachazo:2013hca}.}
\ba
&F_{m,n}(i,j)\!=\!
\frac{(-1)^{n+1}}{s_{12\ldots m+1}} 
\!\sum_{\e_{\hat j}}\!
A(ij -\!\!\hat j)
\!\!\! \sum_{\rho \in S_{m-1}}  
\!
\!
\!
X(s,\rho)
  A\big(\rho(12\ldots i\!-\!1\, i\!+\! 1 \ldots m)\, m{+}1\,m{+}2 \ldots { \hat j} \ldots n\big)\,.
\label{generalFn}
\ea
Here the first summation includes all possible polarization states of the internal gluon \(\hat{j}\), whose momentum conservation requires $
   k_{\hat{j}} = -k_{-\hat{j}} = k_i + k_j,
   $
   with $ s_{ij} \in h_m $ and hence $ s_{ij} = 0 $.
 The second summation is over all permutations \( \rho \in S_{m-1} \) of the gluons \( 1, 2, \ldots, i-1, i+1, \ldots, m \).  Throughout this paper, \( A(\ldots) \) without a superscript refers specifically to a YM amplitude with a particular ordering \( A^{\rm YM}(\ldots) \), as this notation is frequently employed.

The ratios of Mandelstam variables associated with the gluons \( 1, 2, \ldots, m+1 \) are captured by the term \( X(s, \rho) \). As illustrative examples,
\ba
\label{jkfha3;pre}
X(s, \emptyset) = 1 \quad \text{(for \( m=1, i=1 \))}, \quad
X(s, 2) = \frac{s_{12} + s_{23}}{s_{12}} \quad \text{(for \( m=2, i=1 \))}.
\ea 
For general \( \rho \), \( X(s, \rho) \) can be expressed in terms of a sum over shuffles of subsets,
\ba 
\label{Xsrhoori}
X(s, \rho) = s_{12\ldots m{+}1} \sum_{\pi \in \{ 12 \ldots i{-}1 \} \shuffle \{ i{+1} \ldots m \}} g^{-1}[\rho, \pi] \, {\cal B}_{m+1,i}[12 \ldots i{-}1, i{+1} \ldots m | \pi],
\ea 
where \( g^{-1}[\rho, \pi] \) is the matrix inverse of \( g[\rho, \pi] \), defined as
\ba 
\label{eq25g}
g[\rho, \pi] := \sum_{\substack{a=1 \\ a \neq i}}^{m+1} 
\left( \sum_{b \in Y(a,\rho)} s_{ib} +s_{i\,m+1}  \right) 
{\cal B}_{m+1,i}[Y(\rho, a), Y(a, \rho) | \pi].
\ea 
Here, \( Y(\rho, a) \) and \( Y(a, \rho) \) divide the sequence \( \rho \) into parts relative to \( a \). Specifically, \( Y(\rho, a) \) comprises the elements of \( \rho \) preceding \( a \), while \( Y(a, \rho) \) includes \( a \) followed by the remaining elements of \( \rho \). For example, if \( \rho = 1245 \) and \( a = 4 \), then \( Y(\rho, a) = 12 \) and \( Y(a, \rho) = 45 \). For \( a = m+1 \), the definitions simplify to \( Y(\rho, m+1) := \rho \) and \( Y(m+1, \rho) := \emptyset \). The 
$\shuffle$ operation represents a shuffle product, which generates all possible ways to merge two sequences into a single sequence while preserving the relative order of elements within each original sequence.

The coefficients \( {\cal B}_{m+1,i}[\alpha, \beta | \pi] \), appearing in \eqref{Xsrhoori} and \eqref{eq25g}, are derived from the BCJ relations for YM amplitudes \cite{Bern:2019prr, Bern:2008qj}. These coefficients satisfy the relation
\ba  
\label{BCJYMrelations}
A(\alpha \, i \, \beta \, m{+}1 \, \hat n) = \sum_{\pi \in \alpha \shuffle \beta} A(i \, \pi \, m{+}1 \, \hat n) {\cal B}_{m+1,i}[\alpha, \beta | \pi],
\ea 
where \( k_{\hat n} = -k_1 - k_2 - \ldots - k_{m+1} \), determined by momentum conservation, ensures gauge-invariant YM amplitudes when \( k_{\hat n} \) is on-shell. For our purposes, however, we relax this on-shell requirement for \( k_{\hat n} \), while keeping \( {\cal B}_{m+1,i}[\alpha, \beta | \pi] \) independent of \( k_{\hat n} \) and unaffected. Detailed expressions for \( {\cal B}_{m+1,i}[\alpha, \beta | \pi] \) are provided in \cref{BCJco}.

\subsection{ Review of CHY Formulas}

The CHY formulas provide a powerful framework for expressing massless tree-level scattering amplitudes as integrals over the moduli space of punctured Riemann spheres \cite{Cachazo:2013hca, Cachazo:2013iea}. These amplitudes are written as
\ba
\label{chyintegral}
A^{\rm theory} = \int \frac{\prod_{a=1}^n d\sigma_a }{{\rm vol}{\rm SL}(2,{\mathbb C})} \prod_{a=1}^n {}' \delta(E_a) I_n^L I_n^R,
\ea
where the integration is localized by the universal scattering equations \cite{Cachazo:2013iaa},
\ba
\label{SE}
E_a := \sum_{\substack{b=1 \\ b \neq a}}^n \frac{s_{ab}}{\sigma_{ab}} = 0, \quad ~ {\rm with}~ \sigma_{ab} := \sigma_a - \sigma_b,  \quad~{\rm for}~ 1 \leq a \leq n.
\ea
Here, \( \sigma_a \) represent the puncture positions on the sphere, while \( I_n^L \) and \( I_n^R \) are the theory-specific integrands. The SL(2, \(\mathbb{C}\)) invariance is eliminated by fixing three punctures, accompanied by appropriate compensating factors. Similarly, three scattering equations are excluded from the integral \eqref{chyintegral}, with corresponding compensating factors included. 
The CHY integrands for different theories are given as follows
\ba
&m(\mathbb{I}_n|\mathbb{I}_n): \quad \textbf{PT}(\mathbb{I}_n) \, \textbf{PT}(\mathbb{I}_n), 
\nl
\label{YMintegranddef}
&A^{\rm YM}(\mathbb{I}_n): \quad \textbf{PT}(\mathbb{I}_n) \, {\rm Pf}'{\bf \Psi}(\{\epsilon_i, k_i, \sigma_i\}), \ln
&M^{\rm GR}_n: \quad {\rm Pf}'{\bf \Psi}(\{\epsilon_i, k_i, \sigma_i\}) \, {\rm Pf}'{\tilde{{\bf \Psi}}}(\{\tilde{\epsilon}_i, k_i, \sigma_i\}).
\ea
The Parke-Taylor (PT) factor \( \textbf{PT}(\mathbb{I}_n) \) is defined as 
\ba
\textbf{PT}(\mathbb{I}_n) := \frac{1}{\sigma_{12} \sigma_{23} \ldots \sigma_{n1}}.
\ea
The reduced Pfaffian \( {\rm Pf}'{\bf \Psi} \) is constructed from the \( 2n \times 2n \) antisymmetric matrix \({\bf \Psi}\), which encodes the kinematic and polarization data. The matrix \({\bf \Psi}\) has the block structure,
\ba
{\bf \Psi} \!=\! 
\begin{pmatrix}
A & C \\
-C^T & B
\end{pmatrix},
A_{ab} \!=\! 
\begin{cases}
\frac{k_a \cdot k_b}{\sigma_{ab}}, & a \neq b \\
0, & a = b
\end{cases} ,
B_{ab} \!=\! 
\begin{cases}
\frac{\epsilon_a \cdot \epsilon_b}{\sigma_{ab}}, & a \neq b \\
0, & a = b
\end{cases} ,
C_{ab} \!=\!
\begin{cases}
\frac{\epsilon_a \cdot k_b}{\sigma_{ab}}, & a \neq b \\
-\sum_{b' \neq a} \frac{\epsilon_a \cdot k_{b'}}{\sigma_{ab'}}, & a = b
\end{cases}\,.
\ea
For convenience, the columns and rows of \({\bf \Psi}\) correspond to the ordering of \( k_1, k_2, \ldots, k_n, \epsilon_1\), \(\epsilon_2, \ldots, \epsilon_n \), respectively.
While the Pfaffian of $\bf\Psi$ is zero, removing rows $i, j$ and columns $i, j$ with $1 \leq i<j \leq n$ gives rise to a new matrix which we denote as  ${\bf \Psi}_{i \!\!\! \slash  j \!\!\! \slash ;}$ with nonzero Pfaffian. The reduced Pfaffian is defined as ${\rm Pf}^{\prime} {\bf\Psi} := 2 \frac{(-1)^{i+j}}{\sigma_{ij}} {\rm Pf}{\bf \Psi}_{i \!\!\! \slash  j \!\!\! \slash ;} $ which is independent of the choice of $i,j$. CHY integrands for many  other theories can be constructed through simple operations on these integrands (c.f. \cite{Cachazo:2014nsa, Cachazo:2014xea, Azevedo:2017lkz, He:2016iqi,Cachazo:2016njl,Dong:2024klq}). 

The scattering equations \eqref{SE} admit \((n-3)!\) distinct solutions, and the integral in \eqref{chyintegral} reduces to a summation over these solutions,
\ba
A^{\rm theory} = \sum_{\sigma \in {\rm sol}} \frac{I_n^L I_n^R}{{\rm det}'\Phi},
\ea
where the Jacobian \( {\rm det}'\Phi \) arises from the scattering equations. 
The \( n \times n \) matrix \(\Phi\) is defined as 
\ba
\label{PhiPhi}
\Phi_{ab} =
\begin{cases}
\frac{s_{a b}}{\sigma_{ab}^2}, & a \neq b \\
-\sum_{c \neq a} \frac{s_{ac}}{\sigma_{ac}^2}, & a = b
\end{cases}\,.
\ea
To compute \( {\rm det}'\Phi \), we exclude rows corresponding to fixed punctures \( \{i, j, k\} \) and columns corresponding to their associated \( \{\sigma_p, \sigma_q, \sigma_r\} \). The reduced determinant is then given by 
\ba
\operatorname{det}^{\prime} \Phi := \frac{|\Phi|_{pqr}^{ijk}}{(\sigma_{pq} \sigma_{qr} \sigma_{rp})(\sigma_{ij} \sigma_{jk} \sigma_{ki})}\,,
\ea
where \( |\Phi|_{pqr}^{ijk} \) represents the minor obtained after the removal of these rows and columns.

The CHY formalism elegantly encapsulates tree-level amplitudes for biadjoint \(\phi^3\) theory, YM theory, and gravity, providing a unifying structure and computational tool for scattering amplitudes.

\section{Singular Solutions and Universal Proof for the Hidden Zeros\label{sec:sec3}}

We now examine the structure and consequences of singular solutions to the scattering equations under the condition of vanishing Mandelstam variables. Let us parametrize all vanishing Mandelstam variables as 
\ba \label{paras} 
s_{ab} = \tau \hat{s}_{ab}, \quad {\rm with}~ \tau \to 0, \quad {\forall}\, 1 \leq a \leq m, ~ m+2 \leq b \leq n-1. 
\ea

\subsection{Absence of Regular Solutions}

Suppose there exists a regular solution to the scattering equations \eqref{SE}. By dropping all terms at order \(\tau\), the scattering equations simplify to two distinct sets. The first set, involving \(a = 1, 2, \ldots, m\), becomes
\ba
\label{simSE}
\mathring{E}_a = \sum_{\substack{a'=1 \\ a' \neq a}}^{m+1} \frac{s_{aa'}}{\sigma_{aa'}} + \frac{s_{an}}{\sigma_{an}} = 0.
\ea
The second set, involving \(b = m+1, m+2, \ldots, n-1\), is given by
\ba
\label{simSE2}
\mathring{E}_b = \sum_{\substack{b'=m+1 \\ b' \neq b}}^{n-1} \frac{s_{bb'}}{\sigma_{bb'}} + \frac{s_{bn}}{\sigma_{bn}} = 0.
\ea
Here, \(\mathring{E}_c\) denotes the leading-order term of \(E_c\). The equations for \(E_{m+1}\) and \(E_n\) remain unsimplified at this stage.

The simplified scattering equations \(\{\mathring{E}_a\}\) and \(\{\mathring{E}_b\}\) involve \(m+2\) and \(n-m+2\) variables, respectively. Specifically, the variables in the first set are \(\{\sigma_n, \sigma_1, \sigma_2, \ldots, \sigma_{m+1}\}\), while the second set involves \(\{\sigma_{m+1}, \sigma_{m+2}, \ldots, \sigma_n\}\). Importantly, there are no linear dependencies among the \(m\) equations \(\mathring{E}_a\) in the first set or the \(n-m-2\) equations \(\mathring{E}_b\) in the second set.

To proceed, we use the SL(2) gauge freedom to fix three punctures. There are two natural choices for this gauge fixing. The first option is to fix three punctures within one set, for example, setting \(\sigma_1 \to \sigma_1^\star\), \(\sigma_2 \to \sigma_2^\star\), and \(\sigma_3 \to \sigma_3^\star\). The second option is to fix two punctures in one set and one in the other, for example, setting \(\sigma_1 \to \sigma_1^\star\), \(\sigma_2 \to \sigma_2^\star\), and \(\sigma_{m+2} \to \sigma_{m+2}^\star\).

In the first case, where three punctures are fixed within one set, only \(m-1\) free variables remain to solve the \(m\) independent equations \(\mathring{E}_a\) in the first set. This makes it impossible to find a solution. In the second case, where two punctures are fixed in one set and one in the other, all \(m\) independent equations \(\mathring{E}_a\) can be solved, as \(m\) free variables remain. However, since \(\sigma_{m+1}\) and \(\sigma_n\) are already determined, only \(n-m-1\) free variables remain for the \(n-m\) independent equations \(\mathring{E}_b\) in the second set. This again leads to no solutions.

Thus, we have demonstrated that there are no regular solutions to the scattering equations \eqref{simSE} and \eqref{simSE2} under any choice of gauge fixing. As a result, all \((n-3)!\) solutions of the original scattering equations \eqref{SE} are singular.

\subsection{Singular Solutions}

Singular solutions contribute to the scattering equations \( \mathring{E}_a \) (with \( a = 1, 2, \ldots, m \)) at leading order. In addition to the normal terms given in \eqref{simSE}, there may be at most one additional term, \( \frac{\tau \hat{s}_{ab}}{\sigma_{ab}} \), in \( \mathring{E}_a \), where \( b \in \{m+2, \ldots, n-1\} \), that could have a non-vanishing contribution. If more than one such term, e.g., \( \frac{\tau \hat{s}_{ab}}{\sigma_{ab}} \) and \( \frac{\tau \hat{s}_{ab'}}{\sigma_{ab'}} \) with \( b \neq b' \), were to contribute, then three punctures \( \sigma_a, \sigma_b, \sigma_{b'} \) would have to pinch simultaneously at \(\mathcal{O}(\tau)\). However, such a scenario would require \( s_{abb'} \sim \mathcal{O}(\tau) \), which is inconsistent with the parameterization \cite{Cachazo:2013gna,Cachazo:2013hca,Cachazo:2013iea}.

This implies that \(\sigma_a\) and \(\sigma_b\) can pinch at \(\mathcal{O}(\tau)\) but not with additional punctures. More generally, we claim that in each singular solution, up to \(\min(m, n-m-2)\) pairs of punctures may pinch. These solutions take the form,
\ba 
\label{solstr}
\sigma_{a_1b_1} & \sim \mathcal{O} (\tau), \quad 
\sigma_{a_2b_2} \sim \mathcal{O} (\tau), \quad 
\ldots, \quad  
\sigma_{a_rb_r} \sim \mathcal{O} (\tau), \ln
& {\rm with}~ 1 \leq r \leq \min(m, n-m-2), \quad 
a_1 \neq a_2 \neq \ldots \neq a_r \in \{1, 2, \ldots, m\}, \ln
& b_1 \neq b_2 \neq \ldots \neq b_r \in \{m+2, m+3, \ldots, n-1\}.
\ea 
Here, \( a_1, \ldots, a_r \) and \( b_1, \ldots, b_r \) are all distinct.

The corresponding scattering equations for these singular solutions are,
\ba 
\label{LOSE}
& E_{a_t}  \to \sum_{\substack{a'=1 \\ a' \neq a_t}}^{m+1} \frac{s_{a_ta'}}{\sigma_{a_ta'}} + \frac{s_{a_tn}}{\sigma_{a_tn}} + \frac{s_{a_tb_t}}{\sigma_{a_tb_t}} = 0, \quad {\rm for}~ a_t \in \{a_1, a_2, \ldots, a_r\}, \\
& E_a  \to \sum_{\substack{a'=1 \\ a' \neq a}}^{m+1} \frac{s_{aa'}}{\sigma_{aa'}} + \frac{s_{an}}{\sigma_{an}} = 0, \quad {\rm for}~ a \in \{1, 2, \ldots, m\} \setminus \{a_1, a_2, \ldots, a_r\}, \\
& E_{b_t}  \to \sum_{\substack{b'=m+1 \\ b' \neq b_t}}^{n-1} \frac{s_{b_tb'}}{\sigma_{b_tb'}} + \frac{s_{b_tn}}{\sigma_{b_tn}} - \frac{s_{a_tb_t}}{\sigma_{a_tb_t}} = 0, \quad {\rm for}~ b_t \in \{b_1, b_2, \ldots, b_r\}, \\
& E_b  \to \sum_{\substack{b'=m+1 \\ b' \neq b}}^{n-1} \frac{s_{bb'}}{\sigma_{bb'}} + \frac{s_{bn}}{\sigma_{bn}} = 0, \quad {\rm for}~ b \in \{m+2, m+3, \ldots, n-1\} \setminus \{b_1, b_2, \ldots, b_r\}.
\ea 
The remaining scattering equations, \( E_{m+1} = 0 \) and \( E_n = 0 \), retain all terms and are not simplified.

From this analysis, we classify singular solutions based on the number of pinched pairs. A solution where \( r \) pairs of punctures pinch, as described above, is called an \( r \)-{\it {pinch solution}}. In particular, solutions with \( r \geq 2 \) are referred to as {\it {multiple-pinch solutions}}. This categorization will be central to understanding how singular solutions contribute to the CHY integrals and their role in the emergence of hidden zeros.

\subsection{Proof for the Hidden Zeros of Amplitudes}

The hidden zeros of amplitudes in various theories were first studied in \cite{Arkani-Hamed:2023swr}. The central idea is that when all Mandelstam variables \( s_{ab} = k_a \cdot k_b \) in a rectangular configuration with dimensions \( m \) and \( n-m-2 \) are set to zero for \( 1 \leq m \leq n-3 \), along with the corresponding Lorentz invariants such as \( \epsilon_a \cdot k_b \), \( k_a \cdot \epsilon_b \), and \( \epsilon_a \cdot \epsilon_b \), certain amplitudes vanish. This behavior is observed across biadjoint \(\phi^3\), NLSM, sGal, YM, and GR theories.

To formalize this, define \( x_{ij} := \{ \epsilon_i \cdot \epsilon_j, \epsilon_i \cdot k_j, k_i \cdot \epsilon_j \} \) and collect these terms into
\begin{align}
\label{eq:constrhatHm}
  \hat{H}_m := \begin{Bmatrix} 
    x_{1\,m+2}, x_{1\,m+3}, \ldots, x_{1\,n-1}, \\
    x_{2\,m+2}, x_{2\,m+3}, \ldots, x_{2\,n-1}, \\
    \vdots \hspace{1.1cm} \vdots \hspace{1.7cm} \vdots \\
    x_{m\,m+2}, x_{m\,m+3}, \ldots, x_{m\,n-1}
  \end{Bmatrix}.
\end{align}

Similarly, define \( \hat{\tilde{H}}_m := \hat{H}_m |_{\epsilon \to \tilde{\epsilon}} \). Then, the following statements hold (see the CHY integrands for  NLSM and sGal theories in \cite{Cachazo:2014xea} \footnote{A generalized version of NLSM has recently been proposed in \cite{Early:2025ivr} and it would be interesting to study the zero of its amplitudes using Cachazo-Early-Guevara-Mizera formula \cite{Cachazo:2019ngv} analogously.})
\ba
& \text{Setting } h_m = 0 \quad \Rightarrow \quad 
m^{\phi^3}(\mathbb{I}_n | \mathbb{I}_n) = 
m^{\rm NLSM}(\mathbb{I}_n) = 
m^{\rm sGal}_n = 0, 
\label{scalarconre} \\
& \text{Setting } h_m \sqcup \hat{H}_m = 0 \quad \Rightarrow \quad 
A^{\rm YM}(\mathbb{I}_n) = 0, 
\label{YMconre} \\
& \text{Setting } h_m \sqcup \hat{H}_m \sqcup \hat{\tilde{H}}_m = 0 \quad \Rightarrow \quad 
M^{\rm GR}_n = 0. 
\label{GRconre}
\ea

Here we give the proof of vanishing amplitudes. For any singular solution of the form \eqref{solstr}, we observe,
\ba
\label{tauorder}
\frac{1}{\sigma_{a_ib_i}} \sim \mathcal{O}(\tau^{-1}), \quad 
\frac{s_{a_ib_i}}{\sigma_{a_ib_i}} \sim \mathcal{O}(\tau^{0}), \quad 
\frac{s_{a_ib_i}}{\sigma_{a_ib_i}^2} \sim \mathcal{O}(\tau^{-1}), \quad 
\forall i = 1, 2, \ldots, r.
\ea
Now consider the following blocks. 
\begin{enumerate}
    \item  Determinant of the \( A \)-matrix: All elements in the \( A \)-matrix remain finite, so \( \det'A \sim \mathcal{O}(\tau^0) \).

  \item  Reduced Pfaffian of the \( \bf\Psi \)-matrix: New elements such as \( \frac{k_{a_i} \cdot \epsilon_{b_i}}{\sigma_{a_ib_i}} \) arise, but terms like \( k_{a_i} \cdot \epsilon_{b_i}, \epsilon_{a_i} \cdot \epsilon_{b_i}, \epsilon_{a_i} \cdot k_{b_i} \) are zeros under the assumption $\hat H_m=0$. Therefore, all elements in \( \bf\Psi \) are finite or zero, and \( {\rm Pf}'\bf\Psi \sim \mathcal{O}(\tau^0) \). The same holds for \( {\rm Pf}'\tilde{\bf\Psi} \).

  \item  Parke-Taylor factor: The canonical Parke-Taylor factor \( {\rm PT}(\mathbb{I}_n) \sim \mathcal{O}(\tau^0) \).

  \item  Jacobian of the scattering equations:  There are divergent elements $\frac{s_{a_ib_i}}{\sigma_{a_ib_i}^2}$ in the matrix  $\Phi$ given in \eqref{PhiPhi}, and hence the universal Jacobian \( \det'\Phi \) is at order \( \mathcal{O}(\tau^{-r}) \), reflecting the contributions of singular solutions.
\end{enumerate}

Combining these results, the contribution of an \( r \)-pinch singular solution to the amplitudes \( m^{\phi^3}(\mathbb{I}_n | \mathbb{I}_n), m^{\rm NLSM}(\mathbb{I}_n), m^{\rm sGal}_n, A^{\rm YM}(\mathbb{I}_n), M^{\rm GR}_n \) is at order \( \mathcal{O}(\tau^r) \) where \( \tau \) is an infinitesimal variable.  Summing over all singular solutions, the total amplitudes vanish at \( \mathcal{O}(\tau) \) under their corresponding assumptions, proving \eqref{scalarconre}, \eqref{YMconre}, and \eqref{GRconre}.

This derivation extends naturally to many other theories. The universality of the vanishing behavior under these constraints highlights deep structural properties of scattering amplitudes in diverse physical theories.

\subsection{Preliminary Analysis for General Cases}

In the preceding discussion, we proved the vanishing of YM  amplitudes when all entries of \(\hat{H}_m\) are set to zero in addition to $h_m=0$. Now, we consider the scenario where \(h_m = 0\), but the entries of \(\hat{H}_m\) are kept general. In this case, let us examine singular solutions where only \(r\) pairs of punctures pinch. According to \eqref{tauorder}, the Jacobian from the scattering equations remains at order \(\mathcal{O}(\tau^{-r})\), while the PT factor remains at order \(\mathcal{O}(\tau^{0})\). However, for the reduced Pfaffian, new elements like \(\frac{k_{a_i} \cdot \epsilon_{b_i}}{\sigma_{a_ib_i}}\) appear in the \(\Psi\)-matrix, scaling as \(\mathcal{O}(\tau^{-1})\). Consequently, the full reduced Pfaffian \({\rm Pf}'\Psi\) now scales as \(\mathcal{O}(\tau^{-r})\). This means that all singular solutions contribute to the CHY integral, regardless of the number of pinched pairs.

To analyze more explicitly, consider turning on a single entry \(x_{ij} \in \hat{H}_m\). In this case, only 1-pinch solutions, where \(\sigma_{ij} \sim 0\), can contribute non-vanishing terms to the amplitude in the CHY integral. When all entries \(x_{ij}\) in a single row or column of \(\hat{H}_m\) are turned on, the contributions still come exclusively from 1-pinch solutions, though these contributions must now be summed over different sets of 1-pinch solutions.

When two or more entries of \(\hat{H}_m\) that do not belong to the same row or column are turned on, both 1-pinch and multiple-pinch solutions can contribute non-vanishing terms to the amplitudes. This scenario introduces significant complexity, requiring more sophisticated combinatorial tools for analysis.

Interestingly, as we will see, the contributions of \(r\)-pinch solutions can be organized as a linear combination of CHY integrals with \((n-r)\) or fewer points. Specifically, 1-pinch solutions yield \((n-1)\)-point CHY integrals, producing \((n-1)\)-point YM amplitudes, along with lower-point CHY integrals. However, when contributions from 1-pinch and multiple-pinch solutions are summed together, all \((n-2)\)- and lower-point CHY integrals cancel out, leaving only \((n-1)\)-point YM amplitudes. These remaining contributions ultimately compose the formula in \eqref{totalcase}.

A crucial aspect of reducing \(n\)-point YM amplitudes to lower-point ones lies in the induced lower-point scattering equations associated with each singular solution. In the next section, we will delve deeper into this process. On the support of these lower-point scattering equations, we obtain new reduced Pfaffians and PT factors. These induced quantities depend only on fewer punctures, Mandelstam variables and Lorentz products involving polarizations, making the cancellations between contributions from different solutions more transparent.

We will begin with a simple case in \cref{sec:sec5}, where only one entry of \(\hat{H}_m\) is turned on, and then proceed to more complex cases in subsequent sections.

\section{Induced Lower-Point Scattering Equations \label{sec:sec4}}

For \( r \)-pinch singular solutions, at leading order, all \( a_t \) can be effectively replaced by \( b_t \) in \eqref{LOSE}, except for the term \( \frac{s_{a_t b_t}}{\sigma_{a_t b_t}} \). For scattering equations that do not contain this term, substituting \(\sigma_{a_t} \to \sigma_{b_t}\) leads to the following,
\begin{align}
\label{LOSEr1}
E_a \to \mathring{E}_a = \sum_{\substack{a' = 1 \\ a' \neq a_1, a_2, \ldots, a_r}}^{n} \frac{k_a \cdot k_{a'}}{\sigma_{a a'}} \Bigg|_{\substack{k_{b_t} \to k_{\hat{b}_t} = k_{a_t} + k_{b_t} = 0 \\ \forall b_t \in \{b_1, \ldots, b_r\}}}, \quad \forall a \in \{1, 2, \ldots, n\}\setminus\{a_1, b_1, \ldots, a_r, b_r\}.
\end{align}
For the pairs of scattering equations containing \( \frac{s_{a_t b_t}}{\sigma_{a_t b_t}} \), we have,
\begin{align}
\label{LOSEr11}
& E_{a_t} \to \mathring{E}'_{a_t} = 
\underbrace{ \sum_{\substack{a' = 1 \\ a' \neq a_t}}^{m+1} \frac{s_{a_t a'}}{\sigma_{b_t a'}} + \frac{s_{a_t n}}{\sigma_{b_t n}}  }_{:=\mathring{E}_{a_t}}+ \frac{s_{a_t b_t}}{\sigma_{a_t b_t}} = 0, \\
& E_{b_t} \to \mathring{E}'_{b_t} = \sum_{\substack{b' = m+1 \\ b' \neq b_t}}^{n-1} \frac{s_{b_t b'}}{\sigma_{b_t b'}} + \frac{s_{b_t n}}{\sigma_{b_t n}} - \frac{s_{a_t b_t}}{\sigma_{a_t b_t}} = 0.
\end{align}
Adding these two equations removes the term \( \frac{s_{a_t b_t}}{\sigma_{a_t b_t}} \). The resulting equation can be reorganized as
\begin{align}
\label{LOSEr2}
\mathring{E}_{\hat{b}_t} := 
\mathring{E}'_{a_t} + \mathring{E}'_{b_t} = \sum_{\substack{b' = 1 \\ b' \neq a_1, \ldots, a_r}}^{n} \frac{k_{\hat{b}_t} \cdot k_{b'}}{\sigma_{b_t b'}} \Bigg|_{\substack{k_{b_{t'}} \to k_{\hat{b}_{t'}} = k_{a_{t'}} + k_{b_{t'}} \\ \forall b_{t'} \in \{b_1, \ldots, b_r\}}} = 0, \quad \forall b_t \in \{b_1, \ldots, b_r\}.
\end{align}
Here, we use \( k_{\hat{b}_t} \cdot k_{\hat{b}_{t'}} = s_{a_t b_{t'}} + s_{a_{t'} b_t} \) on the support of \( h_m = 0 \).

Equations \eqref{LOSEr1} and \eqref{LOSEr2} collectively form the \((n-r)\)-point scattering equations describing the scattering of \((n-r)\) particles. The particle set is given by \( \{1, 2, \ldots, n\}\setminus\{a_1, b_1, \ldots, a_r, b_r\} \cup \{\hat{b}_1, \hat{b}_2, \ldots, \hat{b}_r\} \), with on-shell momenta \( k_{\hat{b}_t} = k_{a_t} + k_{b_t} \) for all \( t = 1, 2, \ldots, r \) on the support of \( h_m = 0 \). We denote these induced lower-point scattering equations as
\begin{align}
\label{LOSEr3}
SE(a_1, & a_2, \ldots, a_r) \ln 
& := \{\mathring{E}_a = 0 \,|\, a \in \{1, 2, \ldots, n\}\setminus\{a_1, b_1, \ldots, a_r, b_r\}\} \cup \{\mathring{E}_{\hat{b}_t} = 0 \,|\, b_t \in \{b_1, \ldots, b_r\}\}.
\end{align}
The original \(n\)-point scattering equations are a special case, denoted as \( SE(\emptyset) \).

We rewrite the $(n-r)$-point scattering equations, denoted as \(\mathring{E}_a\), in the form of \eqref{LOSEr1}, ensuring they exhibit a consistent structure for all cases: whether \(a \in \{1, 2, \ldots, m\}\setminus\{a_1, \ldots, a_r\}\) or \(a \in \{m+1, m+2, \ldots, n\}\setminus\{b_1, \ldots, b_r\}\). However, on the support of \(h_m = 0\), many terms in \eqref{LOSEr1} drop out, simplifying the equations. Specifically, for \(a \in \{1, 2, \ldots, m\}\setminus\{a_1, \ldots, a_r\}\), the scattering equations reduce to
\ba
\label{shorteq}
\mathring{E}_a = \sum_{\substack{a' = 1 \\ a' \neq a}}^{m+1} \frac{s_{aa'}}{\sigma_{aa'}} \Bigg|_{\substack{\sigma_{a_t} \to \sigma_{b_t} \\ \forall a_t \in \{a_1, \ldots, a_r\}}} + \frac{s_{an}}{\sigma_{an}} = 0 \,, \quad \forall a \in \{1, 2, \ldots, m\}\setminus\{a_1, \ldots, a_r\}.
\ea
It is important to highlight that \(\mathring{E}'_{a_t}\), as defined in \eqref{LOSEr11}, bears a formal resemblance to \(\mathring{E}_a\) above, with one key difference: \(\mathring{E}'_{a_t}\) includes an additional term \(\frac{s_{a_tb_t}}{\sigma_{a_tb_t}}\). To establish a clear relationship, we define \(\mathring{E}_{a_t} := \mathring{E}'_{a_t} - \frac{s_{a_tb_t}}{\sigma_{a_tb_t}}\) for  all \( a_t \in \{a_1, \ldots, a_r\} \), as shown in \eqref{LOSEr11}. However, it is crucial to note that \(\mathring{E}_{a_t} \neq 0\).

When evaluating the contribution of a set of \( r \)-pinch solutions to an \( n \)-point CHY integral, we can replace the original \(n\)-point scattering equations \( SE(\emptyset) \) with the \((n-r)\)-point scattering equations \( SE(a_1, a_2, \ldots, a_r) \), supplemented by \( \mathring{E}'_{a_t} = 0 \) for all \( a_t \in \{a_1, \ldots, a_r\} \). In practice, we first integrate over \(\sigma_{a_t}\) for \( a_t \in \{a_1, \ldots, a_r\} \) using \( \mathring{E}'_{a_t} \) in the \( n \)-point CHY integral. This produces \( \mathring{E}_{a_t} \) as Jacobians, effectively reducing the \( n \)-point CHY integral to a \((n-r)\)-point one with scattering equations given by \( SE(a_1, \ldots, a_r) \).

The \((n-r)\)-point scattering equations describe the scattering of \((n-r)\) particles, allowing us to analyze their kinematics independently. In this system, we find that \( s_{ab} = 0 \) for \( a \in \{1, 2, \ldots, m\}\setminus\{a_1, \ldots, a_r\} \) and \( b \in \{m+2, \ldots, n-1\}\setminus\{b_1, \ldots, b_r\} \). However, \( s_{a \hat{b}_t} = s_{a a_t} \neq 0 \) for \( a \in \{1, 2, \ldots, m\}\setminus\{a_1, \ldots, a_r\} \) and \( b_t \in \{b_1, \ldots, b_r\} \). This indicates that the \((n-r)\)-point system is not simply a lower-point version of the original \(n\)-point system with \( h_m = 0 \). As a result, the formula \eqref{totalcase} cannot be recursively applied to this \((n-r)\)-point system.

Additionally, the \((n-r)\)-point system described by \eqref{LOSEr1} and \eqref{LOSEr2} can have both regular and singular solutions. All solutions of \( SE(a_1, \ldots, a_r) \) can be uplifted to singular solutions of the original \(n\)-point equations \( SE(\emptyset) \), but some singular solutions of \( SE(\emptyset) \) may correspond to regular solutions of \( SE(a_1, \ldots, a_r) \). Later sections will frequently transition between solutions of different scattering equations, leveraging these transformations.

\section{Turning on a Single Entry in  \(\hat{H}_m\) \label{sec:sec5}}

We begin by analyzing the scenario where only one entry \(x_{ij} \in \hat{H}_m\) is turned on. The goal is to establish the following factorization,
\ba
\label{singlexij}
A(\mathbb{I}_n) \xrightarrow[{\rm except}~ x_{ij} \neq 0] {H_m = 0}  
F_{m,n}(i,j), \quad \text{for } 1 \leq i \leq m < m+2 \leq j \leq n-1.
\ea

In this setup, only 1-pinch singular solutions, where \(\sigma_{ij} \sim 0\), contribute significantly to the amplitude in the CHY integral \eqref{chyintegral}. To simplify the analysis, we employ the following gauge fixing,
\ba
\label{gaugezero1}
\sigma_{m+1} \to 0, \, \sigma_j \to 1, \, \sigma_n \to \infty.
\ea

For the 1-pinch solutions, the \(n\)-point scattering equations reduce to a set of \((n-1)\)-point equations, denoted as \(SE(i)\), described by 
\ba
\label{newsem2}
\mathring{E}_b := \sum_{\substack{b'=1 \\ b' \neq i, b}}^{n-1} \frac{s_{bb'}}{\sigma_{bb'}} \Big|_{k_j \to k_i + k_j} = 0, \quad \text{for } b \in \{1, 2, \ldots, n-1\} \setminus \{i, j\}.
\ea
These equations govern the scattering of \((n-1)\) particles, represented by the set \(\{1, 2, \ldots, i-1, i+1, \ldots, j-1, \hat{j}, j+1, \ldots, n\}\), with \(k_{\hat{j}} = k_i + k_j\), satisfying the on-shell condition.

Additionally, according to \eqref{LOSEr11}, the scattering equation for \(i\) in the original system is given by
\ba
\label{eppi}
\mathring{E}'_i := \underbrace{\sum_{\substack{a'=1 \\ a' \neq i}}^{m+1} \frac{s_{i a'}}{\sigma_{ja'}}}_{\mathring{E}_i} + \frac{s_{ij}}{\sigma_{ij}} = 0.
\ea
Here, \(\sigma_i\) is replaced by \(\sigma_j\) in \(\mathring{E}'_i\), except for the last term \(s_{ij}/\sigma_{ij}\). This allows \(\sigma_i\) to be solved in terms of \(\sigma_j\) and other punctures.

As an example of \eqref{shorteq},
for indices \(a \in \{1, 2, \ldots, m\} \setminus \{i\}\), the scattering equations simplify further,
\ba
\label{epasim}
\mathring{E}_a := \sum_{\substack{a'=1 \\ a' \neq a}}^{m+1} \frac{s_{aa'}}{\sigma_{aa'}} \Big|_{\sigma_i \to \sigma_j} = 0, \quad \text{for } a \in \{1, 2, \ldots, m\} \setminus \{i\}.
\ea

The CHY formula for the YM amplitude then takes the form,
\ba
\label{facproofbem2}
A^{\rm YM}(\mathbb{I}_n) =
\int \prod_{\substack{a=1 \\ a \neq i}}^{m} \frac{d\sigma_a}{\underline{\mathring{E}_a}}
\prod_{\substack{b=m+2 \\ b \neq j}}^{n-1} \frac{d\sigma_b}{\underline{\mathring{E}_b}}
\frac{d\sigma_i}{\underline{\mathring{E}'_i}} \, \mathrm{PT}(12\ldots n{-}1) \, \mathrm{Pf}'\Psi,
\ea
where  
\ba
\label{defppt}
& \mathrm{PT}(12\ldots n{-}1) := \lim_{\sigma_n \to \infty} \sigma_n^2 \, \textbf{PT}(12\ldots n) =- \frac{1}{\sigma_{12}\sigma_{23}\ldots \sigma_{n{-}2,n{-}1}},
\nl &
\mathrm{Pf}'\Psi := \lim_{\sigma_n \to \infty} \sigma_n^2 \, \mathrm{Pf}'\mathbf{\Psi}.
\ea
Here the matrix \(\Psi\) is a modified version of \(\mathbf{\Psi}\), with the rows and columns corresponding to \(k_n\) and \(\epsilon_n\) multiplied by \(\sigma_n\) before taking the \(\sigma_n \to \infty\) limit.

In \eqref{facproofbem2}, since \(\sigma_i\) and \(\sigma_j\) pinch, we can replace \(\sigma_i\) by \(\sigma_j\) in the CHY integrand \(\mathrm{PT}(12\ldots n{-}1) \) \( \, \mathrm{Pf}'\Psi\), except in the denominator \(\sigma_{ij}\).

Finally, while the reduced \((n-1)\)-point scattering equations \eqref{epasim} may themselves exhibit singular solutions with respect to \(\{\sigma_1, \sigma_2, \ldots, \sigma_{i-1}, \sigma_{i+1}, \ldots, \sigma_n\}\), these contributions are subleading and can be ignored for the current analysis.

\subsection{Factorization of the Reduced Pfaffian} 

Choosing \(q = 2\) if \(i = 1\) and \(q = 1\) otherwise, we write
\ba
{\rm Pf}'\Psi = 2(-1)^{i+q}\frac{1}{\sigma_{iq}} 
{\rm Pf}\Psi_{i \!\!\!\slash q\!\!\!\slash ;} \cong 2(-1)^{i+q} \frac{1}{\sigma_{jq}} 
{\rm Pf}\Psi_{i \!\!\!\slash q\!\!\!\slash ;}.
\ea
This substitution \(\sigma_i \to \sigma_j\) aligns with the 1-pinch solution structure, where \(\sigma_i\) and \(\sigma_j\) are nearly coincident.
To analyze the reduced Pfaffian \({\rm Pf}'\Psi\), we expand it along the row and column corresponding to \(\epsilon_i\),
\ba
{\rm Pf}'\Psi
\cong
2
\frac{(-1)^{i+q}}{\sigma_{jq}}  \Bigg[
(-1)^{n+i+j} \frac{\epsilon_i \cdot k_j}{\sigma_{ij}} {\rm Pf}\Psi_{i \!\!\!\slash q\!\!\!\slash j\!\!\!\slash ; i \!\!\!\slash} 
+ (-1)^{n+i+n+j+1} \frac{\epsilon_i \cdot \epsilon_j}{\sigma_{ij}} {\rm Pf}\Psi_{i \!\!\!\slash q\!\!\!\slash ; i \!\!\!\slash j\!\!\!\slash} 
+ \mathcal{R}
\Bigg],
\ea
where \(\mathcal{R}\) includes the remaining terms involving sums over indices \(l\) distinct from \(i, j\),
\ba
\mathcal{R} =  &
\sum_{\substack{l=1 \\ l \neq i, j, q}}^{n-1} (-1)^{n+i+l+\theta(i-l)} \frac{\epsilon_i \cdot k_l}{\sigma_{il}} {\rm Pf}\Psi_{i \!\!\!\slash q\!\!\!\slash l\!\!\!\slash ; i \!\!\!\slash} 
+ \sum_{\substack{l=1 \\ l \neq i, j}}^{n-1} (-1)^{n+i+n+l+\theta(i-l)} \frac{\epsilon_i \cdot \epsilon_l}{\sigma_{il}} {\rm Pf}\Psi_{i \!\!\!\slash q\!\!\!\slash ; i \!\!\!\slash l\!\!\!\slash}
\nl & +
 (-1)^{n+i+n+1}
\e_i\cdot k_{n}
{\rm Pf}\Psi_{i \!\!\!\slash q\!\!\!\slash {n} \!\!\!\slash ; i \!\!\!\slash } 
+
 (-1)^{n+i+1}
\e_i\cdot \e_{n}
{\rm Pf}\Psi_{i \!\!\!\slash q\!\!\!\slash  ; i \!\!\!\slash {n} \!\!\!\slash } .
\ea
Here, \(\Psi_{i \!\!\!\slash q\!\!\!\slash j\!\!\!\slash ; i \!\!\!\slash}\) represents the matrix \(\Psi\) with the rows and columns corresponding to \(k_i\), \(k_q\), \(k_j\), and \(\epsilon_i\) removed. $\theta (i-l)$ is the Heaviside step function.

\subsubsection{Expansion of Sub-Pfaffians}

To simplify the terms in \(\mathcal{R}\), consider sub-Pfaffians like \({\rm Pf}\Psi_{i \!\!\!\slash q\!\!\!\slash l\!\!\!\slash ; i \!\!\!\slash}\). Expanding along the row and column for \(\epsilon_j\), the dominant contribution is proportional to  \(C_{jj} \cong -\epsilon_j \cdot k_i / \sigma_{ij}\), giving
\ba
{\rm Pf}\Psi_{i \!\!\!\slash q\!\!\!\slash l\!\!\!\slash ; i \!\!\!\slash}  
\cong  
(-1)^{n+1-\theta(j-l)} \frac{\epsilon_j \cdot k_i}{\sigma_{ij}} {\rm Pf}\Psi_{i \!\!\!\slash j \!\!\!\slash q\!\!\!\slash l\!\!\!\slash ; i \!\!\!\slash j \!\!\!\slash}.
\ea
Similarly, for \({\rm Pf}\Psi_{i \!\!\!\slash q\!\!\!\slash ; i \!\!\!\slash l\!\!\!\slash}\), the dominant term is 
\ba
{\rm Pf}\Psi_{i \!\!\!\slash q\!\!\!\slash ; i \!\!\!\slash l\!\!\!\slash}  
\cong  
(-1)^{n+j-\theta(j-l)+j} \frac{\epsilon_j \cdot k_i}{\sigma_{ij}} {\rm Pf}\Psi_{i \!\!\!\slash j \!\!\!\slash q\!\!\!\slash ; i \!\!\!\slash j \!\!\!\slash l \!\!\!\slash}.
\ea
Plugging these results into \(\mathcal{R}\), we observe that all terms can be reorganized into a single Pfaffian, resulting in
\ba
\mathcal{R} \cong  (-1)^{n-1} \frac{\epsilon_j \cdot k_i}{\sigma_{ij}} \frac{1}{\sigma_{jq}} {\rm Pf}\Psi_{i \!\!\!\slash j \!\!\!\slash q\!\!\!\slash ; j \!\!\!\slash}.
\ea
The reduced Pfaffian thus simplifies to
\ba
\label{exppre}
{\rm Pf}'\Psi
\cong
2(-1)^{j+q} 
\Bigg[
(-1)^n & \frac{\epsilon_i \cdot k_j}{\sigma_{ij}}
\frac{1}{\sigma_{jq}} 
{\rm Pf}\Psi_{i \!\!\!\slash q\!\!\!\slash j\!\!\!\slash ; i \!\!\!\slash} 
\ln 
& - \frac{\epsilon_i \cdot \epsilon_j}{\sigma_{ij}}
\frac{1}{\sigma_{jq}} 
{\rm Pf}\Psi_{i \!\!\!\slash q\!\!\!\slash ; i \!\!\!\slash j\!\!\!\slash} 
+ (-1)^{n-1+i-j} \frac{\epsilon_j \cdot k_i}{\sigma_{ij}} \frac{1}{\sigma_{jq}} {\rm Pf}\Psi_{i \!\!\!\slash j \!\!\!\slash q\!\!\!\slash ; j \!\!\!\slash}
\Bigg].
\ea

\subsubsection{Transition to \((n-1)\)-Point Pfaffian}

The matrix \(\Psi_{i \!\!\!\slash q\!\!\!\slash j\!\!\!\slash ; i \!\!\!\slash}\) is nearly independent of \(\sigma_i\), except for contributions in the diagonal elements of the \(C\)-matrix. These elements can be simplified by substituting \(\sigma_i\) with \(\sigma_j\),
\ba
C_{cc} = - \sum_{\substack{c'=1 \\ c' \neq c}}^{n-1} \frac{\epsilon_c \cdot k_{c'}}{\sigma_{cc'}} \cong - \sum_{\substack{c'=1 \\ c' \neq i, c}}^{n-1} \frac{\epsilon_c \cdot k_{c'}}{\sigma_{cc'}} \Big|_{k_j \to k_i + k_j}, \quad \text{for } c \in \{1,2,\ldots,n-1\} \setminus \{i, j, q\}.
\ea
Here, the substitution combines \(\frac{\epsilon_c \cdot k_i}{\sigma_{cj}} + \frac{\epsilon_c \cdot k_j}{\sigma_{cj}} = \frac{\epsilon_c \cdot (k_i + k_j)}{\sigma_{cj}}\), effectively treating \(k_i + k_j\) as the momentum of the new particle \(\hat{j}\).

We can further express the polarization vector \(\epsilon_j^\mu\) in the row and column corresponding to \(\epsilon_j\) as \(\sum_{\epsilon_{\hat{j}}} (\epsilon_j \cdot \epsilon_{\hat{j}}) \epsilon_{\hat{j}}^\mu\). The scalar factor \(\sum_{\epsilon_{\hat{j}}} (\epsilon_j \cdot \epsilon_{\hat{j}})\) can naturally be factored out, becoming a prefactor of the entire Pfaffian. The resulting matrix, denoted as \(\Psi_{i \!\!\!\slash j\!\!\!\slash \hat{j} ; i \!\!\!\slash j \!\!\!\slash \hat{j}}\), is obtained by deleting the rows and columns associated with \(k_i\) and \(\epsilon_i\), and replacing those associated with \(k_j\) and \(\epsilon_j\) by those of \(k_{\hat{j}}\) and \(\epsilon_{\hat{j}}\). This process can be summarized succinctly as
\ba
{\rm Pf} \Psi_{i \!\!\!\slash q\!\!\!\slash j\!\!\!\slash ; i \!\!\!\slash} = 
\sum_{\epsilon_{\hat{j}}} (\epsilon_j \cdot \epsilon_{\hat{j}}) \, 
{\rm Pf} \Psi_{i \!\!\!\slash q\!\!\!\slash j\!\!\!\slash ; i \!\!\!\slash j \!\!\!\slash \hat{j}}.
\ea

This allows us to interpret \(\frac{1}{\sigma_{jq}} {\rm Pf}\Psi_{i \!\!\!\slash q\!\!\!\slash j\!\!\!\slash ; i \!\!\!\slash}\) as the reduced Pfaffian for the scattering of \((n-1)\) gluons \(1, 2, \ldots, i-1, i+1, i+2, \hat{j}, \ldots, n\) with \(k_{\hat{j}} = k_i + k_j\) on the support of the \((n-1)\)-point scattering equations \eqref{newsem2},
\ba
\frac{1}{\sigma_{jq}}{\rm Pf}\Psi_{i \!\!\!\slash q\!\!\!\slash j\!\!\!\slash ; i \!\!\!\slash} 
\cong \frac{(-1)^{q+j+1}}{2}
\sum_{\epsilon_{\hat{j}}} (\epsilon_j \cdot \epsilon_{\hat{j}}) \,  
{\rm Pf}'\Psi_{i \!\!\!\slash j\!\!\!\slash \hat{j} ; i \!\!\!\slash j \!\!\!\slash \hat{j}}.
\ea

Similarly, for \(\Psi_{i \!\!\!\slash q\!\!\!\slash ; i \!\!\!\slash j\!\!\!\slash}\), we first move the row and column associated with \(k_j^\mu\) to the positions originally occupied by \(\epsilon_j\). This reordering introduces a relative sign factor of \((-1)^{n-1}\) in the Pfaffian. Next, we express \(k_j^\mu\) in its row and column as \(\sum_{\epsilon_{\hat{j}}} (k_j \cdot \epsilon_{\hat{j}}) \epsilon_{\hat{j}}^\mu\). These steps yield the following expression
\ba
\frac{1}{\sigma_{jq}}{\rm Pf}\Psi_{i \!\!\!\slash q\!\!\!\slash ; i \!\!\!\slash j\!\!\!\slash} 
\cong 
\frac{(-1)^{q+j+1+n-1}}{2} 
\sum_{\epsilon_{\hat{j}}} (k_j \cdot \epsilon_{\hat{j}}) \,  
{\rm Pf}'\Psi_{i \!\!\!\slash j\!\!\!\slash \hat{j} ; i \!\!\!\slash j \!\!\!\slash \hat{j}}.
\ea

For \(\Psi_{i \!\!\!\slash q\!\!\!\slash j\!\!\!\slash ; j \!\!\!\slash}\), we replace \(\sigma_i\) with \(\sigma_j\) and rewrite \(\epsilon_i^\mu\) as \(\sum_{\epsilon_{\hat{j}}} (\epsilon_i \cdot \epsilon_{\hat{j}}) \, \epsilon_{\hat{j}}^\mu\). Similarly, this yields
\ba
\frac{1}{\sigma_{jq}}{\rm Pf}\Psi_{i \!\!\!\slash q\!\!\!\slash j\!\!\!\slash ; j \!\!\!\slash} 
\cong 
\frac{(-1)^{q+j+1+j-i}}{2}
\sum_{\epsilon_{\hat{j}}} (\epsilon_i \cdot \epsilon_{\hat{j}}) \,  
{\rm Pf}'\Psi_{i \!\!\!\slash j\!\!\!\slash \hat{j} ; i \!\!\!\slash j \!\!\!\slash \hat{j}}.
\ea

Plugging these results into the expanded expression for \({\rm Pf}'\Psi\) in \eqref{exppre}, we achieve the factorized form of the reduced Pfaffian,
\ba
\label{exppre2}
{\rm Pf}'\Psi
\cong
(-1)^{n+1}
\sum_{\epsilon_j}
\underbrace{
\left(
\epsilon_i \cdot k_j \epsilon_j \cdot \epsilon_{\hat{j}} + 
\epsilon_i \cdot \epsilon_j k_i \cdot \epsilon_{\hat{j}} -
k_i \cdot \epsilon_j \epsilon_i \cdot \epsilon_{\hat{j}}
\right)
}_{A(ij - \hat{j})}
\frac{1}{\sigma_{ij}}
{\rm Pf}'\Psi_{i \!\!\!\slash j\!\!\!\slash \hat{j} ; i \!\!\!\slash j \!\!\!\slash \hat{j}}.
\ea

\subsubsection{ Integration over \(\sigma_i\)}

Using \({\mathring E}'_i\) to integrate \(\sigma_i\), we get
\ba
\label{simpleintegral}
\int \frac{d\sigma_i}{\underline{{\mathring E}'_i}} \frac{1}{\sigma_{ij}} = \frac{1}{\mathring{E}_i}.
\ea
Substituting this result into the CHY formula, we have 
\ba
\label{simpleintegral2dfaa}
A(\mathbb{I}_n) = (-1)^{n+1}
\sum_{\epsilon_{\hat{j}}} A(ij - j)
\int 
\prod_{\substack{a=1 \\ a \neq i}}^m \frac{d\sigma_a}{\underline{{\mathring E}_a}}
\prod_{\substack{b=m+2 \\ b \neq j}}^{n-1} \frac{d\sigma_b}{\underline{{\mathring E}_b}}
\left( {\rm Pf}'\Psi_{i \!\!\!\slash j\!\!\!\slash \hat{j}; i \!\!\!\slash j \!\!\!\slash \hat{j}} {\rm PT}(12\ldots n{-}1) \frac{1}{{\mathring E}_i} \right)_{\sigma_i \to \sigma_j}.
\ea
Note that the CHY integrand in \eqref{simpleintegral2dfaa} introduces a complicated spurious pole \({{\mathring E}_i}\), which resembles a scattering equation term \({{\mathring E}_a}\) but does not correspond to an actual integration contour. This term arises as a Jacobian factor during the integration over \(\sigma_i\) in \eqref{simpleintegral}. To distinguish it from the actual scattering equations, we refer to it as a scattering-equation-like Jacobian. In the following, we demonstrate how to systematically eliminate this term.

\subsection{Induced Lower-Point PT Factors}

To eliminate the spurious pole \({{\mathring E}_i}\), we aim to reduce the term \({\rm PT}(12\ldots m{+}1) / {{\mathring E}_i} \big|_{\sigma_i \to \sigma_j}\) into a linear combination of lower-point PT factors supported by the \(m{-}1\) scattering equations \({\mathring E}_a\) defined in \eqref{epasim}.

To achieve this, we propose the following ansatz, where the \((m-1)!\) coefficients \(X(s, \rho)\) are treated as undetermined functions of the Mandelstam variables,
\ba
\label{ansatz}
s_{12\ldots m{+}1} {\rm PT}(12\ldots i{-}1\, j\, i{+}1 \ldots m{+}1) 
\cong {\mathring E}_i \sum_{\rho \in S_{m-1}} X(s, \rho) {\rm PT}(\rho(12\ldots i{-}1\, i{+}1 \ldots m) m{+}1).
\ea

\subsubsection{Expansion of the Scattering-Equation-Like Jacobian}
For each term in \({\mathring E}_i\) defined in \eqref{eppi}, with \(1 \leq a' \leq m+1\) and \(a' \neq i\), we repeatedly apply partial fraction decomposition,
\ba
\label{shuffle1}
\frac{s_{ia'}}{\sigma_{ja'}} {\rm PT}(\rho(12\ldots i{-}1 \, i{+}1 \ldots m) m{+}1) = s_{ia'} {\rm PT}(Y(\rho, a') \shuffle j, Y(a', \rho)\, m{+}1),
\ea
where \(Y(\rho, a')\) are defined below \eqref{eq25g}. The notation \({\rm PT}(Y(\rho, a') \shuffle j, \ldots)\) abbreviates the sum of PT factors over all shuffles of \(Y(\rho, a')\) with \(j\). 

Substituting the shuffle decomposition \eqref{shuffle1} into the ansatz and reorganizing the terms yields,
\ba
\label{PTkk}
s_{12\ldots m{+}1} & {\rm PT}(12\ldots i{-}1\, j\, i{+}1 \ldots m{+}1) 
\ln &
\cong \sum_{\rho \in S_{m-1}} X(s, \rho) \sum_{\substack{a' = 1 \\ a' \neq i}}^{m+1} 
\left(\sum_{b \in Y(a',\rho)} s_{ib} +s_{i\, m+1} \right)
{\rm PT}(Y(\rho, a') j Y(a', \rho) m{+}1).
\ea

\subsubsection{BCJ Relations and Matching Coefficients}

The \(m{-}1\) independent scattering equations \({\mathring E}_a\) from \eqref{epasim} are analogous to those for the scattering of \(m{+}2\) gluons \(1, 2, \ldots, i-1, j, i+1, \ldots, m, m+1, \hat{n}\), where the off-shell particle \(\hat{n}\) has \(k_{\hat{n}}^2 = (k_1 + k_2 + \ldots + k_{m+1})^2 = 2s_{12\ldots m{+}1} \neq 0\). These equations are sufficient to derive BCJ relations among the PT factors,
\ba
\label{ptbcj1}
{\rm PT}(12 \ldots i{-}1\, j\, i{+}1 \ldots m{+}1) \cong  
\!\!\!\!\!\! 
\sum_{\pi \in \{12 \ldots i{-}1\} \shuffle \{i{+}1 \ldots m\}} 
\!\!\!\!\!\!
{\rm PT}(j\, \pi\, m{+}1) {\cal B}_{m+1,i}[12 \ldots i{-}1, i{+}1 \ldots m | \pi],
\ea
where \({\cal B}_{m+1,i}\) represents a ratio of Mandelstam variables. Similarly, for \({\rm PT}(Y(\rho, a') j Y(a', \rho) m{+}1)\), we have,
\ba 
\label{ptbcj2}
{\rm PT}(Y(\rho, a') j Y(a', \rho) m{+}1) \cong \sum_{\pi \in Y(\rho, a') \shuffle Y(a', \rho)} {\rm PT}(j\, \pi\, m{+}1) {\cal B}_{m+1,i}[Y(\rho, a'), Y(a', \rho) | \pi].
\ea
Plugging \eqref{ptbcj1} and \eqref{ptbcj1} into \eqref{PTkk} and 
 matching coefficients of the basis PT factors \({\rm PT}(j\, \pi\, m{+}1)\) on both sides, we derive,
\ba
\label{equationsfory}
s_{12\ldots m{+}1} {\cal B}_{m+1,i}[12 \ldots i{-}1, i{+}1 \ldots m | \pi] 
= \sum_{\rho \in S_{m-1}} X(s, \rho) g[\rho, \pi], \quad \forall \pi \in S_{m-1},
\ea
where $g[\rho, \pi]$ is given in \eqref{eq25g}. 
Solving this linear system gives rise to \(X(s, \rho)\) expressed in \eqref{Xsrhoori}.

\subsubsection{Structure of \(X(s, \rho)\)}
Although \eqref{Xsrhoori} involves matrix inversion, it has been confirmed (numerically up to \(m = 7\)) that \(X(s, \rho)\) contains only $(m-1) + (i-1)(m-i)$  physical poles involving gluon $i$,
\ba
\label{denominator}
\prod_{
\substack{1 \leq p \leq i \leq q \leq m \\
p < q}
} s_{p\, p+1 \ldots q}.
\ea
Explicit expressions for \(X(s, \rho)\) with polynomial numerators and denominators of the above form have been computed up to \(m = 5\). For specific choices of \(\rho\), clear patterns emerge that suggest general formulas for arbitrary multiplicity. For example,
\ba 
X(s, 23\ldots m)= \frac{-s_{2n}(s_{12}-s_{3n})(s_{123}-s_{4n})\ldots (s_{12\ldots m-1}-s_{mn})}{s_{12}s_{123}\ldots s_{12\ldots m}}\,,
\ea 
which has been numerically validated up to \(m = 7\).

\subsubsection{Factorization Achieved}
Rewriting  \eqref{ansatz} as
\ba
\label{ansatz2}
{\rm PT}(12 \ldots n{-}1) \frac{1}{{\mathring E}_i} 
\overset{SE(i)}{=} 
\frac{1}{s_{12\ldots m{+}1}} \sum_{\rho \in S_{m-1}} X(s, \rho) {\rm PT}(\rho(12 \ldots i{-}1\, i{+}1 \ldots m) m{+}1 \ldots n{-}1),
\ea
and combining it with the \((n-1)\)-point scattering equations \(SE(i)\) and reduced Pfaffian, we recover the \((n-1)\)-point YM amplitude,
\ba
\label{facproofbem3}
A(\mathbb{I}_n) \xrightarrow[{\rm except}~ x_{ij} \neq 0] {H_m = 0}  \frac{(-1)^{n+1}}{s_{12\ldots m{+}1}} 
  \sum_{\epsilon_{\hat j}} & A(ij{-}j) 
\sum_{\rho \in S_{m-1}} X(s, \rho) 
\ln & 
\times A(\rho(12 \ldots i{-}1\, i{+}1 \ldots m) m{+}1 \ldots j{-}1\,\hat{j}\,j {+}1 \ldots n{-}1).
\ea
This directly proves the desired factorization \eqref{singlexij}.

\subsection{Turning on an Entire Row or Column}

When an entire row or column in \(\hat{H}_m\) is turned on, the analysis remains centered on singular solutions where only a single pair of punctures pinch. These solutions contribute individually, allowing the contributions to be summed directly. This leads to the following results
\ba 
&A(\mathbb{I}_n) \xrightarrow [\text{except } x_{ij} \neq 0 ~\forall\, m+2 \leq j \leq n-1 ]  {{H_m = 0}} 
\sum_{m+2 \leq j \leq n-1} F_{m,n}(i,j), \quad \text{where } 1 \leq i \leq m\,,
\nl
&A(\mathbb{I}_n) \xrightarrow [\text{except } x_{ij} \neq 0 ~\forall\, 1 \leq i \leq m  ] {H_m = 0} 
\sum_{1 \leq i \leq m} F_{m,n}(i,j), \quad \text{where } m+2 \leq j \leq n-1\,.
\ea
These expressions summarize the contributions from turning on an entire row or column as simple summations over the corresponding \(F_{m,n}(i,j)\) factors for each \(i\) or \(j\).

\section{Turning on Two Entries Not in the Same Row or Column \label{sec:sec6}}

When ${\rm min}(m,n-m-2)\geq 2$ and we turn on \(x_{i_1j_1}\) and \(x_{i_2j_2}\), where \(i_1 \neq i_2\) and \(j_1 \neq j_2\), our goal is to demonstrate the following,
\ba
\label{goalijij}
A({\mathbb I}_n)\! \xrightarrow[{\rm except}~ x_{i_1j_1}, \, x_{i_2j_2} \neq 0]{H_m=0}
F_{m,n}(i_1,j_1) + F_{m,n}(i_2,j_2)\,,
\ea
with \(i_1, i_2 \in \{1, \ldots, m\}\), \(i_1 \neq i_2\), and \(j_1, j_2 \in \{m+2, \ldots, n-1\}\), \(j_1 \neq j_2\).

In this case, the CHY integral for the YM amplitude receives non-vanishing contributions from three distinct types of singular solutions:
\begin{enumerate}
 \item  Solutions where only \(\sigma_{i_1}\) and \(\sigma_{j_1}\) pinch.
 \item  Solutions where only \(\sigma_{i_2}\) and \(\sigma_{j_2}\) pinch. 
 \item  Solutions where both pairs, \((\sigma_{i_1}, \sigma_{j_1})\) and \((\sigma_{i_2}, \sigma_{j_2})\), pinch simultaneously.

\end{enumerate}

We denote these contributions as \(A_{{\mathbb I}_n}^{\sigma_{i_1j_1} \sim 0}\), \(A_{{\mathbb I}_n}^{\sigma_{i_2j_2} \sim 0}\), and \(A_{{\mathbb I}_n}^{\sigma_{i_1j_1}, \sigma_{i_2j_2} \sim 0}\), respectively. Thus, the total contribution can be expressed as
\ba
\label{conijijijij}
A({\mathbb I}_n)\! \xrightarrow[{\rm except}~ x_{i_1j_1}, \, x_{i_2j_2} \neq 0]{H_m=0}
A_{{\mathbb I}_n}^{\sigma_{i_1j_1} \sim 0} + A_{{\mathbb I}_n}^{\sigma_{i_2j_2} \sim 0} + A_{{\mathbb I}_n}^{\sigma_{i_1j_1}, \sigma_{i_2j_2} \sim 0}.
\ea

In the subsequent analysis, we show how \eqref{conijijijij} leads to \eqref{goalijij}.

Although the contributions from 1-pinch solutions were analyzed in \cref{sec:sec5}, where they exclusively produced \((n-1)\)-point CHY integrals corresponding to \((n-1)\)-point YM amplitudes, additional complexities arise in the current case:

\begin{enumerate}
    \item
The contributions from 1-pinch solutions now include a linear combination of \((n-1)\)-point CHY integrals corresponding to \((n-1)\)-point YM amplitudes and additional \((n-2)\)-point CHY integrals.
    
   \item 
   The contributions from 2-pinch solutions contribute solely to \((n-2)\)-point CHY integrals.
\end{enumerate}

By combining these contributions, we will demonstrate that all \((n-2)\)-point CHY integrals cancel, leaving only the \((n-1)\)-point CHY integrals. These remaining integrals reconstruct the \((n-1)\)-point YM amplitudes required in \eqref{goalijij}.

\subsection{Contributions of 1-Pinch Solutions
\label{sec:con1}}

We analyze the contribution of 1-pinch solutions, where only \(\sigma_{i_1j_1} \sim 0\). In this scenario, the induced \((n-1)\)-point scattering equations are given by \eqref{newsem2}, with \(i\) and \(j\) identified as \(i_1\) and \(j_1\), respectively. Additionally, the equation corresponding to \(i_1\) is provided in \eqref{eppi}.

For these singular solutions, where only \(\sigma_{i_1}\) and \(\sigma_{j_1}\) pinch, the reduced Pfaffian retains the structure shown in \eqref{exppre2}. Therefore, the CHY integral restricted to these solutions can be expressed as
\ba
\label{odafiosd}
A_{{\mathbb I}_n}^{\sigma_{i_1j_1} \sim 0} = (-1)^{n+1} &
\sum_{\e_{\hat j_1}}
A(i_1j_1 - \hat{j}_1) 
\ln &
\times \int 
\prod_{\substack{a=1 \\ a \neq i_1}}^{m} \frac{d\sigma_a}{\underline{{\mathring E}_a}}
\prod_{\substack{b=m+2 \\ b \neq j_1}}^{n-1} \frac{d\sigma_b}{\underline{{\mathring E}_b}}  
\left( {\rm Pf}' \Psi_{i \!\!\!\slash_1 j\!\!\!\slash_1 \hat j_1 ; i \!\!\!\slash_1 j\!\!\!\slash_1 \hat j_1} 
{\rm PT}(12 \ldots n{-}1)  
\frac{1}{{\mathring E}_{i_1}} 
\right)_{\sigma_{i_1} \to \sigma_{j_1}},
\ea
where only \(\sigma_{i_1}\) and \(\sigma_{j_1}\) pinch.

Rewriting \({\rm PT}(12\ldots n{-}1)/{\mathring E}_{i_1}\) using the decomposition \eqref{ansatz2}, we have,
\ba
\label{ansatz23}
\frac{{\rm PT}(12\ldots n{-}1)}{{\mathring E}_{i_1}}  \cong
\frac{1}{s_{12\ldots m+1}}
  \sum_{\rho_{i_1} \in S_{m-1}}  
  X(s, \rho_{i_1}) 
  {\rm PT}(\rho_{i_1} \, m{+}1 \ldots n{-}1),
\ea
where the subscript \(i_1\) in \(\rho_{i_1}\) indicates that it is a permutation of the \(m-1\) labels \(\{1, \ldots, m\} \setminus \{i_1\}\).

In earlier discussions, such as in \eqref{facproofbem2}, the combination of \((n-1)\)-point scattering equations, reduced Pfaffians, and PT factors directly yielded \((n-1)\)-point amplitudes. However, in the present case, this no longer holds due to the interplay with additional singular solutions, as detailed in the subsequent analysis.

\subsubsection{\((n-1)\)-Point YM Amplitude }

To understand the \((n-1)\)-point CHY formula for a YM amplitude, consider the expression,
\ba
\label{np1pchy}
A_{\rho_{i_1} \,  m{+}1 \ldots \hat j_1 \ldots n-1} = 
\int
\prod_{\substack{a=1 \\ a \neq i_1}}^{m} \frac{d\sigma_{a}}{\underline{{\mathring E}_a}}
\prod_{\substack{b=m+2 \\ b \neq j_1}}^{n-1} \frac{d\sigma_{b}}{\underline{{\mathring E}_b}}  
\left( 
{\rm Pf}' \Psi_{i \!\!\!\slash_1 j\!\!\!\slash_1 \hat j_1 ; i \!\!\!\slash_1 j \!\!\!\slash_1 \hat j_1} 
{\rm PT}(\rho_{i_1} \, m{+}1 \ldots n{-}1)
\right)_{\sigma_{i_1} \to \sigma_{j_1}}\,.
\ea
If \(x_{i_2j_2}\) is also turned on, the \((n-1)\)-point CHY integral receives additional contributions from singular solutions where \(\sigma_{i_2}\) and \(\sigma_{j_2}\) pinch. These contributions are absent in the analysis leading to \eqref{odafiosd}, as well as in earlier discussions like \eqref{facproofbem2}, where \(x_{i_2j_2}\) was set to zero.

In this \((n-1)\)-point system, involving particles \(\{1, 2, \ldots, i_1-1, i_1+1, \ldots, m, m+1, \ldots, j_1-1, \hat j_1, j_1+1, \ldots, n\}\), the following properties hold,
\begin{enumerate}
    \item 

 For \(a \in \{1, 2, \ldots, i_1-1, i_1+1, \ldots, m\}\) and \(b \in \{m+2, \ldots, j_1-1, j_1+1, \ldots, n-1\}\),
  \ba\nonumber
  s_{ab} = 0.
  \ea
   \item 
   However, for \(a \in \{1, 2, \ldots, i_1-1, i_1+1, \ldots, m\}\),
  \ba
  \nonumber
  s_{a \hat j_1} = s_{a i_1} \neq 0.
  \ea
\end{enumerate}

This setup does not reduce to a straightforward lower-point case, such as the one described by \eqref{eq:constrhm}, and therefore, the formula \eqref{totalcase} cannot be applied recursively. Importantly, in this \((n-1)\)-point system, both regular and singular solutions exist for the scattering equations.

Following the convention established in \eqref{conijijijij}, the \((n-1)\)-point YM amplitude can be decomposed as
\ba
\label{jafl300}
A_{\rho_{i_1} \, m{+}1 \ldots \hat j_1 \ldots n-1} = 
A_{\rho_{i_1} \, m{+}1 \ldots \hat j_1 \ldots n-1}^{\rm reg} 
+ 
A_{\rho_{i_1} \, m{+}1 \ldots \hat j_1 \ldots n-1}^{\sigma_{i_2  j_2} \sim 0},
\ea
where the terms correspond to the contributions from regular solutions and singular solutions (\(\sigma_{i_2  j_2} \sim 0\)) of the \((n-1)\)-point CHY integral.
The CHY integral in \eqref{odafiosd} captures only the regular contribution,
\ba
A_{{\mathbb I}_n}^{\sigma_{i_1  j_1} \sim 0} =\frac{(-1)^{n+1}}{s_{12\ldots m+1}} 
\sum_{\e_{\hat j_1}}
A(i_1 j_1 - \hat j_1)
\sum_{\rho_{i_1} \in S_{m-1}} 
X(s, \rho_{i_1}) 
A_{\rho_{i_1} \, m{+}1 \ldots \hat j_1 \ldots n-1}^{\rm reg},
\ea
as opposed to the full amplitude \(A_{\rho_{i_1} \, m{+}1 \ldots \hat j_1 \ldots n-1}\). 

From \eqref{generalFn}, the full contribution \(F_{m, n}(i_1, j_1)\) is given by
\ba
F_{m, n}(i_1, j_1) =  \frac{(-1)^{n+1}}{s_{12\ldots m+1}} 
\sum_{\e_{\hat j_1}}
A(i_1 j_1 - \hat j_1)
\sum_{\rho_{i_1} \in S_{m-1}} 
X(s, \rho_{i_1}) 
A_{\rho_{i_1} \, m{+}1 \ldots \hat j_1 \ldots n-1}.
\ea

The discrepancy between \(A_{{\mathbb I}_n}^{\sigma_{i_1  j_1} \sim 0}\) and \(F_{m, n}(i_1, j_1)\) arises from the term \(A_{\rho_{i_1} \, m{+}1 \ldots \hat j_1 \ldots n-1}^{\sigma_{i_2  j_2} \sim 0}\), which, as we will demonstrate,  is itself a combination of \((n-2)\)-point CHY integrals. Specifically,
\ba
\label{odafiosd0}
\Delta_{i_1 j_1}^{i_2 j_2} = 
A_{{\mathbb I}_n}^{\sigma_{i_1  j_1} \sim 0} - F_{m, n}(i_1, j_1) = 
 \frac{(-1)^{n}}{s_{12\ldots m+1}} \sum_{\e_{\hat j_1}}
A(i_1 j_1 - \hat j_1)
\sum_{\rho_{i_1} \in S_{m-1}} 
X(s, \rho_{i_1}) 
A_{\rho_{i_1} \, m{+}1 \ldots \hat j_1 \ldots n-1}^{\sigma_{i_2  j_2} \sim 0}.
\ea
This analysis reveals the critical role of \((n-2)\)-point contributions in resolving discrepancies, as detailed in the subsequent section. 

\subsubsection{\((n-2)\)-Point CHY Integral \label{sec:con12}}

Starting with the \((n-1)\)-point CHY integral \eqref{np1pchy}, we now examine the contributions from its singular solutions where \(\sigma_{i_2j_2} \sim 0\). These singular solutions induce \((n-2)\)-point scattering equations, denoted \(SE(i_1, i_2)\), as given in \eqref{LOSEr3}. These equations involve particles \(\{1, 2, \ldots, n\} \setminus \{i_1, j_1, i_2, j_2\} \cup \{\hat j_1, \hat j_2\}\).

The original \((n-1)\)-point scattering equation with respect to \(i_2\) is provided in \eqref{LOSEr11}. Rewriting it explicitly under the gauge fixing \(\sigma_{m+1} \to 0\), \(\sigma_{j_1} \to 1\), and \(\sigma_n \to \infty\), we have,
\ba
\label{eppi2pp22223}
{\mathring E}'_{i_2} =
\underbrace{\sum_{\substack{a'=1 \\ a' \neq i_1, i_2}}^{m+1} \frac{s_{i_2 a'}}{\sigma_{j_2 a'}} 
- \frac{s_{i_1 i_2}}{\sigma_{j_1 j_2}}}_{{\mathring E}_{i_2}}
+ \frac{s_{i_2 j_2}}{\sigma_{i_2 j_2}} = 0.
\ea
The term \(\frac{s_{i_1 i_2}}{\sigma_{j_1 j_2}}\) in \({\mathring E}'_{i_2}\) originates from \(\frac{s_{\hat j_1 i_2}}{\sigma_{j_1 i_2}}\), which involves particles \(\hat j_1\) and \(i_2\). Here, we recognize that \(s_{\hat j_1 i_2} = s_{i_1 i_2} \neq 0\) on the support of \(h_m = 0\), as \(\hat j_1\) replaces \(i_1\) in the \((n-1)\)-point system.

On the support of the \((n-2)\)-point scattering equations, the \((n-1)\)-point reduced Pfaffian in \eqref{np1pchy} reduces to an \((n-2)\)-point Pfaffian in a manner analogous to \eqref{exppre2},
\ba
\label{exppre2pp}
{\rm Pf}' \Psi_{i \!\!\!\slash_1 j\!\!\!\slash_1 \hat j_1 ; i \!\!\!\slash_1 j \!\!\!\slash_1 \hat j_1} 
\cong (-1)^n \sum_{\e_{\hat j_2}} A(i_2 j_2 - \hat j_2) 
\frac{1}{\sigma_{i_2 j_2}}
{\rm Pf}' \Psi_{i \!\!\!\slash_1 j\!\!\!\slash_1 i \!\!\!\slash_2 j\!\!\!\slash_2 \hat j_1 \hat j_2 ; i \!\!\!\slash_1 j\!\!\!\slash_1 i \!\!\!\slash_2 j\!\!\!\slash_2 \hat j_1 \hat j_2}.
\ea

This reduction implies that the \((n-1)\)-point CHY integral in \eqref{np1pchy} for the singular solutions \(\sigma_{i_2  j_2} \sim 0\) transforms into a combination of \((n-2)\)-point CHY integrals,
\ba
\label{odafiosd1} 
A_{\rho_{i_1} \, m{+}1 \ldots \hat j_1 \ldots n-1}^{\sigma_{i_2  j_2} \sim 0} &= (-1)^n 
\sum_{\e_{\hat j_2}} A(i_2 j_2 - \hat j_2)
\int \prod_{\substack{a=1 \\ a \neq i_1, i_2}}^{m} \frac{d\sigma_a}{\underline{{\mathring E}_a}}
\prod_{\substack{b=m+2 \\ b \neq j_1, j_2}}^{n-1} \frac{d\sigma_b}{\underline{{\mathring E}_b}}
\frac{d\sigma_{j_2}}{\underline{{{\mathring E}}_{\hat j_2}}} 
\nl &
\times
\left( {\rm Pf}' \Psi_{i \!\!\!\slash_1 j\!\!\!\slash_1 i \!\!\!\slash_2 j\!\!\!\slash_2 \hat j_1 \hat j_2 ; i \!\!\!\slash_1 j\!\!\!\slash_1 i \!\!\!\slash_2 j\!\!\!\slash_2 \hat j_1 \hat j_2}
{\rm PT}(12 \ldots i\!\!\!\slash_1 \ldots i\!\!\!\slash_2 \ldots m+1 \ldots n{-}1) \right)_{\substack{\sigma_{i_1} \to \sigma_{j_1} \\ \sigma_{i_2} \to \sigma_{j_2}}}
\nl &
\times \frac{{\rm PT}(\rho_{i_1} m{+}1)}{{\rm PT}(12 \ldots i\!\!\!\slash_1 \ldots i\!\!\!\slash_2 \ldots m{+}1)} 
\frac{1}{{{\mathring E}}_{i_2}} \Bigg|_{\substack{\sigma_{i_1} \to \sigma_{j_1} \\ \sigma_{i_2} \to \sigma_{j_2}}}.
\ea
In the above expression, \({\rm PT}(12 \ldots i\!\!\!\slash_1 \ldots i\!\!\!\slash_2 \ldots m{+}1)\) denotes an \((m+1)\)-point PT factor, with \(i_1\) possibly smaller or larger than \(i_2\).
 The integral involves:
 \begin{itemize}
     \item    
 An \((n-2)\)-point CHY measure, denoted \(d \mu_{[i_1 i_2]}\),
   \item  An \((n-2)\)-point reduced Pfaffian, denoted \({\rm Pf}'\Psi_{[i_1 i_2]}\),
   \item  An \((n-2)\)-point PT factor, denoted \({\rm PT}_{[i_1 i_2]}\).
 \end{itemize}
Despite the structure of \((n-2)\)-point components, the CHY integral fails to yield an \((n-2)\)-point YM amplitude due to the additional coefficient in the last line of \eqref{odafiosd1}.

\subsubsection{Full Expressions for the Discrepancies}

Substituting \eqref{odafiosd1} into \eqref{odafiosd0}, we find that the contribution from the solutions where \(\sigma_{i_1j_1} \sim 0\) can be expressed as 
\ba
\label{examplesim1}
A_{{\mathbb I}_n}^{\sigma_{i_1j_1} \sim 0} = F_{m,n}(i_1, j_1) + \Delta_{i_1 j_1}^{i_2 j_2},
\ea
where the discrepancy term \(\Delta_{i_1 j_1}^{i_2 j_2}\) is given by
\ba
\label{odafiosd00}
\Delta_{i_1 j_1}^{i_2 j_2} = 
 -\sum_{\e_{\hat j_1}} A(i_1 j_1 - \hat j_1) 
\sum_{\e_{\hat j_2}} A(i_2 j_2 - \hat j_2)
\int d\mu_{[i_1 i_2]} {\rm Pf}' \Psi_{[i_1 i_2]} {\rm PT}_{[i_1 i_2]} \, t_{i_1 j_1}^{i_2 j_2},
\ea
with 
\ba
\label{jaflt}
t_{i_1 j_1}^{i_2 j_2} = - 
\frac{1}{s_{12 \ldots m+1}}
\sum_{\rho_{i_1} \in S_{m-1}} 
X(s, \rho_{i_1}) 
\frac{{\rm PT}(\rho_{i_1} m{+}1)}{{\rm PT}(12 \ldots i\!\!\!\slash_1 \ldots i\!\!\!\slash_2 \ldots m+1)}
\frac{1}{{\mathring E}_{i_2}} \Bigg|_{\substack{\sigma_{i_1} \to \sigma_{j_1} \\ \sigma_{i_2} \to \sigma_{j_2}}}.
\ea
Here, \(t_{i_1 j_1}^{i_2 j_2}\) depends only on the \(m+1\) punctures \(\{\sigma_1, \sigma_2, \ldots, \sigma_{m+1}\}\) under the replacements \(\sigma_{i_1} \to \sigma_{j_1}\) and \(\sigma_{i_2} \to \sigma_{j_2}\).

Similarly, for the contributions from the solutions where \(\sigma_{i_2  j_2} \sim 0\), we have
\ba
\label{examplesim2}
A_{{\mathbb I}_n}^{\sigma_{i_2  j_2} \sim 0} = F_{m,n}(i_2, j_2) + \Delta_{i_2 j_2}^{i_1 j_1},
\ea
where the discrepancy \(\Delta_{i_2 j_2}^{i_1 j_1}\) is given by
\ba
\label{odafiosd002}
\Delta_{i_2 j_2}^{i_1 j_1} = -
 \sum_{\e_{\hat j_1}} A(i_1 j_1 - \hat j_1) 
\sum_{\e_{\hat j_2}} A(i_2 j_2 - \hat j_2)
\int d\mu_{[i_1 i_2]} {\rm Pf}' \Psi_{[i_1 i_2]} {\rm PT}_{[i_1 i_2]} \, t_{i_2 j_2}^{i_1 j_1},
\ea
with 
\ba
t_{i_2 j_2}^{i_1 j_1} = 
-\frac{1}{s_{12 \ldots m+1}}
\sum_{\rho_{i_2} \in S_{m-1}} 
X(s, \rho_{i_2}) 
\frac{{\rm PT}(\rho_{i_2} m{+}1)}{{\rm PT}(12 \ldots i\!\!\!\slash_1 \ldots i\!\!\!\slash_2 \ldots m+1)}
\frac{1}{{\mathring E}_{i_1}} \Bigg|_{\substack{\sigma_{i_1} \to \sigma_{j_1} \\ \sigma_{i_2} \to \sigma_{j_2}}}.
\ea

Here are the key observations:

\begin{enumerate}
    \item 
 The discrepancy terms \(\Delta_{i_1 j_1}^{i_2 j_2}\) and \(\Delta_{i_2 j_2}^{i_1 j_1}\) involve \((n-2)\)-point CHY integrals.
    \item The additional factors \(t_{i_1 j_1}^{i_2 j_2}\) and \(t_{i_2 j_2}^{i_1 j_1}\) prevent these terms from directly contributing to the \((n-2)\)-point YM amplitudes.
    \item These factors only depend on the \(m+1\) punctures \(\{\sigma_1, \sigma_2, \ldots, \sigma_{m+1}\}\) under the replacements \(\sigma_{i_1} \to \sigma_{j_1}\) and \(\sigma_{i_2} \to \sigma_{j_2}\), which will play a crucial role in subsequent cancellations.
\end{enumerate}

\subsection{Contributions of 2-Pinch Solutions}

For 2-pinch singular solutions, where both \(\sigma_{i_1j_1} \sim 0\) and \(\sigma_{i_2j_2} \sim 0\), the \(n\)-point scattering equations \(SE(\emptyset)\) can be reorganized into \((n-2)\)-point induced scattering equations \(SE(i_1,i_2)\), given in \eqref{LOSEr3}. These equations involve the particles \(\{1, 2, \ldots, n\} \setminus \{i_1, j_1, i_2, j_2\} \cup \{\hat j_1, \hat j_2\}\), alongside two additional equations corresponding to \(i_1\) and \(i_2\).

These \((n-2)\)-point scattering equations, already discussed in \cref{sec:con12}, can also be derived directly from \(SE(\emptyset)\). While the derivation paths differ—whether starting from \(SE(i_1)\) or \(SE(\emptyset)\)—the resulting equations remain consistent, as they share identical kinematics. The scattering equation for \(i_2\) is formally expressed in \eqref{eppi2pp22223}, though here it should be understood as arising directly from the \(n\)-point system.

Applying a two-step reduction analogous to \eqref{exppre2}, we find that the \(n\)-point reduced Pfaffian simplifies to a combination of \((n-2)\)-point ones,
\ba
\label{exppre2pp2222}
{\rm Pf}' \Psi 
\cong - \sum_{\e_{\hat j_1}} A(i_1j_1 -\hat j_1) 
\sum_{\e_{\hat j_2}} A(i_2j_2 -\hat j_2)
\frac{1}{\sigma_{i_1j_1}\sigma_{i_2j_2}}
{\rm Pf}' \Psi_{i \!\!\!\slash_1 j\!\!\!\slash_1 i \!\!\!\slash_2 j\!\!\!\slash_2 \hat j_1 \hat j_2; i \!\!\!\slash_1 j\!\!\!\slash_1 i \!\!\!\slash_2 j\!\!\!\slash_2 \hat j_1 \hat j_2}.
\ea

We then integrate over \(\sigma_{i_1}\) and \(\sigma_{i_2}\), yielding
\ba
\label{jac222}
\int \frac{d\sigma_{i_1}}{\underline{{\mathring E}'_{i_1}}\sigma_{i_1j_1}} = \frac{1}{{\mathring E}_{i_1}}, \quad 
\int \frac{d\sigma_{i_2}}{\underline{{\mathring E}'_{i_2}}\sigma_{i_2j_2}} = \frac{1}{{\mathring E}_{i_2}}.
\ea

The contribution of these 2-pinch singular solutions to the CHY integral is then
\ba
\label{jafl}
A_{{\mathbb I}_n}^{\sigma_{i_1j_1}, \sigma_{i_2j_2} \sim 0} = - 
\sum_{\e_{\hat j_1}} A(i_1j_1 -\hat j_1) 
\sum_{\e_{\hat j_2}} A(i_2j_2 -\hat j_2)
\int d\mu_{[i_1 i_2]} {\rm Pf}' \Psi_{[i_1 i_2]} {\rm PT}_{[i_1 i_2]} \, t_{i_1j_1, i_2j_2},
\ea
where
\ba
\label{jafltfkalewf}
t_{i_1j_1, i_2j_2} :=  
\frac{{\rm PT}(12 \ldots m+1)}{{\rm PT}(12 \ldots i\!\!\!\slash_1 \ldots i\!\!\!\slash_2 \ldots m+1)}
\frac{1}{{\mathring E}_{i_1} {\mathring E}_{i_2}} \Bigg|_{\substack{\sigma_{i_1} \to \sigma_{j_1} \\ \sigma_{i_2} \to \sigma_{j_2}}}.
\ea

Here are some observations on the structure. 
\begin{enumerate}
    \item 
The integral in \eqref{jafl} features the same \((n-2)\)-point CHY measure, reduced Pfaffian, and PT factors as in \eqref{odafiosd00} and \eqref{odafiosd002}.
   \item  However, the coefficient \(t_{i_1j_1, i_2j_2}\) introduces a significant difference. Like the previously encountered coefficients, \(t_{i_1j_1, i_2j_2}\) depends only on the \(m+1\) punctures \(\{\sigma_1, \sigma_2, \ldots, \sigma_{m+1}\}\) under the specified replacements.
\end{enumerate}

These observations will play a crucial role in identifying and addressing cancellations in subsequent sections.

\subsection{Cancellations }

The three coefficients \( t \)'s appearing in \eqref{jafl}, \eqref{odafiosd00}, and \eqref{odafiosd002} obstruct the \((n-2)\)-point CHY integrals from directly producing \((n-2)\)-point YM amplitudes. Notably, all three coefficients depend only on the \( m+1 \) punctures \(\{\sigma_1, \sigma_2, \ldots, \sigma_{m+1}\}_{\sigma_{i_1}\to \sigma_{j_1}, \sigma_{i_2}\to \sigma_{j_2}}\).

Recall that there are \((m-2)\) independent scattering equations, \({\mathring E}_a = 0\) for all \(a \in \{1, 2, \ldots, m\}\setminus\{i_1, i_2\}\), given explicitly in \eqref{shorteq}. These equations are part of the \((n-2)\)-point scattering equations \(SE(i_1, i_2)\). Importantly, they involve only the same \(m+1\) punctures under the gauge fixing \(\sigma_n \to \infty\).

On the support of these \((m-2)\) scattering equations, a key identity emerges that relates the three coefficients \(t\):
\ba
\label{mp2keyid}
t_{i_1j_1}^{i_2j_2} + t_{i_2j_2}^{i_1j_1} + t_{i_1j_1, i_2j_2} \overset{{\mathring E}_a = 0 \, \forall a \in \{1, 2, \ldots, m\}\setminus\{i_1, i_2\}}{=} 0.
\ea
This identity implies that the three coefficients sum to zero when evaluated on the support of the scattering equations.

Note that the identity \eqref{mp2keyid} holds irrespective of the total number of particles \(n\). It depends only on the structure of the \(m+1\) punctures.
When \(m=2\), the identity \eqref{mp2keyid} holds algebraically, without requiring the support of any scattering equations.  The proof for general  \(m \geq 2\) is postponed to \cref{sec:sec8}. To illustrate these principles, explicit examples with \(m=2\) and \(m=3\) are provided in  \cref{appb22}, offering concrete insights into the mechanics of these cancellations.

Using the identity \eqref{mp2keyid}, the contributions of all \((n-2)\)-point CHY integrals in \eqref{jafl}, \eqref{odafiosd00}, and \eqref{odafiosd002} cancel out when combined. This cancellation is expressed as
\ba
\label{deltacan}
\Delta_{i_1j_1}^{i_2j_2} + \Delta_{i_2j_2}^{i_1j_1} + A_{{\mathbb I}_n}^{\sigma_{i_1j_1}, \sigma_{i_2j_2} \sim 0} = 0.
\ea
As a result, the obstruction caused by the coefficients \(t\)'s is eliminated, leaving only the desired \((n-1)\)-point contributions in \eqref{goalijij}.


\subsection{Turning on Two Entire  Rows or Columns}

Now consider the scenario where two entire rows are turned on, meaning all entries \(x_{i_1 j}\) and \(x_{i_2 j}\) in \({\hat H}_m\) are set to nonzero for all  \(m+2 \leq j \leq n-1\), with \(i_1 \neq i_2 \in \{1, 2, \ldots, m\}\) and ${\rm min}(m,n-m-2)\geq 2$. The amplitude \(A({\mathbb I}_n)\) can be expressed as
\ba
\label{conijijijij2}
A({\mathbb I}_n)\! \xrightarrow[
\substack{{\rm except}~ x_{i_1 j}, \, x_{i_2 j} \neq 0
\\
\forall m+2\leq j\leq n-1 }]{H_m=0}
\sum_{m+2 \leq j \leq n-1}\left( A_{{\mathbb I}_n}^{\sigma_{i_1 j} \sim 0} + A_{{\mathbb I}_n}^{\sigma_{i_2 j} \sim 0} \right) + \sum_{j_1 \neq j_2 \in \{m+2, \ldots, n-1\}} A_{{\mathbb I}_n}^{\sigma_{i_1j_1},\sigma_{i_2j_2} \sim 0}.
\ea
Each term \(A_{{\mathbb I}_n}^{\sigma_{i_1j_1} \sim 0}\) contains a discrepancy compared to \(F_{m,n}(i_1, j_1)\) because the reduced \((n-1)\)-point CHY formula requires singular solutions where \(\sigma_{i_2j_2} \sim 0\) for all \(j_2 \neq j_1\). Denoting this discrepancy as \(\Delta_{i_1j_1}\), we have,
\ba
\Delta_{i_1j_1} := A_{{\mathbb I}_n}^{\sigma_{i_1j_1} \sim 0} - F_{m,n}(i_1, j_1) = \sum_{\substack{j_2 = m+2 \\ j_2 \neq j_1}}^{n-1} \Delta_{i_1j_1}^{i_2j_2}.
\ea

When summing over all contributions associated with the two activated rows,  the discrepancies combine with the contributions from double-pinch solutions \(A_{{\mathbb I}_n}^{\sigma_{i_1j_1}, \sigma_{i_2j_2} \sim 0}\). These terms can be reorganized,
\ba 
& \sum_{  m+2\leq j\leq n-1}\left( {\Delta}_{{i_1j}}+{\Delta}_{{i_2j}} \right)+
\sum_{ j_1\neq j_2 \in \{m+2, \ldots, n-1\}}A_{{\mathbb I}_n}^ {\sigma_{i_1j_1},\sigma_{i_2j_2}\sim 0}
\nl & = 
\sum_{ j_1\neq j_2 \in \{m+2, \ldots, n-1\}} (\Delta_{{i_1j_1}}^{i_2j_2} + \Delta_{{i_2j_2}}^{i_1j_1} + A_{{\mathbb I}_n}^{\sigma_{i_1j_1},\sigma_{i_2j_2}\sim 0} ) =0 .
\ea 
Using the cancellation identity  \eqref{deltacan}, 
the discrepancies vanish entirely, leaving
\ba
\label{conijijijij2xx}
A({\mathbb I}_n)\! \xrightarrow[
\substack{{\rm except}~ x_{i_1 j}, \, x_{i_2 j} \neq 0
\\
\forall m+2 \leq j \leq n-1 }]{H_m=0}
\sum_{m+2 \leq j \leq n-1} \left( F_{m,n}(i_1, j) + F_{m,n}(i_2, j) \right),
\ea
with \(i_1 \neq i_2 \in \{1, 2, \ldots, m\}\).

The formula extends naturally to cases where two entire columns are turned on, following the same logic.

\section{Turning on Three Entries Not in the Same Row or Column \label{sec:sec7}}

As we turn on more entries, ensuring that any two are not aligned within the same row or column, the interplay between different types of singular solutions grows increasingly complex. However, the fundamental strategy remains unchanged, relying on systematic cancellations that require progressively sophisticated combinatorial reasoning.

Consider the case where three entries, \(x_{i_1j_1}\), \(x_{i_2j_2}\), and \(x_{i_3j_3}\), are turned on. These indices satisfy \(i_1 \neq i_2 \neq i_3\) and \(j_1 \neq j_2 \neq j_3\). Our goal is to demonstrate,
\ba
\label{goalijijm3}
A({\mathbb I}_n)\! \xrightarrow[{\rm except}~ x_{i_1j_1}, \, x_{{i_2},{j_2}}, \, x_{{i_3},{j_3}} \neq 0]{H_m=0} 
F_{m,n}(i_1,j_1) + F_{m,n}(i_2,j_2) + F_{m,n}(i_3,j_3),
\ea
where \(i_1, i_2, i_3 \in \{1, \ldots, m\}\) and \(j_1, j_2, j_3 \in \{m+2, \ldots, n-1\}\) with  ${\rm min}(m,n-m-2)\geq 3$.

As before, the amplitude \(A({\mathbb I}_n)\) can be decomposed into contributions from different kinds of singular solutions. Denoting the contribution of singular solutions associated with \(\star\) as \(A_{{\mathbb I}_n}^{\star}\), we write
\ba
\label{goalijijm33}
A({\mathbb I}_n)\! \xrightarrow[{\rm except}~ x_{i_1j_1}, \, x_{{i_2},{j_2}}, \, x_{{i_3},{j_3}} \neq 0]{H_m=0} & 
A_{{\mathbb I}_n}^{\sigma_{i_1j_1} \sim 0} + A_{{\mathbb I}_n}^{\sigma_{i_2j_2} \sim 0} + A_{{\mathbb I}_n}^{\sigma_{i_3j_3} \sim 0} \nl
& + A_{{\mathbb I}_n}^{\sigma_{i_1j_1}, \sigma_{i_2j_2} \sim 0} + A_{{\mathbb I}_n}^{\sigma_{i_1j_1}, \sigma_{i_3j_3} \sim 0} + A_{{\mathbb I}_n}^{\sigma_{i_2j_2}, \sigma_{i_3j_3} \sim 0}
\nl
& + A_{{\mathbb I}_n}^{\sigma_{i_1j_1}, \sigma_{i_2j_2}, \sigma_{i_3j_3} \sim 0}.
\ea

This decomposition mirrors the structure observed in the two-entry case but introduces an additional layer of combinatorial complexity due to the interplay of three pairs of punctures. We now analyze each type of contribution and the resulting cancellations to demonstrate the desired result.

\subsection{Contribution of 1-Pinch Solutions}

Similar to \eqref{odafiosd0}, the contribution from the 1-pinch solution with \(\sigma_{i_1  j_1} \sim 0\) can be expressed as a linear combination of \((n-1)\)-point CHY integrals, which are components of \((n-1)\)-point YM amplitudes,
\ba
\label{afhdj}
A_{{\mathbb I}_n}^{\sigma_{i_1  j_1} \sim 0} =  \frac{(-1)^{n+1}}{s_{12\ldots m+1}}   
 \sum_{\e_{\hat j_1}} A(i_1 j_1 - \hat j_1) 
\sum_{\rho_{i_1} \in S_{m-1}}
X(s, \rho_{i_1})  
A_{\rho_{i_1} \, m+1 \ldots \hat j_1 \ldots n-1}^{\mathrm{reg}}.
\ea 
When \(x_{i_2j_2}\) and \(x_{i_3j_3}\) are also turned on, the CHY integral for the \((n-1)\)-point YM amplitude \(A_{\rho_{i_1} \, m+1 \ldots \hat j_1 \ldots n-1}\) receives additional contributions from singular solutions where either \(\sigma_{i_2  j_2} \sim 0\), \(\sigma_{i_3  j_3} \sim 0\), or both pinch simultaneously, in addition to regular solutions for the \((n-1)\)-point scattering equations \(SE(i_1)\),
\ba
A_{\rho_{i_1} \, m+1 \ldots \hat j_1 \ldots n-1} \!=\! A_{\rho_{i_1} \, m+1 \ldots \hat j_1 \ldots n-1}^{\mathrm{reg}}  
\!+\! A_{\rho_{i_1} \, m+1 \ldots \hat j_1 \ldots n-1}^{\sigma_{i_2  j_2} \sim 0}  
\!+\! A_{\rho_{i_1} \, m+1 \ldots \hat j_1 \ldots n-1}^{\sigma_{i_3  j_3} \sim 0}  
\!+\! A_{\rho_{i_1} \, m+1 \ldots \hat j_1 \ldots n-1}^{\sigma_{i_2  j_2}, \sigma_{i_3  j_3} \sim 0}.
\ea 
Substituting this expansion into \eqref{afhdj} and defining the discrepancy between \(A_{{\mathbb I}_n}^{\sigma_{i_1  j_1} \sim 0}\) and \(F_{m, n}(i_1, j_1)\) caused by a specific type of solution \(\star\) as \(\Delta_{i_1 j_1}^\star\), we find 
\ba
\label{fasjkd1}
A_{{\mathbb I}_n}^{\sigma_{i_1  j_1} \sim 0} - F_{m, n}(i_1, j_1) = \Delta_{i_1 j_1}^{i_2 j_2} + \Delta_{i_1 j_1}^{i_3 j_3} + \Delta_{i_1 j_1}^{i_2 j_2, i_3 j_3}.
\ea  

When \(\sigma_{i_2j_2} \sim 0\), the \((n-1)\)-point CHY integral for \(A_{\rho_{i_1} \, m+1 \ldots \hat j_1 \ldots n-1}^{\sigma_{i_2 j_2} \sim 0}\) simplifies into a linear combination of \((n-2)\)-point CHY integrals. Its expression closely resembles \eqref{odafiosd1},  
\allowdisplaybreaks[0]
\ba
\label{odafiosd1fassihf} 
A_{\rho_{i_1} \, m{+}1 \ldots \hat j_1 \ldots n-1}^{\sigma_{i_2 j_2} \sim 0} &= (-1)^n 
\sum_{\e_{\hat j_2}} A(i_2 j_2 - \hat j_2)
\int   d \mu_{[i_1 i_2]}
   {\rm Pf}'\Psi_{[i_1 i_2]}{\rm PT}_{[i_1 i_2]}  
   \ln & \times 
   \frac{{\rm PT}(\rho_{i_1} m{+}1)}{{\rm PT}(12 \ldots i\!\!\!\slash_1 \ldots i\!\!\!\slash_2 \ldots m{+}1)} 
\frac{1}{{{\mathring E}}_{i_2}} \Bigg|_{\substack{\sigma_{i_1} \to \sigma_{j_1} \\ \sigma_{i_2} \to \sigma_{j_2}}}\,,
\quad \text{where only } \sigma_{i_1j_1}, \sigma_{i_2j_2} \sim 0.
\ea
\allowdisplaybreaks[1]\!\!
However, there are important subtleties. In \eqref{odafiosd1fassihf}, contributions from solutions where \(\sigma_{i_3j_3} \sim 0\) are explicitly excluded. While these contributions are naturally absent in \eqref{odafiosd1} because \(x_{i3j3}\) is set to zero, they must be manually discarded here. 

Correspondingly, the associated discrepancy, \(\Delta_{i_1 j_1}^{i_2 j_2}\), still largely follows \eqref{odafiosd00}, with \(t_{i_1 j_1}^{i_2 j_2}\) defined as in \eqref{jaflt}. However, we emphasize that only the contributions from solutions where \(\sigma_{i_1j_1}, \sigma_{i_2j_2} \sim 0\) are considered. To clarify this distinction, we introduce a refined measure, \(d\mu_{[i_1 i_2]}^{\rm reg}\), which restricts the integration to regular solutions in the \((n-2)\)-point system \(\{1, 2, \ldots, n\} / \{i_1, j_1, i_2, j_2\} \cup \{\hat j_1, \hat j_2\}\),
\ba
\label{odafiosd00flkasjf}
\Delta_{i_1 j_1}^{i_2 j_2} = 
 -\sum_{\e_{\hat j_1}} A(i_1 j_1 - \hat j_1) 
\sum_{\e_{\hat j_2}} A(i_2 j_2 - \hat j_2)
\int d\mu_{[i_1 i_2]} ^{\rm reg} {\rm Pf}' \Psi_{[i_1 i_2]} {\rm PT}_{[i_1 i_2]} \, t_{i_1 j_1}^{i_2 j_2}.
\ea
In principle, we could have used \(d\mu_{[i_1 i_2]}^{\rm reg}\) in \eqref{odafiosd00} as well, but it was unnecessary there because solutions with \(\sigma_{i_3j_3} \sim 0\) contribute nothing. Here, however, this distinction is critical.

Similarly, \(\Delta_{i_1 j_1}^{i_3 j_3}\) follows the same structure and principles as \(\Delta_{i_1 j_1}^{i_2 j_2}\), preserving a consistent form throughout.

  When $\sigma_{i_2j_2}, \sigma_{i_3j_3}\sim 0$,
  the \((n-1)\)-point CHY integral for \(A_{\rho_{i_1} \, m+1 \ldots \hat j_1 \ldots n-1}^{\sigma_{i_2  j_2}, \sigma_{i_3  j_3} \sim 0}\) reduces further to a linear combination of \((n-3)\)-point CHY integrals. This process mirrors \eqref{odafiosd00}, and the resulting discrepancy is   
\ba
\label{delta33333}
\Delta_{i_1 j_1}^{i_2 j_2, i_3 j_3} = (-1)^n \sum_{\e_{\hat j_1}} A(i_1 j_1 - \hat j_1)   
\sum_{\e_{\hat j_2}}& A(i_2 j_2 - \hat j_2)  
\sum_{\e_{\hat j_3}} A(i_3 j_3 - \hat j_3) 
\ln &  \times
\int d\mu_{[i_1 i_2 i_3]} ^{\rm reg}\, {\rm Pf}' \Psi_{[i_1 i_2 i_3]} {\rm PT}_{[i_1 i_2 i_3]} \, t_{i_1 j_1}^{i_2 j_2, i_3 j_3},
\ea  
where    
\ba
\label{t33333}
t_{i_1 j_1}^{i_2 j_2, i_3 j_3} = -\frac{1}{s_{12 \ldots m+1}} \sum_{\rho_{i_1} \in S_{m-1}}  
X(s, \rho_{i_1}) \frac{{\rm PT}(\rho_{i_1} \, m+1)}{{\rm PT}(12 \ldots i \!\!\! \slash_1 \ldots i \!\!\! \slash_2 \ldots i \!\!\! \slash_3 \ldots m+1)}  
\frac{1}{{\mathring E}_{i_2} {\mathring E}_{i_3}} \Bigg|_{\substack{\sigma_{i_1} \to \sigma_{j_1} \\ \sigma_{i_2} \to \sigma_{j_2} \\ \sigma_{i_3} \to \sigma_{j_3}}}.
\ea 
Here we can use either $d\mu_{[i_1 i_2 i_3]} ^{\rm reg}$ or $d\mu_{[i_1 i_2 i_3]} $ in \eqref{delta33333}.

By simple relabeling, we obtain analogous discrepancies, 
\ba
\label{fasjkd2}
A_{{\mathbb I}_n}^{\sigma_{i_2  j_2} \sim 0} - F_{m, n}(i_2, j_2) = \Delta_{i_2 j_2}^{i_1 j_1} + \Delta_{i_2 j_2}^{i_3 j_3} + \Delta_{i_2 j_2}^{i_1 j_1, i_3 j_3},
\\
\label{fasjkd3}
A_{{\mathbb I}_n}^{\sigma_{i_3  j_3} \sim 0} - F_{m, n}(i_3, j_3) = \Delta_{i_3 j_3}^{i_1 j_1} + \Delta_{i_3 j_3}^{i_2 j_2} + \Delta_{i_3 j_3}^{i_1 j_1, i_2 j_2}.
\ea 

This systematic approach demonstrates how the contributions of 1-pinch solutions interact with higher-order pinch solutions through cascading discrepancies.

\subsection{Contribution of 2-Pinch Solutions and  Corresponding Cancellations}

When two pairs of punctures pinch, \(\sigma_{i_1j_1}, \sigma_{i_2j_2} \sim 0\), the \(n\)-point CHY integrals for \(A_{\mathbb{I}_n}\) reduce to a linear combination of \((n-2)\)-point CHY integrals. Consequently, the contribution \(A_{{\mathbb{I}}_n}^{\sigma_{i_1j_1}, \sigma_{i_2j_2} \sim 0}\) is still largely described by \eqref{jafl}, with the coefficient \(t_{i_1j_1, i_2j_2}\) defined in \eqref{jafltfkalewf}. However, similar to \eqref{odafiosd00flkasjf}, we must emphasize that only contributions from solutions where \(\sigma_{i_3}, \sigma_{j_3}\) do not pinch are included in the \((n-2)\)-point CHY integral,
\ba
\label{jaflfasjkdf}
A_{{\mathbb I}_n}^{\sigma_{i_1j_1}, \sigma_{i_2j_2} \sim 0} = - 
\sum_{\e_{\hat j_1}} A(i_1j_1 -\hat j_1) 
\sum_{\e_{\hat j_2}} A(i_2j_2 -\hat j_2)
\int d\mu_{[i_1 i_2]} ^{\rm reg} {\rm Pf}' \Psi_{[i_1 i_2]} {\rm PT}_{[i_1 i_2]} \, t_{i_1j_1, i_2j_2}.
\ea
The identity \eqref{mp2keyid} holds for any singular solutions where only \(\sigma_{i_1j_1}, \sigma_{i_2j_2} \sim 0\). Thus, the cancellation \eqref{deltacan} applies to \eqref{odafiosd00flkasjf}, its relabeling, and \eqref{jaflfasjkdf}. These cancellations ensure that the contributions from 1-pinch and 2-pinch solutions partially offset each other when summed appropriately.

This mechanism highlights the intricate interplay between contributions from different types of singular solutions and ensures that the resulting expressions simplify in alignment with the overarching structure of the CHY formalism.

\subsection{Contribution of 3-Pinch Solutions and Corresponding Cancellations}
When three pairs of punctures pinch $\sigma_{i_1j_1}, \sigma_{{i_2},{j_2}}, \sigma_{{i_3}, {j_3}}\sim 0$, the $n$-pt CHY integrals for $A_{\mathbb I_n}$  reduces to a linear combination of $(n-3)$-pt CHY integrals, 
\ba 
\label{fasjkd4}
A_{{\mathbb I}_n}^{\sigma_{i_1j_1}, \sigma_{{i_2},{j_2}}, \sigma_{{i_3}, {j_3}}\sim 0}  =    (-1)^n\sum_{\e_{\hat j_1}} & A(i_1j_1 -\hat j_1) 
\sum_{\e_{\hat j_2}} 
A(i_2j_2 -{\hat j_2})\sum_{\e_{\hat j_3}}
A(i_3j_3 -{\hat j_3})
\ln
& 
\times
\int  d\mu_{[i_1 i_2 i_3]}^{\rm reg}  {\rm Pf}'\Psi_{[i_1 i_2 i_3]} {\rm PT}_{[i_1i_2 i_3]} t_{i_1j_1,i_2j_2,i_3j_3}\,,
\ea 
with
\ba 
\label{t3332}
t_{i_1j_1,i_2j_2,i_3j_3} =   \frac{ {\rm PT}( 12 \ldots  m+1)}{{\rm PT}(12 \ldots i\!\!\!\slash_1 \ldots i\!\!\!\slash_2 \ldots
i\!\!\!\slash_3 \ldots m+1)}
\frac{1}{{ {\mathring E}}_{i_1} { {\mathring E}}_{i_2} { {\mathring E}}_{i_3}  } \Bigg|_{\substack{\sigma_{i_1}\to \sigma_{j_1}
\\
\sigma_{i_2}\to \sigma_{j_2}
\\
\sigma_{i_3}\to \sigma_{j_3}
}}\,.
\ea 
We see it share the same $(n-3)$-pt measure, reduced Pfaffian and PT factors as those in $\Delta_{i_1j_1}^{i_2j_2, i_3j_3} $ given in  \eqref{delta33333} and their relabelling $\Delta_{i_2j_2}^{i_1j_1, i_3j_3} , \Delta_{i_3j_3}^{i_1j_1, i_2j_3} $  except for a different annoying coefficient $t_{i_1j_1,i_2j_2,i_3j_3}$.  

However, both $t_{i_1j_1,i_2j_2,i_3j_3}$ given in \eqref{t3332} and $t_{i_1j_1}^{i_2j_2,i_3j_3}$ given in \eqref{t33333} and its relabelling  $t_{i_2j_2}^{i_1j_1,i_3j_3}$  and $t_{i_3j_3}^{i_1j_1,i_2j_2}$ still just involves $m+1$ punctures    $\{\sigma_1,\sigma_2,\ldots,\sigma_{m+1}\}_{\sigma_{i_1}\to \sigma_{j_1},\sigma_{i_2}\to \sigma_{j_2}, \sigma_{i_3}\to \sigma_{j_3}}$.  Similar to \eqref{mp2keyid},  on the support of $(m-3)$  scattering equations ${\mathring E}_{a}$ all 
 $a\in \{1,2,\ldots,m\}\setminus\{i_1,i_2, i_3\}$,  which are part of the $(n-3)$-pt scattering equations $SE(i_1,i_2,i_3)$, there is an identity for the four $t$'s
 \ba 
 \label{mp3keyid}
 t_{i_1j_1}^{i_2j_2,i_3j_3} + t_{i_2j_2}^{i_1j_1,i_3j_3}+t_{i_3j_3}^{i_1j_1,i_2j_2} + t_{i_1j_1,i_2j_2,i_3j_3} \overset{{\mathring E}_{a}=0 ~ \forall a\in \{1,2,\ldots,m\}\setminus\{i_1,i_2, i_3\} }{=} 0
 \ea 
  Its proof  is postponed to \cref{sec:sec8}.
Correspondingly, all $(n-3)$-pt CHY integrals in \eqref{fasjkd1}, \eqref{fasjkd2},
 \eqref{fasjkd3} and  \eqref{fasjkd4} cancel each other when we add them up,
 \ba 
 \label{deltacan333}
 \Delta_{i_1j_1}^{i_2j_2,i_3j_3}+  \Delta_{i_2j_2}^{i_1j_1,i_3j_3} + \Delta_{i_3j_3}^{i_1j_1,i_2j_2} +
 A_{{\mathbb I}_n}^{\sigma_{i_1j_1},\sigma_{i_2j_2}, \sigma_{i_3j_3}\sim 0} = 0\,.
 \ea 
 Together with previous cancellations \eqref{deltacan} and their relabellings, we get 
 \eqref{goalijijm3}.

\section{General Cases \label{sec:sec8}}

In this section, we explore the general scenario where \(p\) entries, none of which are in the same row or column, are turned on within the reduced structure of the \(n\)-point YM amplitude \(A(\mathbb{I}_n)\) under the condition \(h_m = 0\). This setup generalizes previous cases by allowing for arbitrary distributions of activated entries across distinct rows and columns, with \(p \leq {\rm min}(m,n - m - 2)\). The resulting CHY integral incorporates contributions from increasingly complex singular solutions, ranging from 1-pinch to \(p\)-pinch configurations. Each type of contribution is systematically organized and analyzed to uncover the combinatorial cancellations that ultimately simplify the amplitude. Finally, we consider the case where an arbitrary number of entire rows or columns  are turned on.

\subsection{Turning on an Arbitrary Number of Entries Not in the Same Row or Column}

We now generalize to cases where \(p\) entries, none of which are in the same row or column, are turned on in the reduced  \(n\)-point YM amplitude \(A(\mathbb{I}_n)\) with \(h_m = 0\). Specifically, we turn on \(x_{i_1j_1}, x_{i_2j_2},\ldots, x_{i_p, j_p} \in \hat{H}_m\), where all \(i_1, \ldots, i_p\) and \(j_1, \ldots, j_p\) are distinct, while keeping all other \(x_{ij} \in \hat{H}_m\) set to zero. This analysis applies for \(2\leq p \leq {\rm min}(m, n-m-2)\). Remind that for $p=1$, there are no discrepancy terms and the factorization formula \eqref{facproofbem3} have been proved to be true for any value of $m$ and $n$ with $1\leq m \leq n-3$.  For \(2\leq p \leq {\rm min}(m, n-m-2)\), the CHY integral for the amplitude incorporates contributions from 1-pinch to \(p\)-pinch solutions, expressed as
\ba
\label{jakfls}
A_{\mathbb{I}_n} \xrightarrow[
\substack{{\rm except}~ x_{i_{1}j_{1}}, \ldots, x_{i_{p}j_{p}} \neq 0
}]{H_m=0}   \sum_{w=1}^p  A_{\mathbb{I}_n}^{\sigma_{i_{w}j_{w}} \sim 0}+  \sum_{u=2}^{p} \sum_{1 \leq r_1 < r_2 < \ldots < r_u \leq p} A_{\mathbb{I}_n}^{\sigma_{i_{r_1}j_{r_1}}, \ldots, \sigma_{i_{r_u}j_{r_u}} \sim 0}.
\ea

\subsubsection{Contribution of 1-Pinch Solutions}

The contribution from a single-pinch solution, where \(\sigma_{i_{w}j_{w}} \sim 0\) with $w=1,2,\ldots,p$, is given by
\ba
A_{\mathbb{I}_n}^{\sigma_{i_{w}j_{w}} \sim 0} = F_{m, n}(i_{w}, j_{w}) +\sum_{u=2}^{p} ~ \sum_{\substack{r_1,r_2,\ldots,r_{v-1},r_{v+1},\ldots,r_u
\\
{\rm with}~ r_v=w
\\
1 \leq r_1<r_2 < \ldots < r_u \leq p}} \Delta_{i_{r_v}j_{r_v}}^{i_{r_1}j_{r_1},\ldots, i\!\!\!\slash_{r_v}j\!\!\!\slash_{r_v} , \ldots,  i_{r_u}j_{r_u}},
\ea
where the second summation runs over \(u-1\) distinct values of \(r_\star\) ranging from \(1\) to \(p\), all different from \(w\). Once these \(r_\star\) are chosen, \(v\) is determined as the position of \(w\) in the ordered sequence of \(\{r_1, r_2, \ldots, r_{v-1}, r_{v+1}, \ldots, r_u, w\}\), sorted from smallest to largest. For convenience, we then set \(r_v = w\). Several illustrative lower-point examples are provided in \eqref{examplesim1}, \eqref{examplesim2}, \eqref{fasjkd1}, \eqref{fasjkd2}, and \eqref{fasjkd3}.

The descrapency  \(\Delta_{i_{r_v}j_{r_v}}^{i_{r_1}j_{r_1},\ldots, i\!\!\!\slash_{r_v}j\!\!\!\slash_{r_v} , \ldots,  i_{r_u}j_{r_u}}\) is computed as
\ba
\label{jakfls2}
\Delta_{i_{r_v}j_{r_v}}^{i_{r_1}j_{r_1},\ldots, i\!\!\!\slash_{r_v}j\!\!\!\slash_{r_v} , \ldots,  i_{r_u}j_{r_u}} & = (-1)^{(2n+3-u)u/2}  
\prod_{v'=1}^u  \sum_{\epsilon_{\hat{j}_{r_{v'}}}} A(i_{r_{v'}}j_{r_{v'}} - \hat{j}_{r_{v'}}) 
\ln 
\times & \int d\mu_{[i_{r_1} i_{r_2} \ldots i_{r_u}]} ^{\rm reg} {\rm Pf}'\Psi_{[i_{r_1} i_{r_2} \ldots i_{r_u}]}
{\rm PT}_{[i_{r_1} i_{r_2} \ldots i_{r_u}]}  t_{i_{r_v}j_{r_v}}^{i_{r_1}j_{r_1},\ldots, i\!\!\!\slash_{r_v}j\!\!\!\slash_{r_v} , \ldots,  i_{r_u}j_{r_u}},
\ea
with
\ba
\label{t33333g}
t_{i_{r_v}j_{r_v}}^{i_{r_1}j_{r_1},\ldots, i\!\!\!\slash_{r_v}j\!\!\!\slash_{r_v} , \ldots,  i_{r_u}j_{r_u}}& = -\frac{1}{s_{12\ldots m+1}} 
\ln   \times &
\sum_{\rho_{i_{r_v}} \in S_{m-1}} \frac{X(s, \rho_{i_{r_v}}) {\rm PT}(\rho_{i_{r_v}} m+1)}{{\rm PT}(12 \ldots i\!\!\!\slash_{r_1} \ldots i\!\!\!\slash_{r_2} \ldots \ldots i\!\!\!\slash_{r_u} \ldots m+1)} \frac{1}{\prod\limits_{\substack{v'=1\\
v'\neq v}}^u \mathring{E}_{i_{r_{v'}}}} \Bigg|_{\substack{\sigma_{i_{r_1}} \to \sigma_{j_{r_1}} \\ \sigma_{i_{r_2}} \to \sigma_{j_{r_2}} \\ \ldots \\ \sigma_{i_{r_u}} \to \sigma_{j_{r_u}}}},
\ea
where \(d\mu_{[i_{r_1} i_{r_2} \ldots i_{r_u}]}^{\rm reg}\), \({\rm Pf}'\Psi_{[i_{r_1} i_{r_2} \ldots i_{r_u}]}\), and \({\rm PT}_{[i_{r_1} i_{r_2} \ldots i_{r_u}]}\) are the \((n-u)\)-point measure, reduced Pfaffian, and Parke-Taylor factor for the particles \(\{1, 2, \ldots, n\} \setminus \{i_{r_1},j_{r_1}, i_{r_2},j_{r_2}, \ldots, i_{r_u},j_{r_u}\}\cup \{ \hat j_{r_1},\hat j_{r_2},\ldots, \hat j_{r_u}\}\), respectively.

\subsubsection{Contribution of \(u\)-Pinch Solutions}

For \(u\)-pinch solutions with \(2 \leq u \leq p\), the contribution is given by
\ba
\label{jakfls3}
A_{\mathbb{I}_n}^{\sigma_{i_{r_1}j_{r_1}}, \ldots, \sigma_{i_{r_u}j_{r_u}}} =(-1)^{(2n+3-u)u/2}   & 
\prod_{v=1}^u \sum_{\epsilon_{\hat{j}_{r_v}}} A(i_{r_v}j_{r_v} - \hat{j}_{r_v}) 
\ln 
\times \int  d\mu_{[i_{r_1} i_{r_2} \ldots i_{r_u}]}^{\rm reg} & 
{\rm Pf}'\Psi_{[i_{r_1} i_{r_2} \ldots i_{r_u}]} {\rm PT}_{[i_{r_1} i_{r_2} \ldots i_{r_u}]}   t_{i_{r_1}j_{r_1}, i_{r_2}j_{r_2}, \ldots, i_{r_u}j_{r_u}},
\ea
where  
\ba
t_{i_{r_1}j_{r_1}, i_{r_2}j_{r_2}, \ldots, i_{r_u}j_{r_u}} = \frac{{\rm PT}(12 \ldots m+1)}{{\rm PT}(12 \ldots i\!\!\!\slash_{r_1} \ldots i\!\!\!\slash_{r_2} \ldots \ldots i\!\!\!\slash_{r_u} \ldots m+1)} \frac{1}{\prod_{v=1}^u \mathring{E}_{i_{r_v}}} \Bigg|_{\substack{\sigma_{i_{r_1}} \to \sigma_{j_{r_1}} \\ \sigma_{i_{r_2}} \to \sigma_{j_{r_2}} \\ \ldots \\ \sigma_{i_{r_u}} \to \sigma_{j_{r_u}}}}.
\ea

\subsubsection{Identities for \(t\)-Coefficients}

The coefficients \(t_{i_{r_v}j_{r_v}}^{i_{r_1}j_{r_1},\ldots, i\!\!\!\slash_{r_v}j\!\!\!\slash_{r_v} , \ldots,  i_{r_u}j_{r_u}}\) and \(t_{i_{r_1}j_{r_1}, i_{r_2}j_{r_2}, \ldots, i_{r_u}j_{r_u}}\) involve only \(m+1\) punctures. On the support of the \((m-u)\) scattering equations \(\mathring{E}_a = 0\) for \(a \in \{1, 2, \ldots, m\} \setminus \{i_{r_1}, i_{r_2}, \ldots, i_{r_u}\}\), which are part of the \((n-u)\)-point scattering equations, the following identity holds,
\ba 
\label{keykeyidid}
\sum _{v=1}^u t_{i_{r_v}j_{r_v}}^{i_{r_1}j_{r_1},\ldots, i\!\!\!\slash_{r_v}j\!\!\!\slash_{r_v} , \ldots,  i_{r_u}j_{r_u}} + t_{i_{r_1}j_{r_1},i_{r_2}j_{r_2},\ldots, i_{r_u}j_{r_u}}   \overset{{\mathring E}_{a}=0 ~ \forall a\in \{1,2,\ldots,m\}\setminus\{i_{r_1},i_{r_2},\ldots, i_{r_u}\} } = 0\,.
\ea 
The proof of this identity is provided in the \cref{appproof}, ensuring cancellations among contributions of various \(u\)-pinch solutions,
\ba 
\label{keykeyidid2}
\sum _{v=1}^u \Delta_{i_{r_v}j_{r_v}}^{i_{r_1}j_{r_1},\ldots, i\!\!\!\slash_{r_v}j\!\!\!\slash_{r_v} , \ldots,  i_{r_u}j_{r_u}} + A_{\mathbb I_n}^{ \sigma_{i_{r_1}j_{r_1}},\sigma_{i_{r_2}j_{r_2}},\ldots, \sigma_{i_{r_u}j_{r_u}}\sim 0}   =0\,. 
\ea 
 As a result, we get
\ba 
A_{\mathbb I_n} \xrightarrow[
\substack{{\rm except}~ x_{i_{1}j_{1}}, \ldots, x_{i_{p}j_{p}} \neq 0
}]{H_m=0}    \sum\limits_{w=1}^p F_{m,n}(i_{w},j_{w})\,.
\ea

\subsection{Turning on an Arbitrary Number of Entire Rows or Columns}

We now consider the case where \(p'\) entire rows are turned on with \(1\leq p' \leq m\) for the \(n\)-point YM amplitude \(A(\mathbb{I}_n)\) under the constraint \(h_m = 0\). This corresponds to turning on all \(x_{ij} \in \hat{H}_m\) for \(i \in \{i_1, i_2, \ldots, i_p\}\) and \(m+2 \leq j \leq n-1\), while keeping all other \(x_{ij} \in \hat{H}_m\) killed off. Here $i_1, i_2, \ldots, i_p$ are all distinct. In this setup, the CHY integral for the amplitude receives contributions from 1-pinch, 2-pinch, up to \(p\)-pinch solutions with $p={\rm min}(p',n-m-2)$. Summing over these contributions, we have,
\ba
A_{\mathbb{I}_n} \to 
\sum_{w=1}^p \sum_{j_w=m+2}^{n-1} 
A_{\mathbb{I}_n}^{\sigma_{i_{w}j_w} \sim 0}
+
\sum_{u=2}^p \,\sum_{2 \leq r_1 < r_2 < \ldots < r_u \leq p} \,\sum_{\substack{m+2 \leq j_{r_1}, j_{r_2}, \ldots, j_{r_u} \leq n-1 \\ j_{r_1}, j_{r_2}, \ldots, j_{r_u} \text{are~all~distinct}}} A_{\mathbb{I}_n}^{\sigma_{i_{r_1}j_{r_1}}, \ldots, \sigma_{i_{r_u}j_{r_u}} \sim 0}.
\ea
The contribution from a single pinched solution, where $\sigma_{i_{w}j_{w}} \sim 0$ with $w=1,2,\ldots,p$,  is given by
\ba
&A_{\mathbb{I}_n}^{\sigma_{i_{w}j_{w}} \sim 0} = F_{m,n}(i_{w}, j_{w}) \ln
&~+ \sum_{u=2}^{p} ~\sum_{\substack{
r_1,r_2,\ldots,r_{v-1},r_{v+1},\ldots,r_u
\\
{\rm with}~
r_v=w
\\
1 \leq r_1< r_2  < \ldots < r_u \leq p }} ~\sum_{\substack{
j_{r_1},j_{r_2},\ldots,j_{r_{v-1}},j_{r_{v+1}},\ldots,j_{r_u}
\\
{\rm with}~m+2 \leq j_{r_1}, j_{r_2}, \ldots, j_{r_u} \leq n-1 \\ j_{r_1},j_{r_2}, \ldots, j_{r_u}  \text{are all distinct}}} ~\Delta_{i_{r_v}j_{r_v}}^{i_{r_1}j_{r_1},\ldots, i\!\!\!\slash_{r_v}j\!\!\!\slash_{r_v} , \ldots,  i_{r_u}j_{r_u}},
\ea
where \(\Delta_{i_{r_v}j_{r_v}}^{i_{r_1}j_{r_1},\ldots, i\!\!\!\slash_{r_v}j\!\!\!\slash_{r_v} , \ldots,  i_{r_u}j_{r_u}}\) is defined as in \eqref{jakfls2}. The contribution from \(u\)-pinch solutions with \(2 \leq u \leq p\) follows the same form as in \eqref{jakfls3}.

Combining all contributions, the amplitude becomes
\ba 
\label{jakfls4}
A_{\mathbb I_n} \to &
\sum_{w=1}^{p} \sum_{j_w=m+2}^{n-1}F_{m,n}(i_w,j_w) + 
\sum_{u=2}^{p}
\sum_{1\leq r_1 < r_2 < \ldots < r_u \leq p }  \sum_{
\substack{m+2\leq  j_{r_1},j_{r_2},\ldots, j_{r_u} \leq n-1
\\
j_{r_1},j_{r_2},\ldots, j_{r_u} {\rm ~are ~all~distinct}
}} 
\nl 
& \times 
(\sum _{v=1}^u \Delta_{i_{r_v}j_{r_v}}^{i_{r_1}j_{r_1},\ldots, i\!\!\!\slash_{r_v}j\!\!\!\slash_{r_v} , \ldots, i_{r_u}j_{r_u}} + A_{\mathbb I_n}^{ \sigma_{i_{r_1}j_{r_1}},\sigma_{i_{r_2}j_{r_2}},\ldots, \sigma_{i_{r_u}j_{r_u}}\sim 0})\,. 
\ea 
Using the identity \eqref{keykeyidid2}, the contributions simplify, and we recover
\ba
\label{jakfls42}
A_{\mathbb{I}_n} \xrightarrow[
\substack{{\rm except}~ x_{i_1 j}, \ldots, x_{i_pj} \neq 0
\\
\forall m+2\leq j\leq n-1 }]{H_m=0}\sum_{w=1}^p \sum_{j_w=m+2}^{n-1} F_{m,n}(i_{w}, j_w).
\ea
Equation \eqref{conijijijij2xx} works as a special case of 
 \eqref{jakfls42} with $p=2$. 

The result \eqref{totalcase} is a special case of \eqref{jakfls42} with \(p = m\), where all rows are turned on. Conversely, \eqref{totalcase} is the most general result, as \eqref{jakfls42} can be directly derived by killing off specific \(x_{ij}\) terms.

\section{Discussion\label{sec:sec9}}

This paper establishes a robust framework for analyzing \(n\)-point Yang-Mills (YM) amplitudes using the Cachazo-He-Yuan (CHY) formalism \cite{Cachazo:2013hca, Cachazo:2013iea}, revealing a consistent recursive structure that decomposes amplitudes into lower-point amplitudes \cite{Zhang:2024iun}. By systematically examining singular solutions at various pinch levels—1-pinch, 2-pinch, and beyond—we have made the hidden zeros of YM amplitudes manifest \cite{Arkani-Hamed:2023swr} and connected them to their broader recursive organization. These results deepen our understanding of the CHY framework's ability to describe gauge theories and illuminate the role of singularities in amplitude decomposition.

The new factorization formulas provide a natural avenue for constructing higher-point YM amplitudes from lower-point ones by recursively applying specific kinematic constraints. Auxiliary variables, similar to those in Britto-Cachazo-Feng-Witten recursion \cite{Britto:2005fq}, and tools like the residue theorem could streamline this process, offering a systematic method for navigating the factorization subspaces. These formulas inherently involve three-point YM amplitudes, which, while ill-defined in four-dimensional spacetime, acquire physical meaning in the collinear limit by producing well-defined prefactors. This behavior aligns with splitting functions in the collinear regime, potentially offering applications in precision collider phenomenology, particularly in computing collinear contributions to scattering amplitudes \cite{Catani:1998nv, Altarelli:1977zs}.

The function \( X(s, \rho) \), central to the recursive structure, encapsulates planar poles dictated by the  factorization conditions of the CHY formalism. Although these poles arise naturally from the framework, their explicit appearance in \( X(s, \rho) \) remains nontrivial and warrants further exploration. Developing efficient methods, such as recursive constructions or symbolic expansions, could shed light on its structure. Additionally, potential Feynman-like rules tailored to \( X(s, \rho) \) might simplify its computation and enhance its interpretative power.

The new factorization formulas also suggest connections to the concept of minimal kinematics, where specific Mandelstam variables are set to zero (cf. \cite{Cachazo:2013iea,Cachazo:2020uup, Early:2018zuw, Early:2024nvf}). This regime introduces regular solutions alongside singular ones in the scattering equations, offering a broader perspective on YM amplitudes. Investigating these conditions could uncover additional structures or symmetries, providing new insights into the recursive patterns revealed by the factorizations.

Another intriguing direction involves extending the new factorizations to amplitudes derived from higher-dimensional operators, such as \( F^3 \) amplitudes in bosonic open string theories \cite{Broedel:2012rc,He:2016iqi,Garozzo:2018uzj}, or the $({\rm DF})^2+{\rm YM}+\phi^3$ effective field theory \cite{Azevedo:2018dgo, He:2018pol, Azevedo:2017lkz}, as well as to GR amplitudes, where reducing the second reduced Pfaffian presents unique challenges. These extensions could generalize the recursive structures to stringy integrals \cite{Mafra:2011nv, Stieberger:2009hq,He:2019drm} and reveal new insights into the interplay between gauge, gravity, and string theories. Additionally, the application of the framework to loop-level CHY formulas (c.f.\cite{Geyer:2015bja, Geyer:2015jch, Cachazo:2015aol,Geyer:2016wjx, Geyer:2019hnn, Geyer:2022cey,He:2017spx,Feng:2022wee,Dong:2023stt}) could uncover novel structures in YM integrands at higher loops.

The curve integral formalism offers a complementary perspective by framing amplitude recursion through moduli spaces of algebraic curves. Originally developed for scalar theories like \(\text{Tr}(\phi^3)\) \cite{Arkani-Hamed:2023swr}, it extends naturally to gauge and gravity amplitudes using a scaffolding procedure \cite{Arkani-Hamed:2023jry}. While our new factorizations focus on specific kinematic constraints in YM amplitudes, combining these approaches could bridge the geometric insights of curve integrals with the recursive structures revealed by the new factorizations, particularly in higher-loop contexts \cite{ Arkani-Hamed:2024nhp, Arkani-Hamed:2024yvu, 
Arkani-Hamed:2023lbd, Arkani-Hamed:2023mvg,  De:2024wsy,Arkani-Hamed:2024tzl}.

\section*{Acknowledgements}
We sincerely appreciate Freddy Cachazo for proposing this study and providing valuable insights and guidance that shaped its direction. His expertise has been an inspiring influence throughout this project. 
We also thank Jin Dong for his comments on the manuscript and Qu Cao, Song He,  Yang Li, Sebastian P\"ogel,  Bruno Umbert, Ellis Ye Yuan, and Kang Zhou for useful discussions.    This research was supported in part by a grant
from the Gluskin Sheff/Onex Freeman Dyson Chair in Theoretical Physics and by Perimeter
Institute. Research at Perimeter Institute is supported in part by the Government of Canada
through the Department of Innovation, Science and Economic Development Canada and by
the Province of Ontario through the Ministry of Colleges and Universities.

\appendix

\section{Coefficients in BCJ Relations \label{BCJco}} 
This appendix provides the explicit form of the coefficients \({\cal B}_{m+1,i}[\alpha, \beta | \pi]\) in the BCJ relations for color-ordered amplitudes. These coefficients are essential for expressing the ratios of Mandelstam variables \(X(s, \rho)\) as defined in \eqref{Xsrhoori}. To provide a clear explanation, we first introduce the simpler case of \({\cal B}_{1,3}[\alpha', \beta' | \pi']\) with $|\alpha'|=|\alpha|$ and  $|\beta'|=|\beta|$ and then generalize it to \({\cal B}_{m+1,i}[\alpha, \beta | \pi]\) through a relabeling procedure.

For an $(m+2)$-point massless scattering system with  two ordered sets \(\alpha'=\{4,5,\ldots,|\alpha|+3 \}\) and \(\beta'=\{|\alpha|+4,|\alpha|+5,\ldots,m+2\}\), the PT factor obeys the following BCJ relation \cite{Bern:2019prr, Bern:2008qj},
\ba
\label{appdefa2}
{\rm PT} (\alpha', 3, \beta', 1) \cong \sum_{\pi' \in \alpha' \shuffle \beta'} {\rm PT}(3, \pi', 1)
 {\cal B}_{1,3}[\alpha', \beta' | \pi' ],
\ea
where the coefficients  are defined as 
\ba 
{\cal B}_{1,3}[\alpha', \beta' | \pi']:= 
\prod_{c=4}^{|\alpha|+3} \frac{\mathcal{F}_c(3, \pi', 1)}{s_{13\, c+1 \,c+2 \ldots m+2}}\,.
\ea 
The numerator $\mathcal{F}_c(3, \pi', 1)$ is defined as
\ba
\mathcal{F}_c(\gamma) :=
\begin{cases}
\sum_{l=t_c}^{m+1} \mathcal{S}_{c, \gamma(l)} & \text{if } t_{c-1} < t_c \\
-\sum_{l=1}^{t_c} \mathcal{S}_{c, \gamma(l)} & \text{if } t_{c-1} > t_c
\end{cases}
+ 
\begin{cases}
s_{13\, c+1 \,c+2 \ldots m+2} & \text{if } t_{c-1} < t_c < t_{c+1} \\
-s_{13\, c+1 \,c+2 \ldots m+2} & \text{if } t_{c-1} > t_c > t_{c+1} \\
0 & \text{otherwise}
\end{cases}
\,.
\ea 
In the above, \(t_c\) is the position of label \(c\) in the ordered set \(\gamma = \{3, \pi', 1\}\), except for specific cases where \(t_3 = t_5\) and \(t_{|\alpha|+4} = 0\). $\gamma(l)$ is the $l$-th entry of the ordered set $\gamma$. The subscripts \(c+1\) and \(c+2\) are understood modulo \(m+2\). The terms \(\mathcal{S}_{a, b}\) are defined as
\ba 
\mathcal{S}_{a, b} :=
\begin{cases}
s_{ab} & \text{if } a < b \text{ or } b \in \{1, 3\}, \\
0 & \text{otherwise}.
\end{cases}
\ea
Note that in these definitions, the label \(2\) is excluded from the indices of the Mandelstam variables. Interestingly, the BCJ relations in \eqref{appdefa2} remain valid even if particle \(2\) becomes massive. More explanations can be found below \eqref{jkfasldkafdl24ex}.

What we require in \eqref{Xsrhoori} and \eqref{eq25g} are the coefficients \({\cal B}_{m+1, i}[\alpha, \beta | \pi]\), where \(\alpha \sqcup \beta = \{1, 2, \ldots, m\} \setminus \{i\}\) and \(\pi\) is a permutation of \(\{1, 2, \ldots, m\} \setminus \{i\}\). These coefficients are derived from \({\cal B}_{1,3}[\alpha', \beta' | \pi']\) as defined in \eqref{appdefa2} through a substitution rule,
\ba
g = \{1, 3, 4, \ldots, m+2\} \to \{m+1, i, \alpha(1),\alpha(2), \ldots, \alpha(|\alpha|),\beta(1),\beta(2), \ldots, \beta(|\beta|)\}   .
\ea
Using this rule, the coefficients are expressed as
\ba
{\cal B}_{m+1, i}[\alpha, \beta | \pi] := {\cal B}_{1,3}\big[\alpha', \beta' \big| \pi|_{g^{-1}}\big]\Big|_{g}.
\ea
where $g^{-1}= \{m+1, i, \alpha,\beta\} \to \{1, 3, 4, \ldots, m+2\}$ such that $\pi|_{g^{-1}}\big|_{g}=\pi$.

\section{Examples of Discrepancy Cancellations  
\label{appb22}}

This appendix provides two detailed examples to illustrate how discrepancies cancel when turning on two entries in the YM amplitude \(A(\mathbb{I}_n)\) under the condition \(h_m = 0\). Specifically, we consider cases where the two entries are not in the same row or column in $\hat H_m$, demonstrating how the cancellations occur when summing over contributions from different types of solutions.

\subsection{General \(n\)-Point Case with \(m=2\) }  

For an \(n\)-point YM amplitude \(A(\mathbb{I}_n)\) with \(h_2=0\) and \(n \geq 6\), the amplitude vanishes when \({\hat H}_2 = 0\). Turning on \(x_{14}, x_{25} \neq 0\) while keeping all other \(x_{ij} \in {\hat H}_2\) set to zero leads to
\ba
A_{\mathbb{I}_n} \to  
A_{\mathbb{I}_n}^{\sigma_{14} \sim 0} + 
A_{\mathbb{I}_n}^{\sigma_{25} \sim 0} + 
A_{\mathbb{I}_n}^{\sigma_{14}, \sigma_{25} \sim 0}.
\ea 
The contributions $A_{\mathbb{I}_n}^{\sigma_{14} \sim 0}$ are given as follows, 
\ba
A_{\mathbb{I}_n}^{\sigma_{14} \sim 0} = F_{2,n}(1,4) - \sum_{\epsilon_{\hat{4}}} A(14 -\hat{4}) \sum_{\epsilon_{\hat{5}}} A(25 -\hat{5}) 
\int d\mu_{[12]} {\rm Pf}' \Psi_{[12]} {\rm PT}(34 \ldots n-1)  t_{14}^{25}\,,
\ea 
with 
\ba 
t_{14}^{25}= -\frac{s_{12} + s_{23}}{s_{12}s_{123}}
\frac{1}{\sigma_{53} \big(\frac{s_{12}}{\sigma_{54}} + \frac{s_{23}}{\sigma_{53}}\big)}\,.
\ea 
Here, \(d\mu_{[12]}\) and \({\rm Pf}'\Psi_{[12]}\) represent the \((n-2)\)-point CHY measure and reduced Pfaffian, respectively, for particles \(\{1, 2, \ldots, n\}\setminus\{1, 2, 4, 5\} \cup \{\hat{4}, \hat{5}\}\).  
Similarly,  
\ba
A_{\mathbb{I}_n}^{\sigma_{25} \sim 0} = F_{2,n}(2,5) - \sum_{\epsilon_{\hat{4}}} A(14 -\hat{4}) \sum_{\epsilon_{\hat{5}}} A(25 -\hat{5}) 
\int d\mu_{[12]} {\rm Pf}' \Psi_{[12]} {\rm PT}(34 \ldots n-1) t_{25}^{14}\,,
\ea  
with 
\ba 
t_{25}^{14}=
\frac{ s_{13}}{s_{12}s_{123}} \frac{1}{\sigma_{43} \big(  \frac{s_{12}}{\sigma_{45}}+ \frac{s_{13}}{\sigma_{43}} 
\big)}\,.
\ea 
The contribution from both pairs of punctures pinching simultaneously is  
\ba
A_{\mathbb{I}_n}^{\sigma_{14}, \sigma_{25} \sim 0} = -\sum_{\epsilon_{\hat{4}}} A(14 -\hat{4}) \sum_{\epsilon_{\hat{5}}} A(25 -\hat{5}) 
\int d\mu_{[12]} {\rm Pf}' \Psi_{[12]} {\rm PT}(34 \ldots n-1) 
\,,
\ea
with
\ba 
t_{14,25}=  \frac{1}{\sigma_{45} \sigma_{53} \big(\frac{s_{12}}{\sigma_{45}} + \frac{s_{13}}{\sigma_{43}}\big) \big(\frac{s_{12}}{\sigma_{54}} + \frac{s_{23}}{\sigma_{53}}\big)}\,.
\ea 
Algebraic cancellations between the terms lead to the identity, $t_{14}^{25}+t_{25}^{14}+ t_{14,25}=0$, i.e.,  
\ba
-\frac{s_{12} + s_{23}}{s_{12}s_{123}} \frac{1}{\sigma_{53} \big(\frac{s_{12}}{\sigma_{54}} + \frac{s_{23}}{\sigma_{53}}\big)} + 
\frac{s_{13}}{s_{12}s_{123}} \frac{1}{\sigma_{43} \big(  \frac{s_{12}}{\sigma_{45}}+\frac{s_{13}}{\sigma_{43}}\big)} + 
\frac{1}{\sigma_{45} \sigma_{53} \big(\frac{s_{12}}{\sigma_{45}} + \frac{s_{13}}{\sigma_{43}}\big) \big(\frac{s_{12}}{\sigma_{54}} + \frac{s_{23}}{\sigma_{53}}\big)} = 0.
\ea 
Thus, \(A_{\mathbb{I}_n} \to F_{2,n}(1,4) + F_{2,n}(2,5)\).

\subsection{General \(n\)-Point Case with \(m=3\)  }

For an \(n\)-point YM amplitude \(A(\mathbb{I}_n)\) with \(h_3=0\) and \(n \geq 7\), turning on \(x_{15}, x_{26} \neq 0\) while keeping all other \(x_{ij} \in {\hat H}_3\) set to zero leads to 
\ba
A_{\mathbb{I}_n} \to 
A_{\mathbb{I}_n}^{\sigma_{15} \sim 0} + 
A_{\mathbb{I}_n}^{\sigma_{26} \sim 0} + 
A_{\mathbb{I}_n}^{\sigma_{15}, \sigma_{26} \sim 0}.
\ea  
Each contribution is expressed as
\ba
&A_{\mathbb{I}_n}^{\sigma_{15} \sim 0} = F_{3,n}(1,5) - \sum_{\epsilon_{\hat{5}}} A(15 -\hat{5}) \sum_{\epsilon_{\hat{6}}} A(26 -\hat{6}) 
\int d\mu_{[12]} {\rm Pf}' \Psi_{[12]} {\rm PT}(34 \ldots n-1) t_{15}^{26},
\nl
&A_{\mathbb{I}_n}^{\sigma_{26} \sim 0} = F_{3,n}(2,6)  - \sum_{\epsilon_{\hat{5}}} A(15 -\hat{5}) \sum_{\epsilon_{\hat{6}}} A(26 -\hat{6})
\int  d\mu_{[12]} {\rm Pf}' \Psi_{[12]} {\rm PT}(34 \ldots n-1) t_{26}^{15},
\nl
&A_{\mathbb{I}_n}^{\sigma_{15}, \sigma_{26} \sim 0} =  - \sum_{\epsilon_{\hat{5}}} A(15 -\hat{5}) \sum_{\epsilon_{\hat{6}}} A(26 -\hat{6})
\int  d\mu_{[12]} {\rm Pf}' \Psi_{[12]} {\rm PT}(34 \ldots n-1) t_{15,26}.
\ea  
The coefficients \(t_{15}^{26}\), \(t_{26}^{15}\), and \(t_{15,26}\) depend on specific puncture and momentum configurations,
\ba 
t_{15}^{26}=&
-\Bigg(
\frac{\left(s_{12}+s_{23}+s_{24}\right)
   \left(s_{123}+s_{34}\right)}{s_{12} s_{123}
   s_{1234}}
   \frac{1}{\sigma_{63}}
   + 
   \frac{s_{24} \left(s_{13}+s_{23}+s_{34}\right)}{s_{12}
   s_{123} s_{1234}} 
     \frac{\sigma_{34}}{\sigma_{36}\sigma_{64}}
   \Bigg)
   \nl 
   &
\times \frac{1}{\big( \frac{s_{12}}{\sigma_{65}} + \frac{ s_{23}}{\sigma_{63}}
   +  \frac{ s_{24}}{\sigma_{64}}\big)}  \,,
\nl 
t_{26}^{15}=&
-\Bigg(
\frac{-\left(\left(s_{13}+s_{14}\right) s_{23}
   \left(s_{13}+s_{23}+s_{34}\right)\right)-s_{12} \left(s_{13}
   s_{23}+s_{14} \left(s_{23}+s_{34}\right)\right)}{s_{12}
   s_{23} s_{123} s_{1234}}
   \frac{1}{\sigma_{53}}
 \nl 
 &  
-\frac{s_{14} \left(s_{12}+s_{23}\right)
   \left(s_{13}+s_{23}+s_{34}\right)}{s_{12} s_{23} s_{123}
   s_{1234}} 
     \frac{\sigma_{34}}{\sigma_{35}\sigma_{54}}
   \Bigg)
\times \frac{1}{\big( \frac{s_{12}}{\sigma_{56}} + \frac{ s_{13}}{\sigma_{53}}
   +  \frac{ s_{14}}{\sigma_{54}}\big)}  \,,
\nl
 t_{15,26}=& \frac{1}{\sigma_{56}\sigma_{63}\big( \frac{s_{12}}{\sigma_{56}} + \frac{ s_{13}}{\sigma_{53}}
   +  \frac{ s_{14}}{\sigma_{54}}\big)\big( \frac{s_{12}}{\sigma_{65}} + \frac{ s_{23}}{\sigma_{63}}
   +  \frac{ s_{24}}{\sigma_{64}}\big)}   \,.
\ea 
 Importantly, the following identity ensures cancellation, \ba
 \label{afdnkdj}
t_{15}^{26} + t_{26}^{15} + t_{15,26} \overset{\frac{s_{31}}{\sigma_{35}} + \frac{s_{32}}{\sigma_{36}} + \frac{s_{34}}{\sigma_{34}} = 0}{=} 0.
\ea
Thus, \(A_{\mathbb{I}_n} \to F_{3,n}(1,5) + F_{3,n}(2,6)\).

\section{Proof of the Key Identities for $t$-Coefficients \eqref{keykeyidid}\label{appproof} }

This appendix provides a rigorous proof of the key identities \(\eqref{keykeyidid}\) for \(t\)-coefficients, ensuring their validity through detailed analysis and derivations.

\subsection{Preliminary Analysis} 

Plugging in the explicit expressions for the \(t\)-coefficients, the key identity \(\eqref{keykeyidid}\) reduces to the following form,
\ba 
\label{jkfasldk}
\sum_{v=1}^{u}
 \sum_{\rho_{i_{r_v}}\in S_{m-1}}
   X ( s,\rho_{i_{r_v}}) & {\rm PT}(    \rho_{i_{r_v}} m+1) {\mathring E}_{i_{r_v}} 
\ln
 -&s_{12\ldots m+1}{
   {\rm PT}(   12\ldots m+1)  
   }
\Bigg|_{\substack{\sigma_{i_{r_1}}\to \sigma_{j_{r_1}}
\\
\sigma_{i_{r_2}}\to \sigma_{j_{r_2}}
\\ 
\ldots
\\
\sigma_{i_{r_u}}\to \sigma_{j_{r_u}}
}}
\overset{{\mathring E}_{a}=0 ~ \forall a\in \{1,2,\ldots,m\}\setminus\{i_{r_1},i_{r_2},\ldots, i_{r_u}\} }  =  0\,,
\ea 
valid for \(2 \leq u \leq p \leq m\).

Here, \(X(s, \rho_{i_{r_v}})\) is a ratio of kinematic invariants \(s\) determined by the 1-pinch solutions. Although we provided an explicit expression for \(X(s, \rho_{i_{r_v}})\) in \eqref{Xsrhoori}, it is originally defined through the relation in \eqref{ansatz}. This definition is particularly useful for our proof. For clarity, we rewrite \eqref{ansatz} here,
\ba
\label{jkfasldk2}
 \sum_{\rho_{i_{r_v}} \in S_{m-1}} 
X(s, \rho_{i_{r_v}}) & {\rm PT}(\rho_{i_{r_v}}, m+1) \mathring{E}_{i_{r_v}}
\ln 
-&
s_{12\ldots m+1} {\rm PT}(12\ldots m+1)
\Bigg|_{\substack{
\sigma_{i_{r_1}} \to \sigma_{j_{r_1}} \\
\sigma_{i_{r_2}} \to \sigma_{j_{r_2}} \\
\ldots \\
\sigma_{i_{r_u}} \to \sigma_{j_{r_u}}
}}
\overset{{\mathring E}_a = 0 \, \forall a \in \{1, 2, \ldots, m\} \setminus \{i_{r_v}\}} = 0,
\ea
for every \(v = 1, 2, \ldots, u\). 

Since \(X(s, \rho_{i_{r_v}})\) depends solely on \(s\), the punctures in \(\eqref{ansatz}\) can be relabeled. This allows \(\eqref{jkfasldk2}\) to formally resemble \(\eqref{jkfasldk}\) by setting \(u = 1\).

The \(\mathring{E}_a\) terms in \(\eqref{jkfasldk}\) and \(\eqref{jkfasldk2}\) can be uniformly expressed as:
\ba
\label{shorteq23r}
\mathring{E}_a = \sum_{\substack{a' = 1 \\ a' \neq a}}^{m+1} \frac{s_{aa'}}{\sigma_{aa'}}
\Bigg|_{\substack{
\sigma_{i_{r_1}} \to \sigma_{j_{r_1}} \\
\sigma_{i_{r_2}} \to \sigma_{j_{r_2}} \\
\ldots \\
\sigma_{i_{r_u}} \to \sigma_{j_{r_u}}}},
\quad \forall a \in \{1, 2, \ldots, m\}.
\ea

These \(\mathring{E}_a\) terms serve two distinct roles:

\begin{enumerate}

    \item  Scattering Equations: Constrain the behavior of the punctures.
    \item Scattering-Equation-Like Jacobian Factors: Arise as coefficients (e.g., in \(\eqref{jac222}\)) but do not impose constraints on the punctures.
\end{enumerate}

While \(\mathring{E}_a\) appears formally identical in both roles, it originates from different contexts, as outlined in \(\eqref{LOSEr11}\) and \(\eqref{shorteq}\).

The current formulation of \(\eqref{jkfasldk}\) and \(\eqref{jkfasldk2}\) involves only the \(m+1\) punctures \(\{\sigma_1, \sigma_2, \ldots, \) \( \sigma_{m+1}\}_{\sigma_{i_{r_1}} \to \sigma_{j_{r_1}}, \ldots, \sigma_{i_{r_u}} \to \sigma_{j_{r_u}}}\). The central objective is to prove \(\eqref{jkfasldk}\) for any \(u \in \{2, 3, \ldots, p\} \subset \{2, 3, \ldots, m\}\), assuming \(\eqref{jkfasldk2}\) holds.

When \(u = m\), the identity \(\eqref{jkfasldk}\) simplifies into an algebraic expression that is independent of the constraints from \(\mathring{E}_a = 0\). Once the identity is established for \(u = m\), imposing additional conditions \(\mathring{E}_a = 0\) directly produces the identities for cases where \(2 \leq u < m\).

Conversely, setting \((u-1)\) additional \(\mathring{E}_a\) terms in \(\eqref{jkfasldk}\) to zero leaves only a single term in the first summation, thereby reducing \(\eqref{jkfasldk}\) to \(\eqref{jkfasldk2}\). This correspondence highlights the relationship between \(\eqref{jkfasldk}\) (for \(u = m\)) and \(\eqref{jkfasldk2}\) (for \(u = 1\)). However, deriving \(\eqref{jkfasldk}\) for \(u = m\) directly from \(\eqref{jkfasldk2}\) is more intricate and requires deeper analysis.

To proceed, the proof of \(\eqref{jkfasldk}\) for \(u = m\) can be reformulated by removing the replacement rules \(\sigma_{i_\star} \to \sigma_{j_\star}\) in \(\eqref{shorteq23r}\), \(\eqref{jkfasldk2}\), and \(\eqref{jkfasldk}\). Moreover, for \(u = m\), the set \(\{i_{r_1}, i_{r_2}, \ldots, i_{r_m}\}\) coincides with \(\{1, 2, \ldots, m\}\). 

Thus, proving the key identities for \(t\)-coefficients \(\eqref{keykeyidid}\) reduces to establishing the following identity,
\ba
\label{jkfasldkafdl}
\sum_{v=1}^{m}
\sum_{\rho_{v} \in S_{m-1}}
X(s, \rho_{v}) {\rm PT}(\rho_{v}, m+1) \mathring{E}_v 
-
s_{12\ldots m+1} {\rm PT}(12\ldots m+1) = 0,
\ea
where \(X(s, \rho_{v})\) for \(v = 1, 2, \ldots, m\) in the remaining part of this appendix 
 is defined by
\ba
\label{jkfasldk2new}
\sum_{\rho_{v} \in S_{m-1}}
X(s, \rho_{v}) {\rm PT}(\rho_{v}, m+1) \mathring{E}_v 
-
s_{12\ldots m+1} {\rm PT}(12\ldots m+1)
\overset{\mathring{E}_a = 0 \, \forall a \in \{1, 2, \ldots, m\} \setminus \{v\}} = 0,
\ea
and \(\mathring{E}_a\) for \(a = 1, 2, \ldots, m\) is consistently expressed as
\ba
\label{shorteq23rnew}
\mathring{E}_a = \sum_{\substack{a' = 1 \\ a' \neq a}}^{m+1} \frac{s_{aa'}}{\sigma_{aa'}}.
\ea

This formulation isolates the essential task: to verify the algebraic consistency of \(\eqref{jkfasldkafdl}\) using the properties of \(\eqref{jkfasldk2new}\).

\subsection{Proof}

We start by revisiting the derivation of the fundamental BCJ relations among PT factors. For given \({\mathring{E}}_1\), we have,
\ba 
\label{jsafkldBCJ}
& {\mathring E}_{1} {\rm PT}(   23\ldots m+1) 
\nl
= &
\sum_{a=2}^{m+1} s_{1\, a}  {\rm PT}(  \{ 23\ldots a-1 \} \shuffle 1, a,a+1,\ldots, m+1) 
\nl  = & \sum_{a=2}^{m+1} (s_{1\,a}+s_{1\,a+1}+\ldots s_{1\,m+1})   {\rm PT}(   23\ldots a-1, 1, a,a+1,\ldots, m+1)\,.
\ea 
When \({\mathring{E}}_1 = 0\), the summation in \eqref{jsafkldBCJ} vanishes, recovering the fundamental BCJ relations. For convenience, we refer to the term \( {\mathring E}_{1} {\rm PT}(23 \ldots m+1) \) as a boundary term, as it vanishes on the support of the scattering equations. This approach will be employed repeatedly in the subsequent derivations.

We also introduce a convention for naming PT factors: \({\rm PT}(12\ldots m+1)\) is referred to as an \((m+2)\)-point PT factor, since its full form is given by \(\mathrm{PT}(12\ldots m{+}1) := \lim_{\sigma_n \to \infty} \sigma_n^2 \, \textbf{PT}(12\ldots \) \( m{+}1\, n)\). Similarly, \({\rm PT}(23\ldots m+1)\) is referred to as an \((m+1)\)-point PT factor. Later, we will frequently encounter the reduction of an \((m+1)\)-point PT factor, combined with a scattering-equation-like Jacobian, to \((m+2)\)-point PT factors.

\subsubsection{Decomposing a Summand with \(v = 1\)}
Using a similar technique as in \eqref{jsafkldBCJ}, we consider a summand in the first term of \eqref{jkfasldkafdl} where \(v = 1\). This summand can be rewritten as a linear combination of \((m+2)\)-point PT factors, with coefficients depending only on Mandelstam variables,
\ba 
\label{jkfasldkafdl22}
 & \sum_{\rho_{{1}}\in S_{m-1}}     X ( s,\rho_{{1}}) {\rm PT}(    \rho_{{1}} m+1) {\mathring E}_{{1}} 
\nl 
  =& \sum_{\rho_{{1}}\in S_{m-1}}     X ( s,\rho_{{1}}) 
   \sum_{\substack{a=2 }
}^{m+1}  \left( \sum_{a'\in {Y}(a,\rho_1) } s_{1a'} + s_{1\, m+1} \right) {\rm PT}(     Y(\rho_{{1}},a) , 1, {Y}(a,\rho_1) , m+1)\,.
\ea 
Here, \(Y(\rho_1, a)\) is the sequence of elements in \(\rho_1\) preceding \(a\), while \(Y(a, \rho_1)\) is the sequence including and following \(a\).  In the next steps, we will expand \({\rm PT}(     Y(\rho_{{1}},a) , 1, {Y}(a,\rho_1) , m+1)\) into a BCJ basis, explicitly including the boundary terms.

\subsubsection{Ansatz for Boundary Terms in a BCJ Expansion}
Consider the following ansatz for boundary terms used to expand a given PT factor \({\rm PT}(\gamma, m+1)\),
\ba 
\label{jkfasldkafdl23}
\sum_{v'=2}^m \sum_{\rho_{v'} \in S_{m-1}} 
{\rm PT}(\rho_{v'}, m+1) {\mathring{E}}_{v'} Z^\star(s, \gamma, \rho_{v'}),
\ea
where \(\rho_{v'}\) is a permutation of the \((m-1)\) labels \(\{1, 2, \ldots, m\} \setminus \{v'\}\). This ansatz involves \((m-1)(m-1)!\) undetermined functions \(Z^\star(s, \gamma, \rho_{v'})\), which are functions of Mandelstam variables. 

The structure resembles the first term in \eqref{jkfasldkafdl} and can also be expanded into \((m+2)\)-point PT factors as follows,
\ba 
\label{jkfasldkafdl24}
 & \sum_{\substack{v'=2}}^m \sum_{\rho_{v'}\in S_{m-1}} 
{\rm PT}(   \rho_{v'}, m+1) {\mathring E}_{v'} Z^\star(s,  \gamma, \rho_{v'}) 
\nl 
  =&  \sum_{\substack{v'=2}}^m  \sum_{\rho_{{v'}}\in S_{m-1}}    Z^\star(s,  \gamma, \rho_{v'})  
   \sum_{\substack{a=1
   \\ a\neq v' }
}^{m+1}  \left( \sum_{a'\in {Y}(a,\rho_{v'}) } s_{v' a'} + s_{v'\, m+1} \right) {\rm PT}(     Y(\rho_{{v'}},a) , v', {Y}(a,\rho_{v'}) , m+1)
\nl
= & \sum_{\pi \in S_m}  {\rm PT}( \pi , m+1 ) \sum_{\substack{v'=2}}^m  Z^\star\big(s,\gamma, \pi_{v\!\!\!\slash' }\big) \left( \sum_{a'\in \tilde Y( v',\pi)  } s_{v' a'} + s_{v'\, m+1} \right) \,.
\ea 
In the final line of \eqref{jkfasldkafdl24}, the summation has been reorganized to explicitly highlight the coefficients of the \(m!\) PT factors. 

Here, \(\pi\) is a permutation of the \(m\) labels \(\{1, 2, \ldots, m\}\), and \(\pi_{v\!\!\!\slash'}\) represents the sequence obtained by removing \(v'\) from \(\pi\). \(\tilde{Y}(v', \pi)\) denotes the subsequence of \(\pi\) following \(v'\) and excluding \(v'\), such that \(Y(v', \pi) = (v', \tilde{Y}(v', \pi))\), and therefore, \(\pi_{v\!\!\!\slash'} = ( Y( \pi, v'), \tilde Y( v',\pi) )\). 

Notably, both \(\rho_{v'}\) and \(\pi_{v\!\!\!\slash'}\) are permutations of \(\{1, 2, \ldots, m\} \setminus \{v'\}\), meaning \(Z^\star(s, \gamma, \rho_{v'})\) and \(Z^\star(s, \gamma, \pi_{v\!\!\!\slash'})\) refer to the same objects.

\subsubsection{Deriving BCJ Expansions for PT Factors Involving Boundary Terms}

We now turn to the PT factor \({\rm PT}(Y(\rho_1, a), 1, Y(a, \rho_1), m+1)\) appearing on the RHS of \eqref{jkfasldkafdl22}. 

When the first entry of the ordering is not \(1\) (i.e., when \(Y(\rho_1, a) = \emptyset\)), we impose constraints on \eqref{jkfasldkafdl24} to set the coefficient of the desired PT factor to \(1\), while ensuring all the coefficients of other PT factors with \(\pi(1) \neq 1\) vanish, where \(\pi(1)\) represents the first entry of the ordering \(\pi\). This procedure yields \((m-1)(m-1)!\) equations, which exactly match the number of undetermined variables \(Z^\star(s, \gamma, \rho_{v'})\),
\ba 
\label{eqzzzz}
\sum\limits_{\substack{v'=2}}^m  Z^\star\big(s,\gamma,\pi_{v\!\!\!\slash' } \big) & \left( \sum\limits_{a'\in \tilde Y( v',\pi)  } s_{v' a'} + s_{v'\, m+1} \right)
 \ln
 =&
 \begin{cases} 
 1
 & \text{if } \pi= (Y(\rho_1, a), 1, Y(a, \rho_1)) \text{ and } \pi(1) \neq 1, 
\\
 0
& \text{if } \pi\neq (Y(\rho_1, a), 1, Y(a, \rho_1)) \text{ and } \pi(1) \neq 1.
\end{cases}
\ea 
This ensures a unique solution for \(Z^\star(s, \gamma, \rho_{v'})\), which we denote as \(Z(s, (Y(\rho_1, a), 1, Y(a, \rho_1)), \rho_{v'})\).

For permutations \(\pi\) where \(\pi(1) = 1\), we denote \(\pi = (1, \pi_1)\), where \(\pi_1\) is a permutation of the \((m-1)\) labels \(\{2,3,\ldots, m\}\). Substituting the solution for \(Z(\ldots)\) into the coefficient of the corresponding PT factors \({\rm PT}(1,\pi_1, m+1)\) in \eqref{jkfasldkafdl24}, we denote these coefficients as \(-\tilde{\mathcal{B}}((Y(\rho_1, a), 1, Y(a, \rho_1)), \pi_1)\), leading to
\ba 
\tilde{\mathcal{B}}((Y(\rho_1, a), 1, Y(a, \rho_1)), \pi_1)
 = &
- \sum_{\substack{v'=2}}^m  Z \big(s,(Y(\rho_1, a), 1, Y(a, \rho_1)), (1, Y( \pi_1, v'), \tilde Y( v', \pi_1) )   \big) 
\nl 
& \times \left( \sum_{a'\in \tilde Y( v', \pi_1)  } s_{v' a'} + s_{v'\, m+1} \right).
\ea 

This construction effectively reproduces the BCJ expansion \cite{Bern:2019prr, Bern:2008qj}, explicitly including boundary terms,
\ba 
\label{jkfasldkafdl24ex}
&{\rm PT}(Y(\rho_1, a), 1, Y(a, \rho_1), m+1) 
\ln
= &
\sum_{v'=2}^m \sum_{\rho_{v'} \in S_{m-1}} 
{\rm PT}(\rho_{v'}, m+1) {\mathring{E}}_{v'} Z(s, (Y(\rho_1, a), 1, Y(a, \rho_1)), \rho_{v'}) 
\nl 
& + 
\sum_{\pi_1 \in S_{m-1}} \tilde{\mathcal{B}}((Y(\rho_1, a), 1, Y(a, \rho_1)), \pi_1)
{\rm PT}(1, \pi_1, m+1),
\nonumber
\ea
where \(\tilde{\mathcal{B}}((Y(\rho_1, a), 1, Y(a, \rho_1)), \pi_1)\) are precisely the coefficients required in the BCJ expansions.

In the special case where \(Y(\rho_1, a) = \emptyset\), the PT factor simplifies to \({\rm PT}(1, \rho_1, m+1)\). This leads to a trivial BCJ expansion: \({\rm PT}(1, \rho_1, m+1) = {\rm PT}(1, \rho_1, m+1)\), where only one \(\tilde{\mathcal{B}}(\ldots)\) coefficient is non-vanishing, and all other \(\tilde{\mathcal{B}}(\ldots)\) and \(Z(\ldots)\) coefficients are zero.

By the way, the BCJ expansion \eqref{jkfasldkafdl24ex}, with boundary terms explicitly included, provides additional insight into related results. For instance, we see that only \((m-1)\) \({\mathring E}_{v'}\) are needed to derive a BCJ expansion. This observation  helps to explain why \eqref{appdefa2} remains valid even when particle \(2\) there  is massive.

\subsubsection{Splitting the RHS into Two Terms}

Plugging \eqref{jkfasldkafdl24ex} into \eqref{jkfasldkafdl22} and subtracting \(s_{12\ldots m+1} {\rm PT}(12 \ldots m+1)\), we obtain
\ba 
\label{asfhjkd}
\sum_{\rho_1 \in S_{m-1}} X(s, \rho_1) {\rm PT}(\rho_1, m+1) {\mathring{E}}_1 - 
s_{12\ldots m+1} {\rm PT}(12 \ldots m+1)
= A + B,
\ea 
where \(A\) and \(B\) are defined as
\ba 
\label{asfhjkdaa}
A = 
-s_{12\ldots m+1} {\rm PT}(12 \ldots m+1) + \sum_{\rho_1 \in S_{m-1}} X(s, \rho_1) 
\sum_{a=2}^{m+1} \Bigg( \sum_{a' \in Y(a, \rho_1)} s_{1a'} + s_{1m+1} \Bigg)
\nl 
\times \sum_{\pi_1 \in S_{m-1}} \tilde{\mathcal{B}}((Y(\rho_1, a), 1, Y(a, \rho_1)), \pi_1) {\rm PT}(1, \pi_1, m+1),
\ea 
and
\ba
\label{asfhjkdbb}
B= & \sum_{\rho_{{1}}\in S_{m-1}}     X ( s,\rho_{{1}}) 
   \sum_{\substack{a=2 }
}^{m+1}  \left( \sum_{a'\in {Y}(a,\rho_1) } s_{1a'} + s_{1\, m+1} \right) 
\nl &  \times 
\sum_{\substack{v'=2}}^m \sum_{\rho_{v'}\in S_{m-1}} 
{\rm PT}(   \rho_{v'}, m+1) {\mathring E}_{v'} Z(s,  (Y(\rho_{{1}},a) , 1, {Y}(a,\rho_1)), \rho_{v'}) 
\,.
\ea 

The term \(A\) represents an expansion of \((m-1)!\) \((m+2)\)-point PT factors \({\rm PT}(1, \pi_1, m+1)\) in a BCJ basis with coefficients determined by the \((m-1)!\) functions \(X(s, \rho_1)\). The term \(B\), on the other hand, is an expansion of \((m-1)(m-1)!\) \((m+1)\)-point PT factors \({\rm PT}(\rho_{v'}, m+1)\), with each term proportional to a function \({\mathring{E}}_{v'}\). 

\subsubsection{Vanishing of \(A\)}

We claim that \(A\) vanishes algebraically even without using the explicit expression for  \(\tilde{\mathcal{B}}((Y(\rho_1, \) \(a), 1, Y(a, \rho_1)), \pi_1)\). 
We observe that setting \({\mathring{E}}_{v'} = 0\) for \(v' = 2, 3, \ldots, m\) causes \(B\) to vanish. Consequently, using the definition of \(X(s, \rho_1)\) given in \eqref{jkfasldk2new}, \(A\) must also vanish under these conditions. Since the \((m-1)!\) \((m+2)\)-point PT factors \({\rm PT}(1, \rho'_1, m+1)\) are in a BCJ basis, their coefficients must vanish algebraically. 

This algebraic vanishing of coefficients implies that \(A\) necessarily vanishes even without imposing the conditions \({\mathring{E}}_{v'} = 0\) for \(v' = 2, 3, \ldots, m\). This observation not only confirms the algebraic nature of the cancellation but also offers a pathway for explicitly determining the \((m-1)!\) functions \(X(s, \rho_1)\) by utilizing the explicit expression for \(\tilde{\mathcal{B}}((Y(\rho_1, a), 1, Y(a, \rho_1)), \pi_1)\).

\subsubsection{Reorganizing \(B\) and Identification of \(X(s, \rho_{v'})\)}

We reorganize \(B\) to explicitly extract the coefficients of each \({\rm PT}(\rho_{v'}, m+1) {\mathring{E}}_{v'}\)
\ba
\label{asfhjkd2}
B = -\sum_{v'=2}^m \sum_{\rho_{v'} \in S_{m-1}} {\rm PT}(\rho_{v'}, m+1) {\mathring{E}}_{v'} X^\star(s, \rho_{v'}),
\ea 
where
\ba 
\label{asfhjkdbb2}
X^\star(s, \rho_{v'}) \!=\! -\!\!\sum_{\rho_1 \in S_{m-1}} \!\! X(s, \rho_1) 
\sum_{a=2}^{m+1} \Bigg( \sum_{a' \in Y(a, \rho_1)} s_{1a'} + s_{1m+1} \Bigg) Z(s, (Y(\rho_1, a), 1, Y(a, \rho_1)), \rho_{v'}).
\ea 

Even without the explicit expression for \(Z(s, (Y(\rho_1, a), 1, Y(a, \rho_1)), \rho_{v'})\), we assert that \(X^\star(s, \rho_{v'}) = X(s, \rho_{v'})\). This follows directly from the definition of \(X(s, \rho_{v'})\) in \eqref{jkfasldk2new}. By setting \({\mathring{E}}_v = 0\) for all \(v \neq v'\), \eqref{asfhjkd} simplifies to
\ba 
\sum_{\rho_{v'} \in S_{m-1}} {\rm PT}(\rho_{v'}, m+1) {\mathring{E}}_{v'} X^\star(s, \rho_{v'}) - s_{12\ldots m+1} {\rm PT}(12 \ldots m+1) \overset{{\mathring{E}}_{v}=0 ~ \forall v \in \{1,2,\ldots,m\}\setminus\{v'\}}{=} 0,
\ea
which exactly matches the definition of \(X(s, \rho_{v'})\).

This observation also provides a systematic pathway for explicitly determining the \((m-1)(m-1)!\) functions \(X(s, \rho_{v'})\) for \(v' = 2, 3, \ldots, m\) by leveraging the explicit expression for \(Z(s, (Y(\rho_1, a), 1, Y(a, \rho_1)), \rho_{v'})\) which is a solution of \eqref{eqzzzz}.

\subsubsection{Conclusion}

Finally, substituting \(B\) from \eqref{asfhjkd2} with \(X^\star(s, \rho_{v'}) = X(s, \rho_{v'})\) and \(A = 0\) into \eqref{asfhjkd}, we obtain the desired identity,
\ba 
\sum_{v=1}^m \sum_{\rho_v \in S_{m-1}} X(s, \rho_v) {\rm PT}(\rho_v, m+1) {\mathring{E}}_v - s_{12\ldots m+1} {\rm PT}(12 \ldots m+1) = 0.
\ea 
This completes the proof.

\subsection{Examples}

Here we present illustrative examples to demonstrate the proof of the key identities \eqref{keykeyidid} for \(t\)-coefficients.

\subsubsection{Example with \(u = m = 2\)}
Let us consider the case 
\(u = m = 2\). Denote
\ba
{\mathring E}_{1}= \frac{s_{12}}{\sigma_{12}} +  \frac{s_{13}}{\sigma_{13}}, \qquad  
{\mathring E}_{2}= \frac{s_{12}}{\sigma_{21}} +  \frac{s_{23}}{\sigma_{23}}.
\ea
The coefficients \(X(s,2)\) and \(X(s,1)\) are defined as ratios of Mandelstam variables, satisfying the following conditions for any \(\sigma_1, \sigma_2, \sigma_3\),
\ba
\label{aflkjd}
X(s,2) {\rm PT}(23) {\mathring E}_{1} \overset { {\mathring E}_{2} =0}{=} s_{123} {\rm PT}(123),
\nonumber
\\
X(s,1) {\rm PT}(13) {\mathring E}_{2} \overset { {\mathring E}_{1} =0}{=}s_{123} {\rm PT}(123).
\ea
The goal is to establish
\ba
\label{appgoal1}
X(s,2) {\rm PT}(23) {\mathring E}_{1}  
+
X(s,1) {\rm PT}(13) {\mathring E}_{2}
=s_{123}
{\rm PT}(123).
\ea

\begin{proof}

We start by expanding \(X(s, 2) {\rm PT}(23) {\mathring E}_{1}\) into 4-point PT factors:
\ba
\label{jkfha}
X(s, 2) {\rm PT}(23) {\mathring E}_{1}
= & 
X(s, 2) \big(
s_{12} {\rm PT}(123) 
+
s_{13} ({\rm PT}(123) + {\rm PT}(213))
\big)
\nl 
= & 
X(s, 2) \big(
(s_{12} + s_{13}) {\rm PT}(123) 
+
s_{13} {\rm PT}(213)
\big).
\ea

From the relationship,
\ba
{\rm PT}(13) {\mathring E}_{2} = (s_{12} + s_{23}) {\rm PT}(213) + s_{23} {\rm PT}(123),
\ea
we isolate \({\rm PT}(213)\) as
\ba
\label{jkfha2}
{\rm PT}(213) = 
\frac{{\rm PT}(13) {\mathring E}_{2} - s_{23} {\rm PT}(123)}{s_{12} + s_{23}}.
\ea

Substituting \eqref{jkfha2} into \eqref{jkfha} and subtracting \(s_{123} {\rm PT}(123)\), we obtain,
\ba
\label{jkfha3}
X(s, 2) {\rm PT}(23) {\mathring E}_{1} - s_{123} {\rm PT}(123)
= &
{\rm PT}(123) \left(
\frac{s_{12} s_{123}}{s_{12} + s_{23}} X(s, 2) - s_{123}
\right)
\nl
& + X(s, 2) \frac{s_{13}}{s_{12} + s_{23}} {\rm PT}(13) {\mathring E}_{2}.
\ea

When \({\mathring E}_{2} = 0\), the second term on the RHS of \eqref{jkfha3} vanishes. According to the condition in \eqref{aflkjd}, the coefficient of \({\rm PT}(123)\) on the RHS of \eqref{jkfha3} must also vanish algebraically. This determines $
X(s, 2) = \frac{s_{12} + s_{23}}{s_{12}}
$.

Now we get 
\ba
\label{jkfha322}
X(s, 2) {\rm PT}(23) {\mathring E}_{1} - s_{123} {\rm PT}(123)
=  X(s, 2) \frac{s_{13}}{s_{12} + s_{23}} {\rm PT}(13) {\mathring E}_{2}.
\ea

Setting ${\mathring E}_{1} = 0$ obviously we have
\ba
-X(s, 2) \frac{s_{13}}{s_{12} + s_{23}} {\rm PT}(13) {\mathring E}_{2} - s_{123} {\rm PT}(123) \overset{{\mathring E}_{1} = 0}{=} 0\,,
\ea
which allows us to identify
$
-X(s, 2) \frac{s_{13}}{s_{12} + s_{23}} = -\frac{s_{13}}{s_{12}}
$
as \(X(s, 1)\), consistent with \eqref{aflkjd}.  Thus, \eqref{jkfha322} reduces to \eqref{appgoal1}, completing the proof.

It is worth noting that in the above demonstration, the explicit expressions of \(X(s, 2)\) and \(X(s, 1)\) were not assumed as inputs for proving \eqref{appgoal1}. Instead, these expressions naturally emerged as byproducts of the proof.
    
\end{proof}

\subsubsection{Example with \(u = m = 3\)}

For \(m=3\), denote
\ba
{\mathring E}_{1}= \frac{s_{12}}{\sigma_{12}} +  \frac{s_{13}}{\sigma_{13}} +  \frac{s_{14}}{\sigma_{14}}, \quad  
{\mathring E}_{2}= \frac{s_{12}}{\sigma_{21}} +  \frac{s_{23}}{\sigma_{23}} +  \frac{s_{24}}{\sigma_{24}}, \quad  
{\mathring E}_{3}= \frac{s_{13}}{\sigma_{31}} +  \frac{s_{23}}{\sigma_{32}} +  \frac{s_{34}}{\sigma_{34}}.
\ea
Six coefficients \(X(s, \rho)\) are defined to satisfy
\ba
\label{aflkjd22}
X(s,23) {\rm PT}(234) {\mathring E}_{1} + X(s,32) {\rm PT}(324) {\mathring E}_{1} 
\overset { {\mathring E}_{2}, {\mathring E}_{3} =0}{=}  s_{1234}{\rm PT}(1234),
\nl
X(s,13) {\rm PT}(134) {\mathring E}_{2} +X(s,31){\rm PT}(314) {\mathring E}_{2} \overset { {\mathring E}_{1}, {\mathring E}_{3} =0}{=} s_{1234}{\rm PT}(1234),
\nl
X(s,12) {\rm PT}(124) {\mathring E}_{3} + X(s,21) {\rm PT}(214) {\mathring E}_{3} \overset { {\mathring E}_{1}, {\mathring E}_{2} =0}{=} s_{1234}{\rm PT}(1234)\,,
\ea
for any value of $\sigma_1,\sigma_2,\sigma_3$ and 
$\sigma_4$.
The task is to prove
\ba
\label{appgoal2}
&\quad X(s,23) {\rm PT}(234) {\mathring E}_{1} + X(s,32) {\rm PT}(324) {\mathring E}_{1}
\nl
&+ X(s,13) {\rm PT}(134) {\mathring E}_{2} + X(s,31) {\rm PT}(314) {\mathring E}_{2}
\nl &
+ X(s,12) {\rm PT}(124) {\mathring E}_{3} + X(s,21) {\rm PT}(214) {\mathring E}_{3} = s_{1234} {\rm PT}(1234).
\ea

\begin{proof}

We first expand \( X(s,23) {\rm PT}(234) {\mathring E}_{1} + X(s,32) {\rm PT}(324) {\mathring E}_{1} \) onto 5-point PT factors,
\ba 
\label{jkfha33}
& X(s,23) {\rm PT}(234) {\mathring E}_{1} + X(s,32) {\rm PT}(324) {\mathring E}_{1}
\nl
= & 
X(s,23) \big[
 (s_{12} + s_{13} + s_{14}) {\rm PT}(1234) 
+ (s_{13} + s_{14}) {\rm PT}(2134)
+ s_{14} {\rm PT}(2314)
\big]
\nl
+ &
X(s,32) \big[
 (s_{12} + s_{13} + s_{14}) {\rm PT}(1324) 
+ (s_{12} + s_{14}) {\rm PT}(3124)
+ s_{14} {\rm PT}(3214)
\big] .
\ea 
Using the following identities for \( {\rm PT}(134) {\mathring E}_{2} \), \( {\rm PT}(314) {\mathring E}_{2} \), \( {\rm PT}(124) {\mathring E}_{3} \), and \( {\rm PT}(214) {\mathring E}_{3} \),
\ba 
& {\rm PT}(134) {\mathring E}_{2} = (s_{12} + s_{23} + s_{24}) {\rm PT}(2134) 
+ (s_{23} + s_{24}) {\rm PT}(1234)
+ s_{24} {\rm PT}(1324) ,
\nl 
& {\rm PT}(314) {\mathring E}_{2} = (s_{12} + s_{23} + s_{24}) {\rm PT}(2314) 
+ (s_{12} + s_{24}) {\rm PT}(3214)
+ s_{24} {\rm PT}(3124) ,
\nl 
& {\rm PT}(124) {\mathring E}_{3} = (s_{13} + s_{23} + s_{34}) {\rm PT}(3124) 
+ (s_{23} + s_{34}) {\rm PT}(1324)
+ s_{34} {\rm PT}(1234) ,
\nl 
& {\rm PT}(214) {\mathring E}_{3} = (s_{13} + s_{23} + s_{34}) {\rm PT}(3214) 
+ (s_{13} + s_{34}) {\rm PT}(2314)
+ s_{34} {\rm PT}(2134) ,
\ea 
we can re-express \( {\rm PT}(2134) \), \( {\rm PT}(2314) \), \( {\rm PT}(3124) \), and \( {\rm PT}(3214) \) in terms of \( {\rm PT}(134) {\mathring E}_{2} \), \( {\rm PT}(314) {\mathring E}_{2} \), \( {\rm PT}(124) {\mathring E}_{3} \), \( {\rm PT}(214) {\mathring E}_{3} \), and the basis terms \( {\rm PT}(1234) \), \( {\rm PT}(1324) \),
\ba 
\label{jkfha233}
&\text{PT}(2134) = \frac{\text{PT}(134)
   {\mathring E}_2}{s_{12}\!+\!s_{23}\!+\!s_{24}}\!-\!\frac{\text{PT}(1324)
   s_{24}}{s_{12}\!+\!s_{23}\!+\!s_{24}}\!-\!\frac{\text{PT}(1234)
   \left(s_{23}\!+\!s_{24}\right)}{s_{12}\!+\!s_{23}\!+\!s_{24}},
   \nl
   &\text{PT}(3124)= 
   \frac{\text{PT}(124) {\mathring E}_3}{s_{13}\!+\!s_{23}\!+\!s_{34}}\!-\!\frac{\text{PT}(1234)
   s_{34}}{s_{13}\!+\!s_{23}\!+\!s_{34}}\!-\!\frac{\text{PT}(1324)
   \left(s_{23}\!+\!s_{34}\right)}{s_{13}\!+\!s_{23}\!+\!s_{34}},
   \nl
   &
   \text{PT}(3214)=
   \frac{ \text{PT}(124){\mathring E}_3\left(s_{13}\!+\!s_{34}\right) s_{24} }{s_{23}
   \left(s_{13}\!+\!s_{23}\!+\!s_{34}\right)
   \left(s_{12}\!+\!s_{13}\!+\!s_{23}\!+\!s_{24}\!+\!s_{34}\right)}\!-\!\frac{\text{PT}(134) {\mathring E}_2
   s_{34} }{s_{23}
   \left(s_{12}\!+\!s_{13}\!+\!s_{23}\!+\!s_{24}\!+\!s_{34}\right)}
   \nl & 
   \!-\!\frac{\text{PT}(314){\mathring E}_2
   \left(s_{13}\!+\!s_{34}\right) }{s_{23}
   \left(s_{12}\!+\!s_{13}\!+\!s_{23}\!+\!s_{24}\!+\!s_{34}\right)}\!+\!\frac{\text{PT}(214){\mathring E}_3
   \left(s_{12}\!+\!s_{23}\!+\!s_{24}\right) }{s_{23}
   \left(s_{12}\!+\!s_{13}\!+\!s_{23}\!+\!s_{24}\!+\!s_{34}\right)}
     \nl & 
   \!-\!\frac{\text{PT}(1324)
   s_{13} s_{24}}{\left(s_{13}\!+\!s_{23}\!+\!s_{34}\right)
   \left(s_{12}\!+\!s_{13}\!+\!s_{23}\!+\!s_{24}\!+\!s_{34}\right)}\!+\!\frac{\text{PT}(1234)
   s_{34}
   \left(s_{13}\!+\!s_{23}\!+\!s_{24}\!+\!s_{34}\right)}{\left(s_{13}\!+\!s_{23}\!+\!s_{34}\right
   ) \left(s_{12}\!+\!s_{13}\!+\!s_{23}\!+\!s_{24}\!+\!s_{34}\right)},
   \nl
   & 
   \text{PT}(2314)=
   \frac{\text{PT}(134) {\mathring E}_2\left(s_{12}\!+\!s_{24}\right) s_{34} }{s_{23}
   \left(s_{12}\!+\!s_{23}\!+\!s_{24}\right)
   \left(s_{12}\!+\!s_{13}\!+\!s_{23}\!+\!s_{24}\!+\!s_{34}\right)}\!+\!\frac{\text{PT}(314) {\mathring E}_2
   \left(s_{13}\!+\!s_{23}\!+\!s_{34}\right) }{s_{23}
   \left(s_{12}\!+\!s_{13}\!+\!s_{23}\!+\!s_{24}\!+\!s_{34}\right)}
     \nl & 
   \!-\!\frac{\text{PT}(124){\mathring E}_3
   s_{24} }{s_{23}
   \left(s_{12}\!+\!s_{13}\!+\!s_{23}\!+\!s_{24}\!+\!s_{34}\right)}\!-\!\frac{\text{PT}(214) {\mathring E}_3
   \left(s_{12}\!+\!s_{24}\right) }{s_{23}
   \left(s_{12}\!+\!s_{13}\!+\!s_{23}\!+\!s_{24}\!+\!s_{34}\right)}
     \nl & 
   \!-\!\frac{\text{PT}(1234)
   s_{12} s_{34}}{\left(s_{12}\!+\!s_{23}\!+\!s_{24}\right)
   \left(s_{12}\!+\!s_{13}\!+\!s_{23}\!+\!s_{24}\!+\!s_{34}\right)}\!+\!\frac{\text{PT}(1324)
   s_{24}
   \left(s_{12}\!+\!s_{23}\!+\!s_{24}\!+\!s_{34}\right)}{\left(s_{12}\!+\!s_{23}\!+\!s_{24}\right
   ) \left(s_{12}\!+\!s_{13}\!+\!s_{23}\!+\!s_{24}\!+\!s_{34}\right)}\,.
\ea 
Plugging \eqref{jkfha233} into the RHS of \eqref{jkfha33} and subtracting \( s_{123} {\rm PT}(123) \), we obtain a combination of terms involving \( {\rm PT}(1234), {\rm PT}(1324) \), and  four boudary terms,
\ba 
&X(s,23) {\rm PT}(234) {\mathring E}_{1} + X(s,32) {\rm PT}(324) {\mathring E}_{1} - s_{123} {\rm PT}(123) 
\nl 
 =& \quad\#_1 {\rm PT}(1234) + \#_2 {\rm PT}(1324) 
\nl 
&+\#_3 
{\rm PT}(134) {\mathring E}_{2} + \#_4 {\rm PT}(314) {\mathring E}_{2} + \#_5 {\rm PT}(124) {\mathring E}_{3} + \#_6 {\rm PT}(214) {\mathring E}_{3}\,.
\ea 

Requiring the coefficients of \( {\rm PT}(1234) \) and \( {\rm PT}(1324) \) to vanish algebraically, , i.e., $\#_1=\#_2=0$,   leads to the following solutions,
\ba 
X(s,23) &= -\frac{s_{2n} (s_{34} + s_{123})}{s_{12}s_{123}}, \quad
X(s,32) = -\frac{s_{3n} s_{24}}{s_{12}s_{123}},
\ea 
where \( s_{2n} = -s_{12} - s_{23} - s_{24} \) on the support of $h_3=0$. 

Next, identifying the opposites of the coefficients of \( {\rm PT}(134) {\mathring E}_{2}, {\rm PT}(314) {\mathring E}_{2}, {\rm PT}(124) {\mathring E}_{3},\) \( {\rm PT}(214) {\mathring E}_{3} \), i.e., $-\#_3,-\#_4, -\#_5, -\#_6$, as \( X(s,13), X(s,31), X(s,12), X(s,21) \),   we find
\ba
X(s,13) &= -\frac{s_{12} s_{14} s_{34} + s_{23} s_{3n} s_{1n} + s_{12} s_{23} (s_{13} + s_{14} + s_{3n})}{s_{12}s_{23}s_{123}}, 
\quad X(s,31) = \frac{s_{14} s_{3n} (s_{12} + s_{23})}{s_{12}s_{23}s_{123}},
\nl
X(s,12) &= -\frac{s_{24} (s_{1n} + s_{123})}{s_{23}s_{123}}, 
\quad X(s,21) =- \frac{s_{14} s_{2n}}{s_{23}s_{123}}.
\ea
These results are consistent with \eqref{Xsrhoori}. Consequently, we have proved \eqref{appgoal2}.

With \eqref{appgoal2} established, setting \( {\mathring E}_{3} = 0 \) yields
\ba
&\quad X(s,23) {\rm PT}(234) {\mathring E}_{1} + X(s,32) {\rm PT}(324) {\mathring E}_{1}
\nl
&+ X(s,13) {\rm PT}(134) {\mathring E}_{2} + X(s,31) {\rm PT}(314) {\mathring E}_{2}
 \overset{{\mathring E}_{3} =0 }= s_{1234} {\rm PT}(1234),
\ea
whose relabeling reproduces \eqref{afdnkdj}.

\end{proof}

\bibliographystyle{jhep}
\bibliography{references}

\providecommand{\href}[2]{#2}\begingroup\raggedright\begin{thebibliography}{10}

\bibitem{Zhang:2024iun}
Y.~Zhang, {\it {New Factorizations of Yang-Mills Amplitudes}},
  \href{http://arxiv.org/abs/2406.08969}{{\tt arXiv:2406.08969}}.

\bibitem{Altarelli:1977zs}
G.~Altarelli and G.~Parisi, {\it {Asymptotic Freedom in Parton Language}},
  {\em Nucl. Phys. B} {\bf 126} (1977) 298--318.

\bibitem{Catani:1998nv}
S.~Catani and M.~Grazzini, {\it {Collinear factorization and splitting
  functions for next-to-next-to-leading order QCD calculations}},  {\em Phys.
  Lett. B} {\bf 446} (1999) 143--152,
  [\href{http://arxiv.org/abs/hep-ph/9810389}{{\tt hep-ph/9810389}}].

\bibitem{Strominger:2017zoo}
A.~Strominger, {\em {Lectures on the Infrared Structure of Gravity and Gauge
  Theory}}.
\newblock 3, 2017.

\bibitem{Pasterski:2021raf}
S.~Pasterski, M.~Pate, and A.-M. Raclariu, {\it {Celestial Holography}},  in
  {\em {Snowmass 2021}}, 11, 2021.
\newblock \href{http://arxiv.org/abs/2111.11392}{{\tt arXiv:2111.11392}}.

\bibitem{Weinberg:1965nx}
S.~Weinberg, {\it {Infrared photons and gravitons}},  {\em Phys.Rev.} {\bf 140}
  (1965) B516--B524.

\bibitem{Strominger:2014pwa}
A.~Strominger and A.~Zhiboedov, {\it {Gravitational Memory, BMS
  Supertranslations and Soft Theorems}},  {\em JHEP} {\bf 01} (2016) 086,
  [\href{http://arxiv.org/abs/1411.5745}{{\tt arXiv:1411.5745}}].

\bibitem{Cao:2024gln}
Q.~Cao, J.~Dong, S.~He, and C.~Shi, {\it {A universal splitting of tree-level
  string and particle scattering amplitudes}},  {\em Phys. Lett. B} {\bf 856}
  (2024) 138934, [\href{http://arxiv.org/abs/2403.08855}{{\tt
  arXiv:2403.08855}}].

\bibitem{Cao:2024qpp}
Q.~Cao, J.~Dong, S.~He, C.~Shi, and F.~Zhu, {\it {On universal splittings of
  tree-level particle and string scattering amplitudes}},  {\em JHEP} {\bf 09}
  (2024) 049, [\href{http://arxiv.org/abs/2406.03838}{{\tt arXiv:2406.03838}}].

\bibitem{Cachazo:2021wsz}
F.~Cachazo, N.~Early, and B.~Gim\'enez~Umbert, {\it {Smoothly splitting
  amplitudes and semi-locality}},  {\em JHEP} {\bf 08} (2022) 252,
  [\href{http://arxiv.org/abs/2112.14191}{{\tt arXiv:2112.14191}}].

\bibitem{GimenezUmbert:2024jjn}
B.~Gim\'enez~Umbert and K.~Yeats, {\it {$\ensuremath{\phi}^p$ amplitudes from
  the positive tropical Grassmannian: Triangulations of extended diagrams}},
  {\em Phys. Rev. D} {\bf 110} (2024), no.~12 125018,
  [\href{http://arxiv.org/abs/2403.17051}{{\tt arXiv:2403.17051}}].

\bibitem{GimenezUmbert:2025ech}
B.~Gim\'enez~Umbert and B.~Sturmfels, {\it {Splitting CEGM Amplitudes}},
  \href{http://arxiv.org/abs/2501.10805}{{\tt arXiv:2501.10805}}.

\bibitem{Arkani-Hamed:2023swr}
N.~Arkani-Hamed, Q.~Cao, J.~Dong, C.~Figueiredo, and S.~He, {\it {Hidden zeros
  for particle/string amplitudes and the unity of colored scalars, pions and
  gluons}},  {\em JHEP} {\bf 10} (2024) 231,
  [\href{http://arxiv.org/abs/2312.16282}{{\tt arXiv:2312.16282}}].

\bibitem{Mafra:2011nw}
C.~R. Mafra, O.~Schlotterer, and S.~Stieberger, {\it {Complete N-Point
  Superstring Disk Amplitude II. Amplitude and Hypergeometric Function
  Structure}},  {\em Nucl. Phys. B} {\bf 873} (2013) 461--513,
  [\href{http://arxiv.org/abs/1106.2646}{{\tt arXiv:1106.2646}}].

\bibitem{Arkani-Hamed:2024fyd}
N.~Arkani-Hamed and C.~Figueiredo, {\it {All-order splits and multi-soft limits
  for particle and string amplitudes}},
  \href{http://arxiv.org/abs/2405.09608}{{\tt arXiv:2405.09608}}.

\bibitem{Arkani-Hamed:2023jry}
N.~Arkani-Hamed, Q.~Cao, J.~Dong, C.~Figueiredo, and S.~He, {\it
  {Scalar-Scaffolded Gluons and the Combinatorial Origins of Yang-Mills
  Theory}},  \href{http://arxiv.org/abs/2401.00041}{{\tt arXiv:2401.00041}}.

\bibitem{Arkani-Hamed:2023lbd}
N.~Arkani-Hamed, H.~Frost, G.~Salvatori, P.-G. Plamondon, and H.~Thomas, {\it
  {All Loop Scattering as a Counting Problem}},
  \href{http://arxiv.org/abs/2309.15913}{{\tt arXiv:2309.15913}}.

\bibitem{Arkani-Hamed:2023mvg}
N.~Arkani-Hamed, H.~Frost, G.~Salvatori, P.-G. Plamondon, and H.~Thomas, {\it
  {All Loop Scattering For All Multiplicity}},
  \href{http://arxiv.org/abs/2311.09284}{{\tt arXiv:2311.09284}}.

\bibitem{Arkani-Hamed:2024nhp}
N.~Arkani-Hamed, Q.~Cao, J.~Dong, C.~Figueiredo, and S.~He, {\it {Nonlinear
  Sigma model amplitudes to all loop orders are contained in the
  Tr(\ensuremath{\Phi}3) theory}},  {\em Phys. Rev. D} {\bf 110} (2024), no.~6
  065018, [\href{http://arxiv.org/abs/2401.05483}{{\tt arXiv:2401.05483}}].

\bibitem{Arkani-Hamed:2024yvu}
N.~Arkani-Hamed and C.~Figueiredo, {\it {Circles and Triangles, the NLSM and
  Tr($\Phi^3$)}},  \href{http://arxiv.org/abs/2403.04826}{{\tt
  arXiv:2403.04826}}.

\bibitem{Arkani-Hamed:2024tzl}
N.~Arkani-Hamed, Q.~Cao, J.~Dong, C.~Figueiredo, and S.~He, {\it {Surface
  Kinematics and ''The'' Yang-Mills Integrand}},
  \href{http://arxiv.org/abs/2408.11891}{{\tt arXiv:2408.11891}}.

\bibitem{Rodina:2024yfc}
L.~Rodina, {\it {Hidden Zeros Are Equivalent to Enhanced Ultraviolet Scaling,
  and Lead to Unique Amplitudes in Tr(\ensuremath{\phi}3) Theory}},  {\em Phys.
  Rev. Lett.} {\bf 134} (2025), no.~3 031601,
  [\href{http://arxiv.org/abs/2406.04234}{{\tt arXiv:2406.04234}}].

\bibitem{Backus:2025hpn}
J.~V. Backus and L.~Rodina, {\it {Emergence of Unitarity and Locality from
  Hidden Zeros at One-Loop}},  \href{http://arxiv.org/abs/2503.03805}{{\tt
  arXiv:2503.03805}}.

\bibitem{Zhou:2024ddy}
K.~Zhou, {\it {Understanding zeros and splittings of ordered tree amplitudes
  via Feynman diagrams}},  \href{http://arxiv.org/abs/2411.07944}{{\tt
  arXiv:2411.07944}}.

\bibitem{Huang:2025blb}
H.~Huang, Y.~Yang, and K.~Zhou, {\it {Note on hidden zeros and expansions of
  tree-level amplitudes}},  \href{http://arxiv.org/abs/2502.07173}{{\tt
  arXiv:2502.07173}}.

\bibitem{Kawai:1985xq}
H.~Kawai, D.~C. Lewellen, and S.~H.~H. Tye, {\it {A Relation Between Tree
  Amplitudes of Closed and Open Strings}},  {\em Nucl. Phys.} {\bf B269} (1986)
  1--23.

\bibitem{Bern:2008qj}
Z.~Bern, J.~J.~M. Carrasco, and H.~Johansson, {\it New relations for
  gauge-theory amplitudes},  {\em Phys. Rev. D} {\bf 78} (2008) 085011,
  [\href{http://arxiv.org/abs/0805.3993}{{\tt arXiv:0805.3993}}].

\bibitem{Bartsch:2024amu}
C.~Bartsch, T.~V. Brown, K.~Kampf, U.~Oktem, S.~Paranjape, and J.~Trnka, {\it
  {Hidden Amplitude Zeros From Double Copy}},
  \href{http://arxiv.org/abs/2403.10594}{{\tt arXiv:2403.10594}}.

\bibitem{Li:2024qfp}
Y.~Li, D.~Roest, and T.~ter Veldhuis, {\it {Hidden Zeros in Scaffolded General
  Relativity and Exceptional Field Theories}},
  \href{http://arxiv.org/abs/2403.12939}{{\tt arXiv:2403.12939}}.

\bibitem{Cachazo:2013hca}
F.~Cachazo, S.~He, and E.~Y. Yuan, {\it {Scattering of Massless Particles in
  Arbitrary Dimensions}},  {\em Phys.Rev.Lett.} {\bf 113} (2014), no.~17
  171601, [\href{http://arxiv.org/abs/1307.2199}{{\tt arXiv:1307.2199}}].

\bibitem{Cachazo:2013iea}
F.~Cachazo, S.~He, and E.~Y. Yuan, {\it {Scattering of Massless Particles:
  Scalars, Gluons and Gravitons}},  {\em JHEP} {\bf 1407} (2014) 033,
  [\href{http://arxiv.org/abs/1309.0885}{{\tt arXiv:1309.0885}}].

\bibitem{Bern:2019prr}
Z.~Bern, J.~J. Carrasco, M.~Chiodaroli, H.~Johansson, and R.~Roiban, {\it {The
  duality between color and kinematics and its applications}},  {\em J. Phys.
  A} {\bf 57} (2024), no.~33 333002,
  [\href{http://arxiv.org/abs/1909.01358}{{\tt arXiv:1909.01358}}].

\bibitem{Cachazo:2013iaa}
F.~Cachazo, S.~He, and E.~Y. Yuan, {\it {Scattering in Three Dimensions from
  Rational Maps}},  {\em JHEP} {\bf 10} (2013) 141,
  [\href{http://arxiv.org/abs/1306.2962}{{\tt arXiv:1306.2962}}].

\bibitem{Cachazo:2014nsa}
F.~Cachazo, S.~He, and E.~Y. Yuan, {\it {Einstein-Yang-Mills Scattering
  Amplitudes From Scattering Equations}},  {\em JHEP} {\bf 01} (2015) 121,
  [\href{http://arxiv.org/abs/1409.8256}{{\tt arXiv:1409.8256}}].

\bibitem{Cachazo:2014xea}
F.~Cachazo, S.~He, and E.~Y. Yuan, {\it {Scattering Equations and Matrices:
  From Einstein To Yang-Mills, DBI and NLSM}},  {\em JHEP} {\bf 07} (2015) 149,
  [\href{http://arxiv.org/abs/1412.3479}{{\tt arXiv:1412.3479}}].

\bibitem{Azevedo:2017lkz}
T.~Azevedo and O.~T. Engelund, {\it {Ambitwistor formulations of R$^{2}$
  gravity and (DF)$^{2}$ gauge theories}},  {\em JHEP} {\bf 11} (2017) 052,
  [\href{http://arxiv.org/abs/1707.02192}{{\tt arXiv:1707.02192}}].

\bibitem{He:2016iqi}
S.~He and Y.~Zhang, {\it {New Formulas for Amplitudes from Higher-Dimensional
  Operators}},  {\em JHEP} {\bf 02} (2017) 019,
  [\href{http://arxiv.org/abs/1608.08448}{{\tt arXiv:1608.08448}}].

\bibitem{Cachazo:2016njl}
F.~Cachazo, P.~Cha, and S.~Mizera, {\it {Extensions of Theories from Soft
  Limits}},  {\em JHEP} {\bf 06} (2016) 170,
  [\href{http://arxiv.org/abs/1604.03893}{{\tt arXiv:1604.03893}}].

\bibitem{Dong:2024klq}
J.~Dong, X.~Li, and F.~Zhu, {\it {Pions from higher-dimensional gluons: general
  realizations and stringy models}},  {\em JHEP} {\bf 07} (2024) 149,
  [\href{http://arxiv.org/abs/2404.11648}{{\tt arXiv:2404.11648}}].

\bibitem{Cachazo:2013gna}
F.~Cachazo, S.~He, and E.~Y. Yuan, {\it {Scattering equations and
  Kawai-Lewellen-Tye orthogonality}},  {\em Phys. Rev.} {\bf D90} (2014), no.~6
  065001, [\href{http://arxiv.org/abs/1306.6575}{{\tt arXiv:1306.6575}}].

\bibitem{Early:2025ivr}
N.~Early, {\it {The CEGM NLSM}},  \href{http://arxiv.org/abs/2502.08016}{{\tt
  arXiv:2502.08016}}.

\bibitem{Cachazo:2019ngv}
F.~Cachazo, N.~Early, A.~Guevara, and S.~Mizera, {\it {Scattering Equations:
  From Projective Spaces to Tropical Grassmannians}},  {\em JHEP} {\bf 06}
  (2019) 039, [\href{http://arxiv.org/abs/1903.08904}{{\tt arXiv:1903.08904}}].

\bibitem{Britto:2005fq}
R.~Britto, F.~Cachazo, B.~Feng, and E.~Witten, {\it {Direct proof of tree-level
  recursion relation in Yang-Mills theory}},  {\em Phys.Rev.Lett.} {\bf 94}
  (2005) 181602, [\href{http://arxiv.org/abs/hep-th/0501052}{{\tt
  hep-th/0501052}}].

\bibitem{Cachazo:2020uup}
F.~Cachazo and N.~Early, {\it {Minimal Kinematics: An All $k$ and $n$ Peek into
  ${\rm Trop}^+{\rm G}(k,n)$}},  {\em SIGMA} {\bf 17} (2021) 078,
  [\href{http://arxiv.org/abs/2003.07958}{{\tt arXiv:2003.07958}}].

\bibitem{Early:2018zuw}
N.~Early, {\it {Generalized permutohedra in the kinematic space}},  {\em JHEP}
  {\bf 06} (2024) 072, [\href{http://arxiv.org/abs/1804.05460}{{\tt
  arXiv:1804.05460}}].

\bibitem{Early:2024nvf}
N.~Early, A.~Pfister, and B.~Sturmfels, {\it {Minimal Kinematics on
  $\mathcal{M}_{0,n}$}},  \href{http://arxiv.org/abs/2402.03065}{{\tt
  arXiv:2402.03065}}.

\bibitem{Broedel:2012rc}
J.~Broedel and L.~J. Dixon, {\it {Color-kinematics duality and double-copy
  construction for amplitudes from higher-dimension operators}},  {\em JHEP}
  {\bf 10} (2012) 091, [\href{http://arxiv.org/abs/1208.0876}{{\tt
  arXiv:1208.0876}}].

\bibitem{Garozzo:2018uzj}
L.~M. Garozzo, L.~Queimada, and O.~Schlotterer, {\it {Berends-Giele currents in
  Bern-Carrasco-Johansson gauge for $F^3$- and $F^4$-deformed Yang-Mills
  amplitudes}},  {\em JHEP} {\bf 02} (2019) 078,
  [\href{http://arxiv.org/abs/1809.08103}{{\tt arXiv:1809.08103}}].

\bibitem{Azevedo:2018dgo}
T.~Azevedo, M.~Chiodaroli, H.~Johansson, and O.~Schlotterer, {\it {Heterotic
  and bosonic string amplitudes via field theory}},  {\em JHEP} {\bf 10} (2018)
  012, [\href{http://arxiv.org/abs/1803.05452}{{\tt arXiv:1803.05452}}].

\bibitem{He:2018pol}
S.~He, F.~Teng, and Y.~Zhang, {\it {String amplitudes from field-theory
  amplitudes and vice versa}},  {\em Phys. Rev. Lett.} {\bf 122} (2019), no.~21
  211603, [\href{http://arxiv.org/abs/1812.03369}{{\tt arXiv:1812.03369}}].

\bibitem{Mafra:2011nv}
C.~R. Mafra, O.~Schlotterer, and S.~Stieberger, {\it {Complete N-Point
  Superstring Disk Amplitude I. Pure Spinor Computation}},  {\em Nucl. Phys. B}
  {\bf 873} (2013) 419--460, [\href{http://arxiv.org/abs/1106.2645}{{\tt
  arXiv:1106.2645}}].

\bibitem{Stieberger:2009hq}
S.~Stieberger, {\it {Open $\&$ Closed vs. Pure Open String Disk Amplitudes}},
  \href{http://arxiv.org/abs/0907.2211}{{\tt arXiv:0907.2211}}.

\bibitem{He:2019drm}
S.~He, F.~Teng, and Y.~Zhang, {\it {String Correlators: Recursive Expansion,
  Integration-by-Parts and Scattering Equations}},  {\em JHEP} {\bf 09} (2019)
  085, [\href{http://arxiv.org/abs/1907.06041}{{\tt arXiv:1907.06041}}].

\bibitem{Geyer:2015bja}
Y.~Geyer, L.~Mason, R.~Monteiro, and P.~Tourkine, {\it {Loop Integrands for
  Scattering Amplitudes from the Riemann Sphere}},  {\em Phys. Rev. Lett.} {\bf
  115} (2015), no.~12 121603, [\href{http://arxiv.org/abs/1507.00321}{{\tt
  arXiv:1507.00321}}].

\bibitem{Geyer:2015jch}
Y.~Geyer, L.~Mason, R.~Monteiro, and P.~Tourkine, {\it {One-loop amplitudes on
  the Riemann sphere}},  {\em JHEP} {\bf 03} (2016) 114,
  [\href{http://arxiv.org/abs/1511.06315}{{\tt arXiv:1511.06315}}].

\bibitem{Cachazo:2015aol}
F.~Cachazo, S.~He, and E.~Y. Yuan, {\it {One-Loop Corrections from Higher
  Dimensional Tree Amplitudes}},  {\em JHEP} {\bf 08} (2016) 008,
  [\href{http://arxiv.org/abs/1512.05001}{{\tt arXiv:1512.05001}}].

\bibitem{Geyer:2016wjx}
Y.~Geyer, L.~Mason, R.~Monteiro, and P.~Tourkine, {\it {Two-Loop Scattering
  Amplitudes from the Riemann Sphere}},  {\em Phys. Rev. D} {\bf 94} (2016),
  no.~12 125029, [\href{http://arxiv.org/abs/1607.08887}{{\tt
  arXiv:1607.08887}}].

\bibitem{Geyer:2019hnn}
Y.~Geyer, R.~Monteiro, and R.~Stark-Much\~ao, {\it {Two-Loop Scattering
  Amplitudes: Double-Forward Limit and Colour-Kinematics Duality}},  {\em JHEP}
  {\bf 12} (2019) 049, [\href{http://arxiv.org/abs/1908.05221}{{\tt
  arXiv:1908.05221}}].

\bibitem{Geyer:2022cey}
Y.~Geyer and L.~Mason, {\it {The SAGEX review on scattering amplitudes Chapter
  6: Ambitwistor Strings and Amplitudes from the Worldsheet}},  {\em J. Phys.
  A} {\bf 55} (2022), no.~44 443007,
  [\href{http://arxiv.org/abs/2203.13017}{{\tt arXiv:2203.13017}}].

\bibitem{He:2017spx}
S.~He, O.~Schlotterer, and Y.~Zhang, {\it {New BCJ representations for one-loop
  amplitudes in gauge theories and gravity}},  {\em Nucl. Phys.} {\bf B930}
  (2018) 328--383, [\href{http://arxiv.org/abs/1706.00640}{{\tt
  arXiv:1706.00640}}].

\bibitem{Feng:2022wee}
B.~Feng, S.~He, Y.~Zhang, and Y.-Q. Zhang, {\it {One-loop diagrams with
  quadratic propagators from the worldsheet}},  {\em JHEP} {\bf 08} (2022) 240,
  [\href{http://arxiv.org/abs/2204.13659}{{\tt arXiv:2204.13659}}].

\bibitem{Dong:2023stt}
J.~Dong, Y.-Q. Zhang, and Y.~Zhang, {\it {One-loop Bern-Carrasco-Johansson
  numerators on quadratic propagators from the worldsheet}},  {\em Phys. Rev.
  D} {\bf 109} (2024), no.~10 L101905,
  [\href{http://arxiv.org/abs/2312.01580}{{\tt arXiv:2312.01580}}].

\bibitem{De:2024wsy}
S.~De, A.~Pokraka, M.~Skowronek, M.~Spradlin, and A.~Volovich, {\it
  {Surfaceology for colored Yukawa theory}},  {\em JHEP} {\bf 09} (2024) 160,
  [\href{http://arxiv.org/abs/2406.04411}{{\tt arXiv:2406.04411}}].

\end{thebibliography}\endgroup

\end{document}